\documentclass[reprint,aps,longbibliography,nofootinbib,floatfix,nobalancelastpage,superscriptaddress]{revtex4-2}
\usepackage{amsmath}
\usepackage{amsfonts}
\usepackage{amssymb}
\usepackage{mathtools}
\usepackage{graphicx}
\usepackage[dvipsnames]{xcolor}
\usepackage[colorlinks=True,citecolor=gray,linkcolor=myRed,urlcolor=gray]{hyperref}
\usepackage{hypcap}
\usepackage{yfonts}
\usepackage{bm}
\usepackage[normalem]{ulem}
\usepackage{dsfont}
\usepackage{braket}

\newcommand\eq[1]{\begin{align}#1\end{align}}

\newcommand{\pd}{{\phantom\dagger}}

\newcommand{\ltwo}{\mathcal{L}_2}

\newcommand{\lmean}{\mathcal{L}_{q}^{\mathrm{mean}}}
\newcommand{\ltyp}{\mathcal{L}_{q}^{\mathrm{typ}}}
\newcommand{\Lq}{\mathcal{L}_{q}}
\newcommand{\Lqp}{\mathcal{L}_{q}^{\pd}}
\newcommand{\ltwomean}{\mathcal{L}_{2}^{\mathrm{mean}}}
\newcommand{\ltwotyp}{\mathcal{L}_{2}^{\mathrm{typ}}}

\newcommand{\vc}{V_{\mathrm{c}}}

\newcommand{\eipd}{\mathcal{E}_{i}^{\pd}}
\newcommand{\er}{\mathcal{E}_{r}}
\newcommand{\erpd}{\mathcal{E}_{r}^{\pd}}

\newcommand{\ttil}{\tilde{t}}
\newcommand{\hh}{\hat{H}}
\newcommand{\hhcal}{\hat{\mathcal{H}}}
\newcommand{\hhgam}{\hat{H}_{\Gamma}}

\definecolor{myBlue}{RGB}{31,119,180}
\definecolor{myOrange}{RGB}{255,127,14}
\definecolor{myGreen}{RGB}{44,160,44}
\definecolor{myRed}{RGB}{214,39,40}
\definecolor{myPurple}{RGB}{148,103,189}

\makeatletter
\def\p@figure{\color{myBlue}}
\def\p@equation{\color{myRed}}
\makeatother

\begin{document}

\title{Multifractality in high-dimensional graphs induced by correlated radial disorder}

\author{David E. Logan}
\email{david.logan@chem.ox.ac.uk}
\affiliation{University of Oxford, Physical and Theoretical Chemistry Laboratory, South Parks Road, Oxford OX13QZ, United Kingdom}

\author{Sthitadhi Roy}
\email{sthitadhi.roy@icts.res.in}
\affiliation{International Centre for Theoretical Sciences, Tata Institute of Fundamental Research, Bengaluru 560089, India}

\begin{abstract}
We introduce a class of models containing robust and analytically demonstrable multifractality induced by disorder correlations. Specifically, we investigate the statistics of eigenstates of disordered tight-binding models on two classes of rooted, high-dimensional graphs -- trees and hypercubes -- with a form of strong disorder correlations we term `radial disorder'. In this model, site energies on all sites equidistant from a chosen root are identical, while  those at different distances are independent random variables (or their analogue for a  deterministic but incommensurate potential, a case of which is also considered). Analytical arguments, supplemented by numerical results, are used to establish that this setting hosts robust and unusual multifractal states. The distribution of multifractality, as encoded in the inverse participation ratios (IPRs), is shown to be exceptionally broad. This leads to a qualitative difference in scaling with system size between the mean and typical IPRs, with the latter the appropriate quantity to characterise the multifractality. The existence of this multifractality is shown to be underpinned by an emergent fragmentation of the graphs into effective one-dimensional chains, which themselves exhibit conventional Anderson localisation. The interplay between the exponential localisation of states on these chains, and the exponential growth of the number of sites with distance from the root, is the origin of the observed multifractality. 
\end{abstract}

\maketitle
% \tableofcontents

%\twocolumngrid
%\tableofcontents
%\twocolumngrid

%%%%%%%%%%%%%%%%%%%%%%%%%%%%%%%%%%%%%%%%%%%%%%%%%%%%%%%%%%%%%%%%%%

\section{Introduction}

The anatomy of eigenstates of a single quantum particle hopping on disordered, high-dimensional graphs is central to understanding a range of fundamental phenomena in condensed matter and statistical physics. On the one hand, it
lies at the heart of the theory of one-body Anderson localisation~\cite{anderson1958absence,abou-chacra1973self,abrahams1979scaling,lee1985disordered,evers2008anderson}, and continues to reveal rich insights into the underlying physics 
some seven decades after its inception~\cite{garciamata2017scaling,garciamata2020two,garciamata2022critical,vanoni2024renormalisation,altshuler2024renormalization}.
On the other hand, the issue has gained renewed significance with the advent of many-body localisation (MBL), which addresses the fate of Anderson localisation in the presence of interactions and constitutes a novel form of ergodicity breaking in quantum many-body systems~\cite{basko2006metal,gornyi2005interacting,oganesyan2007localisation,nandkishore2015many,abanin2017recent,abanin2019colloquium,SierantReview2025}.

The latter reflects the fact that the problem of MBL in archetypal models, such as disordered spin- or fermionic chains, can be mapped exactly onto that of a fictitious single particle on the high-dimensional Fock-space graph of the 
model~\cite{logan1990quantum,altshuler1997quasiparticle,MonthusGarel2010PRB,pietracaprina2016forward,biroli2017delocalized,altland2017field,logan2019many,roy2020fock,tikhonov2021eigenstate,detomasi2020rare,roy2021fockspace,tarzia2020manybody,sutradhar2022scaling,roy2023anatomy,roy2023diagnostics,herre2023ergodicity,schiro2020toy,scoquart2024role,ghosh2024scaling} (see Refs.~\cite{tikhonov2021anderson,roy2024fock} for reviews and further references); 
e.g.\ the Fock-space graph of the disordered tilted-field Ising model with
$L$ spins is an $L$-dimensional hypercube~\cite{roy2024fock}.
Although the problems of one- and many-body localisation are inherently different,
much cross-pollination of ideas between them has occurred, which has 
enriched our understanding of MBL~\cite{deluca2013ergodicity,pietracaprina2016forward,biroli2017delocalised,biroli2020anomalous,tarzia2020manybody,roy2020localisation,roy2021fockspace,garciamata2022critical,biroli2024largedeviation}. Importantly, this has also pinpointed key features of Fock-space graphs, and the structure of eigenstates thereon, which distinguish  MBL from conventional Anderson localisation on high-dimensional graphs.

One of the most significant such features is the presence of correlations in the effective disorder on the Fock-space graph~\cite{roy2020fock}.
Canonical models of MBL show `maximally correlated disorder' on the Fock-space graph, which was argued to be a key requirement for the stability of MBL~\cite{roy2020fock}. By this is meant that the disordered site energies on the Fock-space graph, rescaled appropriately to admit a well-defined thermodynamic limit, are completely slaved to each other for sites separated by a finite (Hamming) distance in the thermodynamic limit. The naive expectation would be that 
correlations leading to increased similarity between nearby site energies
would weaken localisation, yet such correlations act to stabilise the MBL phase.
The essential message here is that strong correlations in the effective disorder in high-dimensional graphs can render the problem fundamentally different from that of conventional Anderson localisation, and as such provides fruitful ground for further exploration.

From the viewpoint of eigenstate structure, an important understanding  is that 
MBL eigenstates are multifractal on the Fock-space graph~\cite{deluca2013ergodicity,luitz2015many,mace2019multifractal,detomasi2020rare,roy2021fockspace,tikhonov2021eigenstate,sutradhar2022scaling,roy2023diagnostics,roy2023anatomy}.  Wavefunctions of a single particle, fictitious or otherwise, can be broadly classified into three kinds: delocalised, localised, and multifractal. 
Delocalised states occupy a finite fraction of sites on the graph, whereas localised ones occupy only a finite number of sites, and hence a vanishing fraction in the thermodynamic limit. Multifractal states are essentially intermediate in character, where  wavefunctions occupy a number of sites which 
grows unboundedly but a fraction which vanishes in the thermodynamic limit.
In genuine single-particle settings, multifractality is somewhat special as it typically appears at criticality,\footnote{Of course, we implicitly assume systems with locality and exclude effective random matrix models with long (infinite)-ranged hopping, such as power-law banded matrices, the Rosenzweig-Porter model or random Erd\"os-R\'enyi graphs, which do host robust multifractality~\cite{rosezweig1960repulsion,mirlin1996transition,kravtsov2015random,evers2008anderson,CugliandoloPRB2024}.} such as Anderson transitions and integer quantum Hall plateau transitions~\cite{chalker1988scaling,chalker1990scaling,ludwig1994integer,huckenstein1995scaling,evers2001multifractality,mirlin2000multifractality,evers2008anderson,tikhonov2019critical}.
Quite tellingly though, correlations in the disorder or hopping matrix elements in such settings have turned out to be a source of robust multifractality~\cite{nosov2018correlation,roy2020localisation,duthie2022anomalous}.
On the other hand, in the many-body setting on the Fock space, multifractality is generically robust and often taken as a defining signature of an ergodicity-broken phase~\cite{deluca2013ergodicity,luitz2015many,mace2019multifractal,detomasi2020rare,roy2021fockspace,tikhonov2021eigenstate,sutradhar2022scaling,roy2023diagnostics,roy2023anatomy}. 

These considerations illustrate that disorder correlations and the multifractality of wavefunctions -- features that arise naturally in many-body systems when represented on the Fock space -- are central to a wealth of phenomena in disordered systems. This prompts a natural 
question: can one construct models of disorder correlations on high-dimensional graphs that not only exhibit robust multifractality but also yield 
analytical insights into the mechanisms underpinning its emergence? This is 
the question we address here. We show explicitly that two classes of high-dimensional graphs -- trees and hypercubes -- can host robust multifractal eigenstates when endowed with a specific class of disorder correlations, which we refer to as {\it radial disorder} for reasons that will become clear shortly.
We demonstrate that radial disorder induces a hierarchical structure on the graph and leads to its fragmentation into an emergent family of effective one-dimensional chains. The interplay between Anderson localisation on these chains, and the structure that relates them back to the original graph, lies at the heart of the resultant multifractality. 
These are the central qualitative results of the present work.

%%%%%%%%%%%%%%%%%%%%%%%%%%%%%%%%%%%%%%%%%%%%%%%%%%%%%%%%%%%%%%%%

\subsection{Tree and hypercube graphs}\label{section:Graphs}

\begin{figure} [!t]
\includegraphics[width=\linewidth]{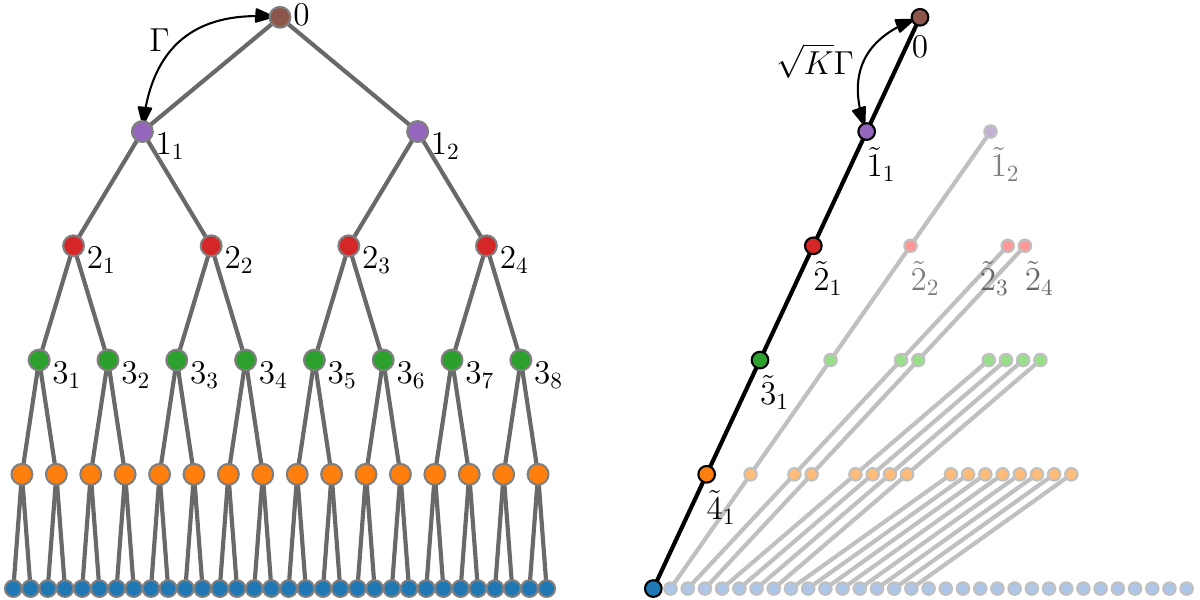}
\caption{\label{fig:treegraph}  
\emph{Left panel}: Rooted Cayley tree, illustrated for connectivity $K=2$. 
 The graph has $L+1$ rows, with $N_{r}=K^{r}$ sites/vertices on row $r$
(and $r=0$ the root site). Hoppings (bonds/edges) are as indicated, each of strength $\Gamma$. With radial disorder, all sites on a given row have the same site energies
(indicated by a common colour); while sites on different rows have distinct site energies.
\emph{Right panel}: Corresponding tree under the symmetrised basis transformation specified in Sec.\ \ref{section:CTs}. Hoppings are again as indicated, all being
of strength $\sqrt{K}\Gamma$. The tree has fragmented into a disconnected set of 1d chains. Full discussion in text (see Sec.\ \ref{section:Puretree}).
}
\end{figure}

Before proceeding, it is expedient to summarise the essential properties of the two graphs considered, and lay out some definitions. For the tree graph, we consider a rooted Cayley tree with branching number $K$, containing $L$ generations and hence $L+1$ rows denoted  $r=0,1,\cdots L$ (with $r=0$ the root site). The number of sites on row $r$ is  $N_{r}=K^{r}$, and each such is a Hamming distance of $r$ from the root. The total number of sites/vertices is $N=\sum_{r=0}^{L}N_{r}\propto K^{L}$, exponentially large in $L$. Since 
sites on the graph are naturally arranged in rows, we denote the sites on
any given row $r$ as $r_1,r_2,\cdots r_{N_r}$.
The tree is inherently loopless (cycle-free), and just a single direct hopping path  connects any pair of vertices. Fig.\ \ref{fig:treegraph} (left panel) illustrates a rooted tree with connectivity  $K=2$, and sites on each row labelled as indicated.

\begin{figure}[!t]
\includegraphics[width=0.9\linewidth]{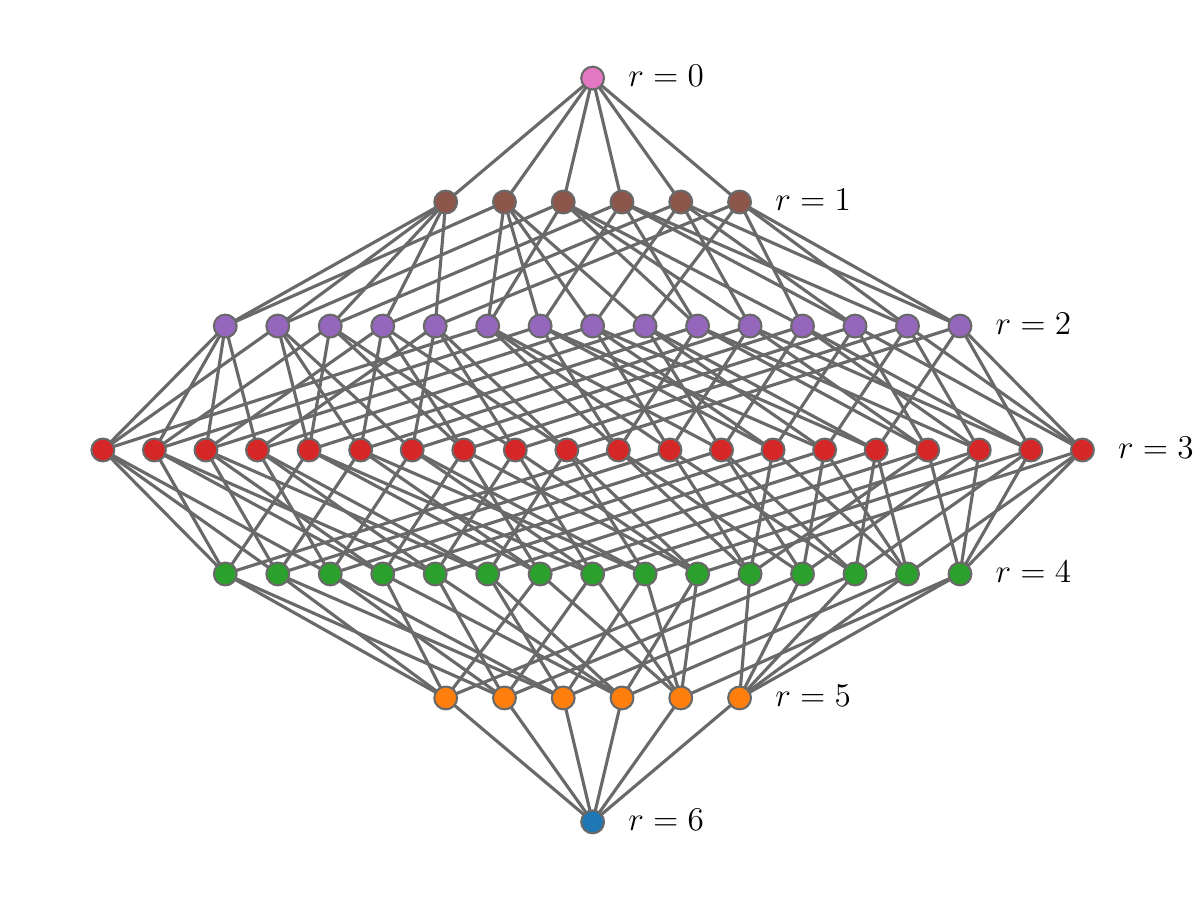}
\caption{$L$-dimensional hypercube graph (illustrated for $L=6$). 
The graph has $L+1$ rows, with $N_{r}=\binom{L}{r}$ sites/vertices on row $r$
($r=0$ is the apex site). 
Under $\hh_{\Gamma}$, hoppings (of strength $\Gamma$) connect only sites on adjacent rows. Each site on row $r$ is connected under $\hh_{\Gamma}$  to $L$ others: to $r$ distinct sites on row $(r-1)$,  and to $(L-r)$  sites on row $(r+1)$.
With radial disorder, all sites on a given row have the same site energies
(indicated by a common colour); while sites on different rows have distinct site energies. For full discussion  see Sec.\ \ref{section:hcgraph}.
}
\label{fig:hypercubegraph}
\end{figure}

The hypercube graph, of dimension $L$, is also an $L$-regular graph, with 
$N=2^L$ sites. Fig.\ \ref{fig:hypercubegraph} illustrates one with $L=6$.
Note that the dimensionality of the hypercube is also the diameter of the graph 
(the maximum value of the shortest path length between any pair of sites on the graph) or, equivalently, the number of generations in the graph. Assigning any one site to be the apical site of the hypercube, with $r=0$, the 
remaining sites can again be arranged in rows $r \in [1,L]$, with any site on row $r$ lying a Hamming distance $r$ from the apex. Each row now contains 
$N_{r}=\binom{L}{r}$ sites, whence a total of $N= 2^{L}$ sites.
Any given site on row $r$ is connected by $r$ edges to $r$ distinct sites in row $r-1$, and by $L-r$ edges to $L-r$ distinct sites in  row $r+1$. In contrast to the tree, any site on row $r$ is connected to the apex site by $r!$ distinct paths. As with the tree, we denote the set of sites on any given row $r$ as $\{r_n\}$ with $n=1,2,\cdots N_r$.

Given the above topology of the tree and hypercube graphs, we will consider tight-binding models (TBM) defined on them. 
They are described the Hamiltonian
\begin{equation}
\label{eq:HTBM}
\hh =\sum_{i} \eipd |i\rangle\langle i| +\Gamma {\sum_{(i,j)}}|i\rangle\langle j|
~\equiv ~\hh_{0}^{\pd}+\hh_{\Gamma}^{\pd}\,,
\end{equation}
where nearest neighbour pairs of sites (denoted by $(i,j)$) have a 
constant hopping matrix element $\Gamma$, and the  site energies $\{{\cal E}_i\}$are disordered. Note that the topology of the graphs guarantees that for a pair of sites to be nearest neighbours, they must reside on consecutive rows.

Two remarks are in order here. In the limit where all the $N$
site-energies ${\cal E}_i$ are i.i.d.\ random variables, the model on the tree reduces to the extensively studied Anderson localisation problem on a tree~\cite{abou-chacra1973self,zirnbauer1986localisation,chalker1990anderson,mirlin1991localisation,derrida1993anderson,monthus2008anderson,monthus2011anderson,Tikhonov2016CayleyTree,kravtsov2018nonergodic,savitz2019anderson,parisi2020anderson} which hosts a localisation transition at a finite critical disorder strength and, more importantly in the present context, a fully delocalised phase at sufficiently weak but non-vanishing disorder.
In the same uncorrelated limit, the problem on the hypercube~\cite{roy2020fock} is essentially the quantum random energy model~\cite{derrida1980random,laumann2014many,baldwin2016manybody}, which is ergodic except at the edges of the many-body spectrum for all disorder strengths. By contrast, with radially correlated disorder, 
 we find the emergence of multifractality for any non-zero disorder strength for both the tree and  hypercube graphs; and throughout the range of parameters in the incommensurate case which corresponds to the localised phase of the 1d 
Aubry-Andr\'e model.
This attests to the emergent multifractality being induced by the radial disorder.

%%%%%%%%%%%%%%%%%%%%%%%%%%%%%%%%%%%%%%%%%%%%%%%%%%%%%%%%%%%

\subsection{Radial disorder}

Let us now define concretely the notion of radial disorder. 
The notation for labelling the set of sites on row $r$ as $\{r_n\}$ implies that the diagonal part of the Hamiltonian in Eq.~\ref{eq:HTBM}, for both the tree and the hypercube, can be expressed as
\begin{equation}
\label{eq:H0rowresolve}
\hh_{0}^{\pd} ~=~ \sum_{r=0}^{L}\hh_{0,r}^{\pd}
~=~\sum_{r=0}^{L}\sum_{n=1}^{N_{r}} \mathcal{E}_{r_{n}}^{\pd}|r_{n}^{\pd}\rangle\langle r_{n}^{\pd}|,
\end{equation}
with $N_{r}=K^{r}$ for the tree and $N_{r}=\binom{L}{r}$ for the hypercube.
Radial disorder refers to a particular form of correlation in the site energies 
$\{{\cal E}_{r_n}\}$, whereby the site energies on all sites of a given row are the same.  Formally,
\eq{\label{eq:radcor-def}
{\cal E}_{r_n}^{\pd} = {\cal E}_r^{\pd}~\forall n \in [1,N_r]\,,
}
where the set $\{{\cal E}_r\}$ are treated as 
independent, quenched random variables.
In addition we will also consider (Sec.\ \ref{section:AAtree}) the case where, instead of quenched disorder, the $\{\er \}$ are given by a deterministic but incommensurate (in $r$) potential, as per the 
Aubry-Andr\'e model~\cite{aubry1980analyticity}.

%%%%%%%%%%%%%%%%%%%%%%%%%%%%%%%%%%%%%%%%%%%%%%%%%%%%%%%%%%%%%%%%%%

\subsection{Measures of multifractality}

It is prudent at this stage to make a general remark about the multifractality 
found in this work.  Multifractality of wavefunctions is most commonly studied using the scaling of the inverse participation ratio (IPR) with the system size, $N$. The $q^{\rm th}$-IPR for a state $\ket{\psi}=\sum_{i=1}^N \psi_i\ket{i}$ is defined as 
\eq{
{\cal L}_q^{\pd} = \sum_{i=1}^N |\psi_i^{\pd}|^{2q}\,.
}
In the context of disordered systems, it is quite common to consider the average IPR, ${\cal L}_q^{\rm mean} = \braket{{\cal L}_q}$ where  $\braket{\cdots}$ denotes an average over disorder realisations. The multifractal exponent is then
defined by the scaling of ${\cal L}_q^{\rm mean}\sim N^{-\tau_q^{\rm mean}}$ with $N$.
An implicit assumption here, one that is often justified, is that the IPRs have a 
well-behaved distribution.

However, as will be shown explicitly, this is not the case for the models considered in this work: the IPR distributions are found to be very broad
(with a mean $\lmean$ that is not in fact remotely representative of the distribution).
As a result, it is the typical IPRs, defined via the geometric mean
\eq{
{\cal L}_q^{\rm typ} = \exp\left[\braket{\ln {\cal L}_q^{\pd}}\right]\,,
}
which are 
the relevant quantities to study; with multifractal exponents defined
via the scaling ${\cal L}_q^{\rm typ}\sim N^{-\tau_q}$.

%%%%%%%%%%%%%%%%%%%%%%%%%%%%%%%%%%%%%%%%%%%%%%%%%%%%%%%%%%%%%%%%%%
 
\subsection*{Organisation of the paper}

The remainder of the paper is organised as follows. 
In Sec.~\ref{section:CTs} we present in detail the case of tree graphs. The fragmentation of the graph, understood most easily for the disorder-free case, is discussed in Sec.~\ref{section:Puretree}, and the persistence of fragmentation in the presence of radial disorder is shown in Sec.~\ref{section:radialtree};  the resultant connection to a Krylov basis is also 
pointed out.  Analytical results for the multifractal statistics, based on the 
interplay between localisation of  states on the effective 1d chain and the hierarchical nature of the original graph, are obtained in 
Sec.~\ref{section:analyticstree}; while numerical calculations from exact diagonalisation (ED), which corroborate these results, are  given in 
Sec.~\ref{section:TreeNumerics}.
The case of the incommensurate, Aubry-Andr\'e-like potential for ${\cal E}_r$ on the tree graph comprises Sec.~\ref{section:AAtree}. Analytical considerations in this context are discussed in Sec.~\ref{section:AACalcs}, followed by 
ED results in Sec.~\ref{section:AAnumerics}.

Sec.~\ref{section:hcgraph} is dedicated to the hypercube graph.
Despite its obvious difference in topology compared to the Cayley tree,
we show explicitly that the graph again fragments, first for the disorder-free 
case and then with radial disorder. Using this, analytical results for the multifractal statistics are presented in Sec.~\ref{section:HypercubeCalcs}, whereas Sec.~\ref{section:Numericshyper} contains numerical corroborations from ED. We conclude with a summary and some remarks in Sec.~\ref{sec:conclusion}.

%%%%%%%%%%%%%%%%%%%%%%%%%%%%%%%%%%%%%%%%%%%%%%%%%%%%%%%%%%%%

\section{Cayley trees}
\label{section:CTs}

We turn now to Cayley trees. 
In considering such, of natural interest are properties which involve the root site, such as the nature of the eigenstates in which it participates, or the return probability $P_{0}(t)$. Before considering the radial disorder, let us look first at the case $\hh_{0} \equiv 0$ -- where the site  energies are all the same, so that $\hh \equiv \hh_{\Gamma}$ involves solely the hopping parts of the TBM. This may seem (and is) simple, but it is also revealing.

%%%%%%%%%%%%%%%%%%%%%%%%%%%%%%%%%%%%%%%%%%%%%%%%%%%%%%%

\subsection{Disorder-free tree, $\hat{H} \equiv \hat{H}_{\Gamma}$}
\label{section:Puretree}

All relevant points we want to make can be illustrated by considering a rooted tree of connectivity $K=2$, shown in Fig.\ \ref{fig:treegraph} (left).
From the basis states in each row/generation, alternative orthonormal sets 
can be constructed, starting from row $1$. For this row, define
\begin{equation}
\label{eq:1tilorbs}
|\tilde{1}_{1}^{\pd}\rangle ~=~\tfrac{1}{\sqrt{2}}|1_{1}^{\pd}+1_{2}^{\pd}\rangle,
~~~~
|\tilde{1}_{2}^{\pd}\rangle ~=~\tfrac{1}{\sqrt{2}}|1_{1}^{\pd}-1_{2}^{\pd}\rangle
\end{equation}
(with notation $|1_{1}+1_{2}\rangle \equiv|1_{1}\rangle+|1_{2}\rangle$ etc).
The basic idea is to start from the definition $|\tilde{1}_{1}\rangle$ of
the totally symmetric combination of orbitals on row $1$, because it is \emph{solely}
this combination to which the root orbital couples under $\hh_{\Gamma}$; with
$|\tilde{1}_{2}\rangle$ then taken to be orthogonal to $|\tilde{1}_{1}\rangle$.
From Eq.\ \ref{eq:1tilorbs} the matrix element of $\hh_{\Gamma}$ coupling the 
$|0\rangle$ and $|\tilde{1}_{1}\rangle$ orbitals is clearly
$\langle 0|\hh|\tilde{1}_{1}\rangle =\tfrac{1}{\sqrt{2}}\langle 0|\hh_{\Gamma}|1_{1}+1_{2}\rangle =\sqrt{2}\Gamma$.
The generalisation of the totally symmetric $|\tilde{1}_{1}\rangle$ to arbitrary connectivity is 
\begin{equation}
\label{eq:me2}
|\tilde{1}_{1}^{\pd}\rangle ~=~\tfrac{1}{\sqrt{K}}\sum_{n=1}^{K}| 1_{n}^{\pd}\rangle 
~~\implies~~
\langle 0|\hh|\tilde{1}_{1}^{\pd}\rangle ~=~\sqrt{K} \Gamma
\end{equation}
with corresponding matrix element $\sqrt{K}\Gamma$.
For row $r=2$ (Fig.\ \ref{fig:treegraph}) one can construct an orthonormal set of 
$K^{2}=4$ orbitals $|\tilde{2}_{n}\rangle$,
\begin{equation}
\label{eq:2tilorbs}
\begin{split}
|\tilde{2}_{1}^{\pd}\rangle ~=&~ \tfrac{1}{2}| 2_{1}^{\pd}+2_{2}^{\pd}+2_{3}^{\pd}+2_{4}^{\pd}\rangle
\\
|\tilde{2}_{2}^{\pd}\rangle ~=&~\tfrac{1}{2}| 2_{1}^{\pd}+2_{2}^{\pd}-2_{3}^{\pd}-2_{4}^{\pd}\rangle
\\
|\tilde{2}_{3}^{\pd}\rangle ~=&~\tfrac{1}{2}| 2_{1}^{\pd}-2_{2}^{\pd}+2_{3}^{\pd}-2_{4}^{\pd}\rangle
\\
|\tilde{2}_{4}^{\pd}\rangle ~=&~ \tfrac{1}{2}| 2_{1}^{\pd}-2_{2}^{\pd}-2_{3}^{\pd}+2_{4}^{\pd}\rangle .
\end{split}
\end{equation}
Once again, $|\tilde{2}_{1}\rangle$ is the totally symmetric combination of row-$2$ orbitals; and it is this combination alone which couples to the $|\tilde{1}_{1}\rangle$ orbital under $\hh =\hhgam$, with matrix element 
$\langle \tilde{1}_{1}|\hhgam|\tilde{2}_{1}\rangle =\sqrt{2}\Gamma$ again.
Similarly, it is solely the $|\tilde{2}_{2}\rangle$ orbital to which $|\tilde{1}_{2}\rangle$ (Eq.\ \ref{eq:1tilorbs}) couples under $\hhgam$, again with 
$\langle \tilde{1}_{2}|\hhgam|\tilde{2}_{2}\rangle =\sqrt{2}\Gamma$. By contrast, neither $|\tilde{2}_{3}\rangle$ nor $|\tilde{2}_{4}\rangle$ couple under $\hhgam$ to the row-$1$ orbitals. The process continues in an obvious manner. Row $r=3$ 
(Fig.\ \ref{fig:treegraph}) has an orthonormal set of  $K^{3}=8$ orbitals, four of which are 
\begin{equation}
\label{eq:3tilorbs}
\begin{split}
|\tilde{3}_{1}^{\pd}\rangle =& \tfrac{1}{2\sqrt{2}}|3_{1}^{\pd}+3_{2}^{\pd}+3_{3}^{\pd}+3_{4}^{\pd}
+3_{5}^{\pd}+3_{6}^{\pd}+3_{7}^{\pd}+3_{8}^{\pd}\rangle
\\
|\tilde{3}_{2}^{\pd}\rangle =& \tfrac{1}{2\sqrt{2}}|3_{1}^{\pd}+3_{2}^{\pd}+3_{3}^{\pd}+3_{4}^{\pd}
-3_{5}^{\pd}-3_{6}^{\pd}-3_{7}^{\pd}-3_{8}^{\pd}\rangle
\\
|\tilde{3}_{3}^{\pd}\rangle =& \tfrac{1}{2\sqrt{2}}|3_{1}^{\pd}+3_{2}^{\pd}-3_{3}^{\pd}-3_{4}^{\pd}
+3_{5}^{\pd}+3_{6}^{\pd}-3_{7}^{\pd}-3_{8}^{\pd}\rangle
\\
|\tilde{3}_{4}^{\pd}\rangle =& \tfrac{1}{2\sqrt{2}}|3_{1}^{\pd}+3_{2}^{\pd}-3_{3}^{\pd}-3_{4}^{\pd}
-3_{5}^{\pd}-3_{6}^{\pd}+3_{7}^{\pd}+3_{8}^{\pd}\rangle.
\end{split}
\end{equation}
The totally symmetric combination $|\tilde{3}_{1}\rangle$ couples under $\hhgam$ solely to the totally symmetric $|\tilde{2}_{1}\rangle$ orbital of row $2$.
Similarly, $|\tilde{3}_{n}\rangle$ for given $n=2,3,4$ is the only combination which couples to  $|\tilde{2}_{n}\rangle$, again with 
$\langle\tilde{2}_{n}|\hhgam|\tilde{3}_{n}\rangle =\sqrt{2}\Gamma$. By contrast, the 
$|\tilde{3}_{n}\rangle$'s with $n=5-8$ (not given here) do not couple under $\hhgam$ to any of the row-$2$ orbitals; and if this was the terminal generation ($L$) of the tree, they would be completely unconnected under $\hhgam$.

Fig.\ \ref{fig:treegraph} (right panel) shows the resultant connections under $\hhgam$ between the transformed basis states. And this generalises in an obvious way
to arbitrary $K$, all non-zero matrix elements under $\hhgam$ being $\sqrt{K}\Gamma$ (as e.g.\  in Eq.\ \ref{eq:me2}).

With this transformation, the problem has clearly fragmented into a disconnected set of  1d chains, whence knowledge of the pure 1d case ($K=1$) suffices to understand the behaviour of a tree with arbitrary $K$. In particular, the sector containing the $0$-orbital is a 1d chain in terms of the set $\{|\tilde{r}_{1}\rangle \}$ of totally
symmetric basis orbitals associated with any row $r$ ($\in [0,L]$) of the tree,
\begin{equation}
\label{eq:mtil1basis}
|\tilde{r}_{1}^{\pd}\rangle ~=~ \frac{1}{\sqrt{N_{r}}} \sum_{n=1}^{N_{r}^{\pd}} |r_{n}^{\pd}\rangle
\end{equation}
(with $N_{r}=K^{r}$, and notation  
$|\tilde{0}_{1}\rangle \equiv |0_{1}\rangle \equiv |0\rangle$ for $r=0$);
but with a nearest neighbour hopping matrix element which is $\sqrt{K}\Gamma$ rather
than $\Gamma$ itself. The associated separable part of the Hamiltonian, denoted  
$\hhcal_{\Gamma}$, is 
\begin{equation}
\label{eq:Hgamma0orb}
\hhcal_{\Gamma}^{\pd} 
~=~\sqrt{K}\Gamma \sum_{r=1}^{L} \Big(
\big|\widetilde{(r-1)}_{1}^{\pd}\big\rangle\big\langle \tilde{r}_{1}^{\pd}\big| ~+~\mathrm{h.c.} \Big).
\end{equation}
So e.g.\ the return probability $P_{0}(t)$ is consequently of form 
$P_{0}(t) =F(\sqrt{K}\Gamma t)$, with $F(\Gamma t)$ the return probability for a 1d chain with $K=1$. From this, on calculating the underlying root-site propagator 
$G_{00}(t)$ (such that $P_{0}(t)=|G_{00}(t)|^{2}$), one obtains 
straightforwardly the exact return probability for $L\to \infty$,
 \begin{equation}
\label{eq:returnprob}
P_{0}^{\pd}(\tilde{t}) ~=~ \Big[\frac{1}{\tilde{t}}J_{1}^{\pd}(2\tilde{t})\Big]^{2}
~\overset{\tilde{t}\gg 1}{\sim}~\frac{1}{\pi\tilde{t}^{3}}
\mathrm{cos}^{2}\big[2\tilde{t}-\tfrac{3\pi}{4}\big]
\end{equation}
with $\tilde{t}=\sqrt{K}\Gamma t$ (and $J_{1}(x)$ a first-kind Bessel function).
As expected physically, $P_{0}(\ttil)$ vanishes as $\tilde{t}\to \infty$; 
and does so with a power-law decay $\propto \tilde{t}^{-3}$ superimposed on an oscillatory background. Generalising this by the same means 
to the probability $P_{r}(\ttil)$ that the system will be found on row $r$ of the chain (with $P_{r}(\ttil =0) =\delta_{r,0}$), gives
\begin{equation}
\label{eq:Prresult}
P_{r}^{\pd}(\tilde{t})~=~
\left[\frac{(1+r)}{\tilde{t}}J_{1+r}^{\pd}(2\tilde{t})\right]^{2} 
~~~~:~\tilde{t}=\sqrt{K}\Gamma t,
\end{equation}
recovering Eq.\ \ref{eq:returnprob} for $r=0$.
Note also that in terms of the original tree basis $\{|r_{n}\rangle\}$, and
due to the above fragmentation, 
$P_{r}(\ttil)\equiv \sum_{n=1}^{K^{r}}P_{r_{n}}(\ttil)$ is 
equivalently the total probability that the system will be found on 
any of the $N_{r}$ sites$\{r_{n}\}$ on row $r$ of the tree.

The system's mean position $\overline{r}(\ttil)$ on the chain/tree,
following initiation on the root site at $\ttil=0$, is simply
\begin{equation}
\label{eq:ballistic}
\overline{r}(\tilde{t})~=~\sum_{r=0}^{\infty}r~P_{r}^{\pd}(\tilde{t})
~\overset{\tilde{t}\gtrsim 1}{\sim}~ v \tilde{t}.
\end{equation}
Using Eq.\ \ref{eq:Prresult}, this gives $\overline{r}(\ttil)\sim \ttil^{2}$ for short times $\ttil \lesssim 1$; but for $\ttil \gtrsim 1$
it rapidly crosses over to the ballistic transport $\overline{r}(\ttil)=v \ttil$,
with dimensionless velocity $v =1.70..$.

%%%%%%%%%%%%%%%%%%%%%%%%%%%%%%%%%%%%%%%%%%%%%%%%%%%%%%%%%%%%

\subsubsection{Connection to Krylov basis}
\label{sec:Krylov1}

The findings above are closely  connected to the notion of a Krylov basis, 
and the associated Krylov 
Hamiltonian~\cite{bala2022quantum,VBalasubEtAlPRD2023,gautam2024spread,NANDY2025}.
Quite generally, to define a Krylov space requires an initial state $|k_{0}\rangle$ and an operator, here the Hamiltonian $\hh$. The space is constructed by repeated application of $\hh$ on $|k_{0}\rangle$, to generate 
$|\tilde{k}_{r}\rangle = \hh^{r}|k_{0}\rangle$. Applying the Gram-Schmidt procedure to mutually orthonormalise all $|\tilde{k}_{r}\rangle$ then generates an orthonormal basis $|k_{r}\rangle $,  the Krylov basis 
$\mathcal{K}(\hh,|k_{0}\rangle)=\mathrm{span}\{ |k_{r}\rangle: r=0,1,2,\cdots\}$.
Starting from $|k_{0}\rangle$, the $\{|k_{r}\rangle\}$ are obtained from the
linear recurrence 
\begin{equation}
\label{eq:Krylovrecurr}
b_{r+1}^{\pd}|k_{r+1}^{\pd}\rangle ~=~
\big(\hh -a_{r}^{\pd}\big)|k_{r}^{\pd}\rangle 
- b_{r}^{\pd}|k_{r-1}^{\pd}\rangle
\end{equation}
with $b_{0}=0$. The Hamiltonian in the Krylov basis -- the Krylov Hamiltonian --
is by construction tridiagonal  (its non-zero elements being the Lanczos 
coefficients), i.e.\ it has the form of a 1d chain, with off-diagonal matrix elements
$b_{r}=\langle k_{r-1}|\hh|k_{r}\rangle$ and diagonals 
$a_{r}=\langle k_{r}|\hh|k_{r}\rangle$.

In the present context, taking $|k_{0}\rangle =|0\rangle$ as the root site of
the Cayley tree, it is readily shown using Eq.\ \ref{eq:Krylovrecurr} with 
$\hh \equiv \hhgam$ that the resultant Krylov basis is  precisely the set 
$\{|\tilde{r}_{1}\rangle\}$ of totally symmetric orbitals 
(Eq.\ \ref{eq:mtil1basis}), and that the Krylov Hamiltonian is what we have denoted by $\hhcal_{\Gamma}$ in Eq.\ \ref{eq:Hgamma0orb} (with $b_{r}=\sqrt{K}\Gamma$ and 
$a_{r}=0$). Further, $\overline{r}(\ttil)$ given in Eq.\ \ref{eq:ballistic} is 
precisely the so-called Krylov spread complexity, quantifying spatiotemporal spread of a wavefunction following initiation on $|0\rangle$;
and study of which has seen a recent surge of popularity in a wide range of 
physical contexts (see e.g.\  the review \cite{ NANDY2025}).

%%%%%%%%%%%%%%%%%%%%%%%%%%%%%%%%%%%%%%%%%%%%%%%%%%%%%%%%%%

\subsection{Radially disordered tree}
\label{section:radialtree}

So far, the `pure hopping'  case $\hh \equiv \hhgam$ has been considered.  
Now we include the radial disorder. Here,  as mentioned in 
Sec.\ \ref{section:Graphs}, the site energies in any given row $r$ are (by definition)
so strongly correlated that they are  completely slaved to each other, 
viz\ $\mathcal{E}_{r_{n}}=\er$ independently of $n$.
Eq.\ \ref{eq:H0rowresolve} for $\hh_{0}$ then reads
$\hh_{0} =\sum_{r=0}^{L}\er \hat{1}_{r}$ with 
$\hat{1}_{r}= \sum_{n=1}^{N_{r}}|r_{n}\rangle\langle r_{n}|$ the projector
for row $r$.
Since, for any $r$ and $m$, the states $|\tilde{r}_{m}\rangle$ of Sec.\ \ref{section:Puretree} are composed solely of tree orbitals
$\{|r_{n}\rangle\}$ from row $r$, they satisfy 
$|\tilde{r}_{m}\rangle =\hat{1}_{r}|\tilde{r}_{m}\rangle$;
and thus
$\hh_{0}|\tilde{r}_{m}\rangle=\er |\tilde{r}_{m}\rangle$ and
$\langle \tilde{r}^{\prime}_{m^{\prime}}|\hh_{0}|\tilde{r}_{m}\rangle=
\er\delta_{r^{\prime}r}\delta_{m^{\prime}m}$.

Hence, exactly the same fragmentation into 1d chains as arises under $\hhgam$ alone persists in the presence of radial disorder. In particular, the separable part of the Hamiltonian which contains the root site, denoted by 
$\hhcal =\hhcal_{0}+\hhcal_{\Gamma}$,  is
\begin{equation}
\label{eq:corr1dtbm}
\hhcal =  
\sum_{r=0}^{L} 
\er^{\pd}|\tilde{r}_{1}^{\pd}\rangle\langle \tilde{r}_{1}^{\pd}|
+
\sqrt{K}\Gamma \sum_{r=1}^{L} \Big(
|\widetilde{(r-1)}_{1}^{\pd}\rangle\langle \tilde{r}_{1}^{\pd}| +\mathrm{h.c.} \Big).
\end{equation}
And once again, with $|k_{0}\rangle =|0\rangle$, it follows from  
Eq.\ \ref{eq:Krylovrecurr} using the full $\hh =\hh_{0}+\hhgam$ that the
Krylov basis is the set  $\{|\tilde{r}_{1}\rangle\}$ of totally symmetric orbitals
(Eq.\ \ref{eq:mtil1basis}), and  that Eq.\ \ref{eq:corr1dtbm} is 
the Krylov Hamiltonian (now with $a_{r}=\er$ and $b_{r}=\sqrt{K}\Gamma$).

We focus on $\hhcal$ -- a 1d tight-binding chain in the $\tilde{r}_{1}$-basis,
 with constant nearest neighbour hopping $\sqrt{K}\Gamma$, and  a total of $L+1$ site energies $\er$. Taking the $\{\er\}$ as i.i.d.\ random variables
with distribution $P_{\mathcal{E}}(\mathcal{E})$, all states are 
in consequence exponentially localised on the chain for any disorder strength 
$W>0$ (and any finite $K$). As elaborated below, rather than enhancing delocalisation -- as one might intuitively have guessed -- introducing strong correlations in the site-energies of the Cayley tree in this way in fact does just the opposite.

In the following, our interest will centre on the localisation characteristics of
the eigenstates $|\psi_{E}\rangle$ of $\hhcal$ on the tree  -- i.e.\ in the 
tree basis $\{|r_{n}\rangle \}$ -- as embodied in the corresponding inverse participation ratios (IPRs) $\Lq$ specified below. In the  $\tilde{r}_{1}$-basis of totally symmetric tree orbitals (Eq.\ \ref{eq:mtil1basis}),  localisation properties on the associated 1d chain are of course quite simple, since
\begin{equation}
\label{eq:eigexp1}
|\psi_{E}^{\pd}\rangle ~=~ \sum_{r=0}^{L} A_{E,r}^{\pd}|\tilde{r}_{1}^{\pd}\rangle
\end{equation}
where $|A_{E,r}|^{2} \propto e^{-|r-r_{E}|/\xi}$
decays exponentially about its localisation centre $r= r_{E}$ with a finite localisation length, $\xi$. As a result, IPRs in the
$\tilde{r}_{1}$-basis, $L_{q} =\sum_{r=0}^{L}|A_{E,r}|^{2q}$ (on which we touch
briefly below), are all finite and $L$-independent.

But equally, one can decompose $|\psi_{E}\rangle$ in the tree basis,
\begin{equation}
\label{eq:eigexp2}
|\psi_{E}^{\pd}\rangle ~=~ \sum_{r=0}^{L}\sum_{n=1}^{N_{r}} 
A_{E,r_{n}^{\pd}}^{\pd}|r_{n}^{\pd}\rangle ,
\end{equation}
with associated IPRs
\begin{equation}
\label{eq:Lqdef1}
\Lq(r_{E}^{\pd}) ~=~\sum_{r=0}^{L}\sum_{n=1}^{N_{r}}|A_{E,r_{n}^{\pd}}^{\pd}|^{2q}.
\end{equation}
Using Eq.\ \ref{eq:mtil1basis} in Eq.\ \ref{eq:eigexp1} then relates
$A_{E,r_{n}}$ to $A_{E,r}$, viz.
\begin{equation}
\label{eq:amprels}
A_{E,r_{n}^{\pd}}^{\pd} ~=~A_{E,r}^{\pd}\big/\sqrt{N_{r}}
\end{equation}
(for all $n\in [1,N_{r}]$), and hence
\begin{equation}
\label{eq:Lqdef2}
\Lq(r_{E}^{\pd}) 
~=~\sum_{r=0}^{L}\frac{|A_{E,r}^{\pd}|^{2q}}{N_{r}^{q-1}}.
\end{equation}
From a numerical viewpoint, direct calculations on trees are limited
(by the tree dimension $N \sim K^{L}$) to rather modest values of $L \lesssim 20$
or so. Eq.\ \ref{eq:Lqdef2} by contrast allows much larger $L$'s to be reached,
by first obtaining the $\{A_{E,r}\}$ for a 1d chain of length $L$, and then
performing the sums to give $\Lq$. We employ this in later sections.
 
More importantly, Eq.\ \ref{eq:Lqdef2} enables some insight to be obtained  analytically, as now considered.

%%%%%%%%%%%%%%%%%%%%%%%%%%%%%%%%%%%%%%%%%%%%%%%%%%%%%%%%%%

\subsection{Analytical considerations}
\label{section:analyticstree}

As mentioned above, the square amplitude
\begin{equation}
\label{eq:Calc1}
|A_{E,r}^{\pd}|^{2} ~= ~\frac{1}{\mathcal{N}} ~e^{-|r -r_{E}^{\pd}|/\xi} 
\end{equation}
decays exponentially about its localisation centre $r_{E}$, with 
localisation length $\xi$. The normalisation constant $\mathcal{N}$ then
follows from $\sum_{r=0}^{L}|A_{E,r}|^{2}=1$, 
\begin{equation}
\mathcal{N} ~=~\frac{e^{1/\xi} +1 - e^{-r_{E}^{\pd}/\xi} - e^{-(L-r_{E}^{\pd})/\xi}}{e^{1/\xi}-1}.
\end{equation}
In the 1d chain of length $L\gg 1$, localised states of given energy can of course coexist for a given disorder realisation, being centred on different parts of the chain and thus having different localisation centres $r_{E}$. Over an ensemble of disorder realisations, it is the distribution of $r_{E}$'s that is sampled when considering all states of some energy $E$. We assume that each such $r_{E}$ 
is equally probable, i.e.\  its distribution is 
$P_{r_{E}}(r_{E})=L^{-1}\Theta(r_{E})\Theta(L-r_{E})$ (with $\Theta(x)$ the unit step function). Given this, the probability that $r_{E}$ lies within a distance $\mathcal{O}(1)$ of the chain ends ($r=0,L$) is $\propto 1/L$ and thus neglectable as $L\to \infty$, while the corresponding probability that $r_{E}$ is of order $L$ is unity. In that case, the above expression for $\mathcal{N}$ reduces in practice to
\begin{equation}
\label{eq:Calc2}
\mathcal{N} ~= ~\frac{e^{1/\xi} +1}{e^{1/\xi}-1} 
~=~ \mathrm{coth}\Big(\frac{1}{2\xi}\Big),
\end{equation}
which we generally use in the following.
With the same reasoning, the IPRs in the $\tilde{r}_{1}$-basis can be obtained, and are given by
\begin{equation}
\label{eq:Calcins1}
L_{q}^{\pd} 
= \sum_{r=0}^{L} |A_{E,r}^{\pd}|^{2q} 
~=~\Big[\mathrm{tanh}\Big(\frac{1}{2\xi}\Big)\Big]^{q}
\mathrm{coth}\Big(\frac{q}{2\xi}\Big).
\end{equation}
As noted earlier, the $L_{q}$ are finite and $L$-independent
(because eigenstates in the $\tilde{r}_{1}$-basis are localised
on the 1d chain on the lengthscale $\xi\sim \mathcal{O}(1)$).

Now consider the main quantities of interest, the IPRs in the original
tree basis, Eq.\ \ref{eq:Lqdef2}. This gives $\Lq(r_{E})$ for an eigenstate with localisation centre $r_{E}$. Over an ensemble of disorder realisations, and hence over the distribution of $r_{E}$, $\Lq$ will be characterised by its distribution 
$P_{\Lq}(\Lq)$, the form of which we obtain below. Two particular measures of that distribution will be considered, namely the typical value $\ltyp$ (geometric mean) and the arithmetic mean $\lmean$; calculated respectively from
\begin{equation}
\label{eq:Calc7}
\begin{split}
\ln \ltyp ~=~&\frac{1}{L+1}\sum_{r_{E}=0}^{L}
\ln \mathcal{L}_{q}^{\pd} (r_{E}^{\pd})
\equiv \big\langle \ln \mathcal{L}_{q}^{\pd} (r_{E}^{\pd})
\big\rangle_{r_{E}^{\pd}}^{\pd}
\\
\lmean ~=&~ \frac{1}{L+1}\sum_{r_{E}=0}^{L}\mathcal{L}_{q}^{\pd} (r_{E}^{\pd})
~\equiv ~\big\langle \mathcal{L}_{q}^{\pd} (r_{E}^{\pd})\big\rangle_{r_{E}^{\pd}}^{\pd}
\end{split}
\end{equation}
(with the right hand sides an obvious shorthand).

With $|A_{E,r}|^{2}$ from Eq.\ \ref{eq:Calc1}, Eq.\ \ref{eq:Lqdef2} for $\Lq(r_{E})$ can be evaluated in full (recall that $N_{r}=K^{r}$).
This yields
\begin{equation}
\label{eq:Calc4}
\begin{split}
&\mathcal{N}^{q} \mathcal{L}_{q}^{\pd}(r_{E}^{\pd}) ~=~
\frac{ K^{(q-1)} e^{-\frac{q r_{E}^{\pd}}{\xi}}- e^{\frac{q}{\xi}}e^{-r_{E}^{\pd}(q-1)\ln K}}{\big[ K^{(q-1)} -e^{\frac{q}{\xi}}\big]}
\\
&~~+~\frac{e^{-(q-1) r_{E}^{\pd}\ln K}}{\big[K^{(q-1)} e^{\frac{q}{\xi}} -1 \big]}
\left[
1 - e^{-(L-r_{E}^{\pd})[(q-1)\ln K +\frac{q}{\xi}]}
\right].
\end{split}
\end{equation}
For use below, let us first look at this in the two cases of $\xi \ll 1$ and 
$\xi \gg 1$, corresponding physically to strong and weak disorder strengths
respectively; and considering that $r_{E}$ is $\propto L$ with probability unity (as discussed above). For the present, we also have in mind $q>1$ ($q=2$ being the most commonly encountered IPR). Consider first $\xi \ll 1$ (more precisely 
$\xi \ll q/[(q-1)\ln K])$. Here,  Eq.\ \ref{eq:Calc4} reduces asymptotically to
\begin{equation}
\label{eq:Calc5}
\mathcal{L}_{q}^{\pd}(r_{E}^{\pd})~\overset{\xi \ll 1}{\sim}~
e^{-r_{E}^{\pd}(q-1)\ln K}
\left[1+\mathcal{O}\big(e^{-q/\xi}\big)\right] 
\end{equation}
(using also $\mathcal{N} \sim 1$ for $\xi \ll 1$, see Eq.\ \ref{eq:Calc2}).
For $\xi \gg 1$ by contrast, Eq.\ \ref{eq:Calc4} gives to leading order that 
\begin{equation}
\label{eq:Calc6}
\mathcal{L}_{q}^{\pd}(r_{E}^{\pd})~\overset{\xi \gg 1}{\sim}~
\frac{1}{(2\xi)^{q}}
\frac{1}{[1-K^{-(q-1)}]} \exp\Big(-\frac{q r_{E}^{\pd}}{\xi}\Big) 
\end{equation}
(with $\mathcal{N} =2\xi +\mathcal{O}(1)$ in this case).

\begin{figure} 
\includegraphics[width=\linewidth]{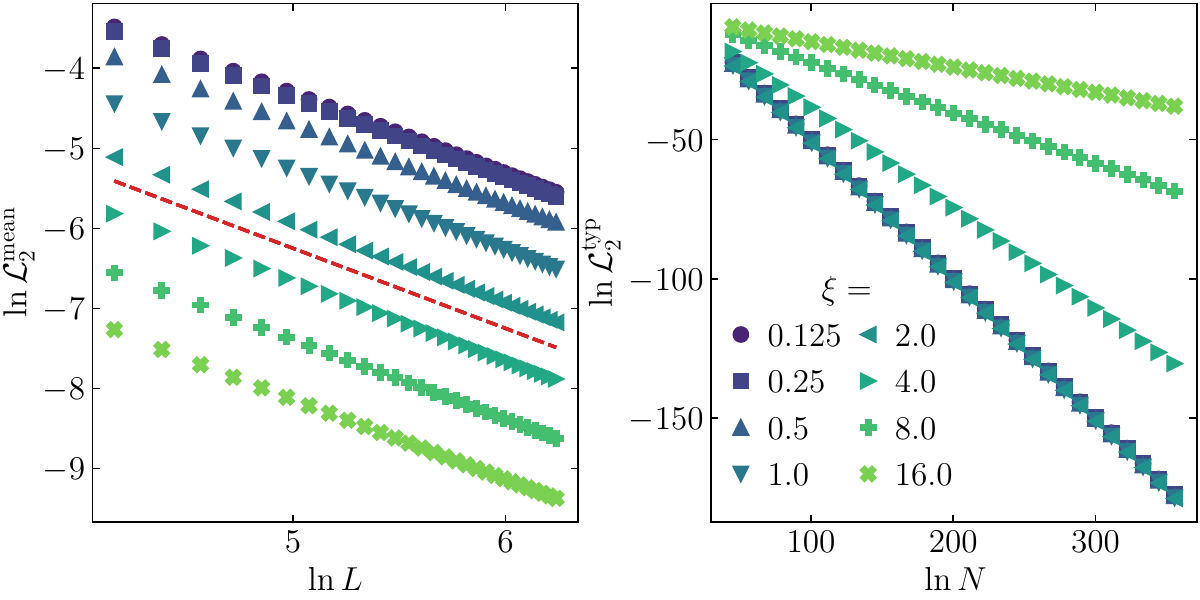}
\caption{Mean $\ltwomean$ (left) and typical $\ltwotyp$  (right) IPR for $q=2$ 
(with $K=2$), computed from the full analytical expressions 
in Eqs.~\ref{eq:Calc4},\ref{eq:Calc7} over the range of $\xi$ indicated.
$\ltwomean$ is plotted \emph{vs} the total number of tree generations, $L$, 
while $\ltwotyp$ is plotted against the total number of sites on the tree, 
$N\propto K^{L}$. 
Red dashed line in the $\ln \ltwomean$ plot indicates a slope of $-1$. 
}
\label{fig:analytical}
\end{figure}

Fig.\ \ref{fig:analytical} shows the mean and typical IPRs for
$q=2$ (with $K=2$), computed using Eq.\ \ref{eq:Calc7} and the \emph{full}
Eq.\ \ref{eq:Calc4} for $\Lq(r_{E})$, over the wide range of  $\xi$'s indicated. 
The mean, $\ltwomean$, is plotted against the total number of tree generations, 
$L$; while the typical, $\ltwotyp$, is shown \emph{vs} the total number of sites on the tree $N$, given by
\begin{equation}
\label{eq:Calc8}
N ~=~ \sum_{r=0}^{L} K^{r} 
~\overset{L\gg 1}\sim~ \frac{K}{K-1}K^{L}.
\end{equation}
Three features in Fig.\ \ref{fig:analytical} stand out:
\begin{enumerate}
  \item $\ltwomean$ decays as a power-law in the number of
generations $L$, specifically $\ltwomean \propto L^{-1}$. In contrast,
$\ltwotyp$ decays as a power-law in the total number of tree sites $N$.
  \item $\ltwotyp$ is many orders of magnitude smaller than $\ltwomean$,
suggesting a very broad distribution of IPRs.
  \item With increasing localisation length $\xi$, the state spreads out more on the chain, so one expects the IPR to decrease. $\ltwomean$ indeed shows this trend. By contrast, $\ltwotyp$ quite remarkably shows the opposite trend, increasing with increasing $\xi$.
\end{enumerate}

To understand these features, let us analyse the behaviour in the two limits of large and small $\xi$.
For $\xi \ll 1$, using the leading behaviour $\mathcal{L}_{q}(r_{E}) \sim K^{-(q-1)r_{E}}$ of Eq.\ \ref{eq:Calc5} in Eq.\ \ref{eq:Calc7} gives
\begin{subequations}
\label{eq:Calc10}
\begin{align}
\lmean &\overset{L\gg 1}{\sim} \frac{1}{[1-K^{-(q-1)}]} \frac{1}{L}
~\propto \frac{1}{L}~~~~:\xi \ll 1
\label{eq:Calc10a}
\\
\ltyp \overset{L\gg 1}{\sim}&\exp\Big[-\frac{1}{2}(q-1)L \ln K\Big]
~\sim ~ N^{-\frac{1}{2}(q-1)};
\label{eq:Calc10b}
\end{align}
\end{subequations}
while for $\xi \gg 1$, Eq.\ \ref{eq:Calc6} 
used in Eq.\ \ref{eq:Calc7} gives
\begin{subequations}
\label{eq:Calc11}
\begin{align}
\lmean \overset{L\gg 1}{\sim}& \frac{1}{(2\xi)^{q}}\frac{1}{[1-K^{-(q-1)}]} 
\frac{1}{[1-e^{-q/\xi}]}\frac{1}{L}
\propto \frac{1}{L}
\label{eq:Calc11a}
\\
\ltyp \overset{L\gg 1}{\sim}&\frac{1}{(2\xi)^{q}}\frac{1}{[1-K^{-(q-1)}]}\frac{1}{L}
\exp\left[-\frac{q}{2\xi}L\right]
\nonumber
\\
~\sim & ~ N^{-q/(2\xi \ln K)}
~~~~~~:~\xi \gg 1.
\label{eq:Calc11b}
\end{align}
\end{subequations}

\begin{figure}
\includegraphics[width=\linewidth]{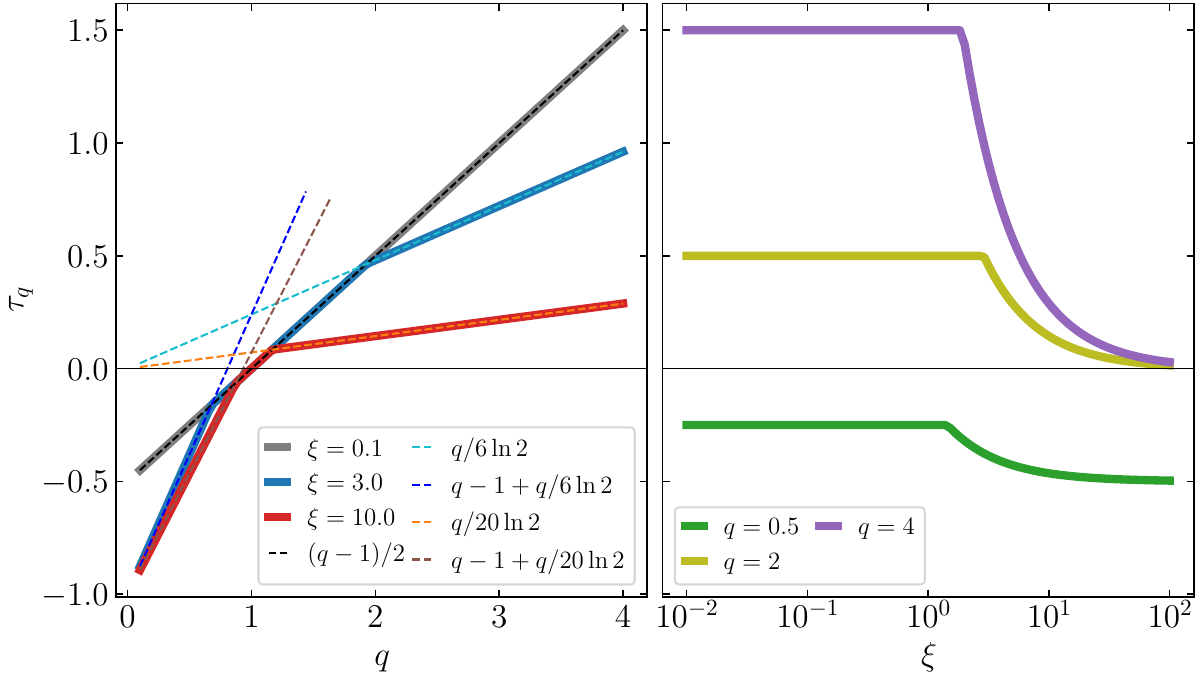}
\caption{\emph{Left panel}: multifractal exponent $\tau_{q}$ as a function of $q$, shown for $\xi =0.1, 3$ and $10$ (with $K=2$). Solid lines show results  computed numerically via Eqs.\ \ref{eq:Calc9},\ref{eq:Calc7} using the full sum expression in  Eq.~\ref{eq:Calc4}, while dashed lines show the exact  results given in Sec.\ \ref{section:tauq} (Eq.\ \ref{eq:mf1} for $\xi =3,10$, and Eq.\ \ref{eq:mf3} for $\xi =0.1$);  the two are indistinguishable. \emph{Right panel}: $\tau_{q}$  \emph{vs} $\xi$, shown for $q=4,2$ and $0.5$,and given exactly by Eq.\ \ref{eq:mf4} for $q=2,4$ and Eq.\ \ref{eq:mf5} for $q=1/2$.
}
\label{fig:tau2-analytical}
\end{figure}

Whether $\xi$ is large or small, Eqs.\ \ref{eq:Calc10a},\ref{eq:Calc11a} 
show that $\lmean \propto 1/L$, as indeed seen in Fig.\ \ref{fig:analytical}
for any $\xi$. As also seen in Fig.\ \ref{fig:analytical}, 
Eqs.\ \ref{eq:Calc10b},\ref{eq:Calc11b} likewise show
that $\ltyp$ scales as a power-law in the total number of tree sites $N$,
\begin{equation}
\label{eq:Calc9}
\ltyp~=~N^{-\tau_{q}^{\pd}} .
\end{equation}
This of course is multifractal behaviour of the eigenstates, with 
multifractal exponent $\tau_{q}$; such behaviour  being manifest in the typical value of $\Lq$, but \emph{not} in the mean $\lmean$. The exponents $\tau_{q}$ can be read off from Eqs.\ \ref{eq:Calc10b},\ref{eq:Calc11b}, 
\begin{equation}
\label{eq:Calc12}
\tau_q(\xi)^{\pd} = \begin{cases}
\frac{1}{2}(q-1)\,&\quad :\xi\ll 1\\
\frac{q}{2\xi\ln K}\,&\quad :\xi \gg 1 .
\end{cases}\,
\end{equation}
This behaviour is indeed seen clearly in  Fig.\ \ref{fig:tau2-analytical}
(right panel), where the $\xi$-dependence of $\tau_{q}$ is shown for three different
$q$'s. Eq.\ \ref{eq:Calc12} is in fact more general than its asymptotic 
derivation here. Within the present framework, the exact $\tau_{q}(\xi)$ can  be obtained and are given in Sec.\ \ref{section:tauq}.
In particular, for any $q\geq 1$, the form  Eq.\ \ref{eq:Calc12} is more generally correct:  $\tau_{q}=\tfrac{1}{2}(q-1)$ holds  for \emph{all} 
$\xi \ln K \leq q/(q-1)$, and likewise $\tau_{q}=q/(2\xi \ln K)$ for 
all $\xi \ln K >q/(q-1)$ (Eq.\ \ref{eq:mf4}).

As noted in point 2 above,  Fig.\ \ref{fig:analytical} shows $\ltwotyp$
to be many orders of magnitude  smaller than $\ltwomean$, suggesting a very broad distribution of IPRs. To understand this, note from 
Eqs.\ \ref{eq:Calc5},\ref{eq:Calc6} that, whether $\xi \ll 1$ or $\xi \gg 1$, 
the IPR as a function of $r_{E}$ is of form
\begin{equation}
\label{eq:Calc13}
\mathcal{L}_{q}^{\pd}(r_{E}^{\pd}) ~ = ~ c_{1}^{\pd} e^{-c_{2}^{\pd}r_{E}^{\pd}}
\end{equation}
where $c_{1}, c_{2}$ (each positive for $q\geq 1$ and with $c_{1}\leq 1$)
depend on $q, K$ and $\xi$. Given that each $r_{E}\in [0,L]$ is equally probable over an ensemble of disorder realisations, the probability distribution of IPRs,
$P_{\Lq}(\Lq)=\int dr~P_{r_{E}}(r)\delta\big(\Lq -c_{1}e^{-c_{2}r}\big)$,
follows as
\begin{equation}
\label{eq:Calc15}
P_{\mathcal{L}_{q}}^{\pd}(\mathcal{L}_{q}^{\pd}) =\frac{1}{\mathcal{L}_{q}}~ 
(c_{2}^{\pd}L)^{-1} \Theta\big(\mathcal{L}_{q}^{\pd} - c_{1}^{\pd} e^{-c_{2}^{\pd}L}\big)
\Theta\big(c_{1}^{\pd} - \mathcal{L}_{q}^{\pd}\big).
\end{equation}
The important point here is the universal $1/\mathcal{L}_{q}$ decay in the 
distribution,\footnote{$P_{\Lq}(\Lq)$ has of course bounded support: 
$\mathcal{L}_{q}$ is positive by definition, and for $q\geq 1$ no $\mathcal{L}_{q}$ can exceed $1$, so the distribution is bounded from above by 
$c_{1} \sim \mathcal{O}(1)$, and from below by a value exponentially small in $L$.} reflecting the broad distribution of localisation centres $r_{E}$.
As a result of this heavy-tailed behaviour, the average value
$\lmean \equiv \int d\mathcal{L}_{q}~\mathcal{L}_{q}P_{\mathcal{L}_{q}}(\mathcal{L}_{q})$
$=(c_{2}L)^{-1}[c_{1}-c_{1}e^{-c_{2}L}] \sim (c_{1}/c_{2})L^{-1}$
is controlled by the $\mathcal{O}(1)$ values of 
$\mathcal{L}_{q}$ (and is thus overall $\propto 1/L$).
But $\mathcal{O}(1)$ values of $\mathcal{L}_{q}$ have a vanishing probability of occurrence as $L\to \infty$, so the mean is thus dominated by occurrences which are statistically irrelevant.
Typical occurrences by contrast are those for which
$\mathcal{L}_{q}$ is exponentially small in $L$.
One sees this clearly by looking at the fraction of the distribution which lies between $x$ and its upper limit of $c_{1}$, namely
\begin{equation}
\label{eq:Calc16}
F_{\mathcal{L}_{q}^{\pd}}(x) ~=~
\int_{x}^{c_{1}} d\mathcal{L}_{q}^{\pd}~P_{\mathcal{L}_{q}^{\pd}}^{\pd}(\mathcal{L}_{q}^{\pd})
~=~\frac{1}{c_{2}L} \ln \left[\frac{c_{1}}{x}\right].
\end{equation}
If $x$ is $\mathcal{O}(1)$, then $F_{\mathcal{L}_{q}}(x) \propto 1/L$ thus vanishes as $L\to \infty$. By contrast, $F_{\mathcal{L}_{q}}(x)$  is finite as 
$L\to \infty$ only if $x\propto e^{-\alpha L}$ is exponentially small in
$L$. In other words, the bulk of the distribution 
$P_{\mathcal{L}_{q}}(\mathcal{L}_{q})$ occurs for exponentially
small values of $\mathcal{L}_{q}$. It is of course for this reason that the geometric mean $\ltyp$ is 
representative of the distribution, while the arithmetic mean $\lmean$ is not.

%%%%%%%%%%%%%%%%%%%%%%%%%%%%%%%%%%%%%%%%%%%%%%%%%%%%%%%%%%%%%%%%%%%%%%%%%%%
\subsubsection{Multifractal exponents $\tau_{q}$}
\label{section:tauq}

We have so far  considered $q > 1$, but would obviously like to know the 
full $q$-dependence of $\tau_{q}$ for $q \geq 0$ and with arbitrary $\xi$.
 This can in fact be determined exactly within the framework employed.
The basic reasons are as follows. Recall that $\tau_{q}$ is defined by 
Eq.\ \ref{eq:Calc9}, viz.\ $\tau_{q} = -\ln\ltyp/\ln N$; and we want the finite limit of this as $L\to \infty$. So
$\tau_{q} \equiv -\ln\ltyp/(L\ln K)$, and from Eq.\ \ref{eq:Calc7}, 
$\ln \ltyp \equiv \langle \ln\Lq(r_{E})\rangle_{r_{E}}$
(such that e.g.\ $\langle r_{E}\rangle_{r_{E}} =L/2$). 
Now Eq.\ \ref{eq:Calc4}  gives  the full expression for 
$\mathcal{N}^{q}\Lq(r_{E})$, and in determining $\tau_{q}$ the normalisation factor 
$\mathcal{N}$ plays no role ($\ln\mathcal{N}^{q}/L$ vanishes as $L\to \infty$).
All that matters is which of the  $r_{E}$-dependent exponentials appearing 
in Eq.\ \ref{eq:Calc4} dominates the $r_{E}$ sum in the expression for $\tau_{q}$; it being the logarithm of the dominant term(s) which determines $\tau_{q}$
as $L\to \infty$. 

The upshot is that the  $q$-dependence of $\tau_{q}$ depends on $\xi$ and $K$ solely in the combination $\xi \ln K$, and divides into two regimes:
(a) For $\xi\ln K >1$,
\begin{equation}
\label{eq:mf1}
	\tau_{q}^{\pd} ~=~\begin{cases}
 						(q-1)+\frac{q}{2\xi \ln K} \quad& :0\leq q \leq q_{*}^{-} \\
 						\tfrac{1}{2}(q-1) \quad& :q_{*}^{-}\leq q \leq q_{*}^{+}\\
 						\frac{q}{2\xi \ln K} \quad& :q_{*}^{+} \leq q
 					\end{cases}
\end{equation}
\label{eq:mf2}
with $q_{*}^{-}<1<q_{*}^{+}$ given by
\begin{equation}
q_{*}^{\pm} ~=~ \frac{1}{1\mp \frac{1}{\xi \ln K}},
\end{equation}
and with $\tau_{q}$ continuous as $q$ crosses both $q_{*}^{-}$ and $q_{*}^{+}$.
As $\xi \ln K \to 1+$, $q_{*}^{+} \to \infty$, and the final regime in
Eq.\ \ref{eq:mf1} is correspondingly absent. In this case:
(b) For $\xi\ln K <1$,
\begin{equation}
\label{eq:mf3}
	\tau_{q}^{\pd} ~=~\begin{cases}
 						(q-1)+\frac{q}{2\xi \ln K} \quad& :0\leq q \leq q_{*}^{-} \\
 						\tfrac{1}{2}(q-1) \quad& :q_{*}^{-}\leq q 
 					\end{cases}
\end{equation}
with $\tau_{q}$ again continuous as $q$ crosses $q_{*}^{-}<1$.
As they must, Eqs.\ \ref{eq:mf1},\ref{eq:mf3} recover the trivial limits
$\tau_{q=1}=0$ (from wavefunction normalisation) and $\tau_{q=0}=-1$.

For any fixed $q$, Eqs.\ \ref{eq:mf1},\ref{eq:mf3} can of course  be inverted, to
give $\tau_{q}$ as a function of $\xi\ln K$. For $q \geq 1$ this yields
\begin{equation}
\label{eq:mf4}
	\tau_{q}^{\pd} ~=~\begin{cases}
 						\frac{1}{2}(q-1) \quad& :0\leq \xi \ln K \leq \frac{q}{(q-1)} \\
 						\frac{q}{2\xi \ln K} \quad& :\xi \ln K\geq \frac{q}{(q-1)}
 					\end{cases}			,	
\end{equation}
while for any given  $q \in [0,1]$, 
\begin{equation}
\label{eq:mf5}
	\tau_{q}^{\pd} ~=~\begin{cases}
 						\frac{1}{2}(q-1) \quad& :0\leq \xi \ln K \leq \frac{q}{(1-q)} \\
 						(q-1) +\frac{q}{2\xi \ln K} \quad& :\xi \ln K\geq \frac{q}{(1-q)}
 					\end{cases}			.
\end{equation}

The results above are illustrated in Fig.\ \ref{fig:tau2-analytical},
which shows both the $q$-dependence of $\tau_{q}$ for three
representative $\xi$'s (left panel), and its $\xi$-dependence for three representative
$q$'s (right panel). Eq.\ \ref{eq:mf1} for the $q$-dependence of $\tau_{q}$, consisting as it does of three continuously connect regimes of linear $q$-dependence for $\xi\ln K >1$, is exemplified by the $\xi=3,10$  examples;
while the $\xi=0.1$ example illustrates Eq.\ \ref{eq:mf3} (in this case, 
$q_{*}^{-}$ is so small that only the $\tau_{q}=\tfrac{1}{2}(q-1)$ behaviour is 
shown).

Likewise, Eqs.\ \ref{eq:mf4},\ref{eq:mf5} for the $\xi$-dependence of $\tau_{q}$ are
exemplified in the right panel of Fig.\ \ref{fig:tau2-analytical}.
In particular, $\tau_{q}(\xi)$ for any given $q$ is $\xi$-independent up to
$\xi =\xi_{c} =q/(|q-1|\ln K)$, and decreases thereafter as $\xi$ increases;
whence $\ltyp =N^{-\tau_{q}}$ increases with increasing $\xi$ (i.e.\ 
\emph{decreasing} disorder strength), which is the behaviour 
remarked upon in Fig.\ \ref{fig:analytical} above for $\ln\ltwotyp$.

One further observation is worth noting here. For $q<1$ the exponent
$\tau_{q}$ is  negative, corresponding to an
$\ltyp =N^{|\tau_{q}|}\propto e^{|\tau_{q}|L\ln K}$ which \emph{increases} exponentially with the number of tree generations $L$.
In that case, $c_{2}$ in Eq.\ \ref{eq:Calc13} for $\Lq(r_{E})$ is negative. The 
resultant  $P_{\Lq}(\Lq)$, rather than being given by Eq.\ \ref{eq:Calc15}, 
is instead  given by
\begin{equation}
\label{eq:Calc17}
P_{\Lq}^{\pd}(\Lq)=\frac{1}{\Lq} ~ (|c_{2}|L)^{-1}\Theta(\Lq -c_{1}^{\pd})
\Theta(c_{1}^{\pd}e^{|c_{2}^{\pd}|L}-\Lq)
\end{equation}
with $c_{1}$  the \emph{lower} limit, and the upper limit $c_{1}e^{|c_{2}|L}$ now
exponentially large in $L$; such that the $\propto 1/\Lq$ fat tail in 
 $P_{\Lq}(\Lq)$ extends up to exponentially large values of $\Lq$ (albeit with exponentially small weight). The mean value of this distribution remains governed
by its upper limit (as for $q>1$), but  $\lmean$ is now exponentially 
large in $L$. As such, $\lmean$ itself could legitimately be viewed as exhibiting
multifractal behaviour (with a calculable exponent $\tilde{\tau}_{q}$).
Yet this behaviour is particular to the $q<1$ regime, and in marked contrast to 
all $q>1$  where $\lmean \propto 1/L$ and multifractality does not arise in $\lmean$.
It is only with the typical value $\ltyp$  that robust multifractality arises for both $q>1$ and $q<1$; and which thus provides the appropriate touchstone for it.

%%%%%%%%%%%%%%%%%%%%%%%%%%%%%%%%%%%%%%%%%%%%%%%%%%%%%%%%%%%%%%%%%%%%%%%%%%%
\subsection{Numerical results: Cayley tree }
\label{section:TreeNumerics}

\begin{figure}
\includegraphics[width=\linewidth]{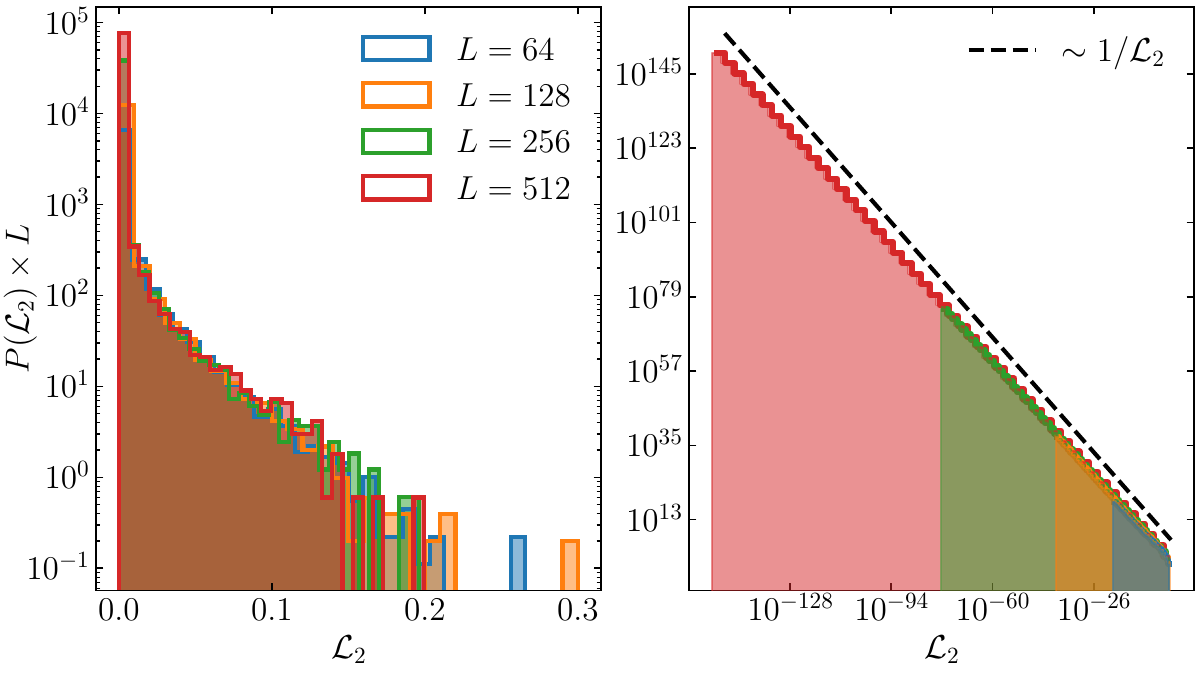}
\caption{
Distribution $P_{\ltwo}(\ltwo)$ of $\ltwo$ obtained from ED
(shown for $W=1$ with $K=2$), for the number of generations $L$ as indicated.
For clarity, $L\times P_{\ltwo}(\ltwo)$ \emph{vs} $\ltwo$ is plotted 
(as explained in text). Left panel shows data with linear bins, right panel with logarithmic bins. In right panel, the black dashed line shows the 
$P_{\ltwo}(\ltwo)  \propto 1/{\cal L}_2$ behaviour, Eq.\ \ref{eq:Calc15}.
}
\label{fig:L2-dist}
\end{figure}

We now turn briefly to numerical calculations, obtained via exact 
diagonalisation (ED) on the 1d Krylov Hamiltonian Eq.\ \ref{eq:corr1dtbm}.
For the i.i.d.\ disorder in the $\{\er\}$, a box distribution 
$P_{\mathcal{E}}(\mathcal{E}) =(2W)^{-1}\Theta(W-|\mathcal{E}|)$ is taken.
$\Lq(r_{E})$ is calculated directly using  Eq.\ \ref{eq:Lqdef2}, with the 
$\{|A_{E,r}|^{2}\}$ obtained from ED on the chain of length $L$.
The calculations here extend up to  $L\sim 500$, which corresponds for $K=2$ to a Cayley tree  of $\sim 500$ generations and dimension  
$N \sim K^{L}\sim 10^{150}$.\footnote{As far as ED on the 1d Krylov chain is concerned, it is of course straightforward to go to much larger values of $L$; but 
the smallness of terms in the $r$-sum in Eq.\ \ref{eq:Lqdef2} for $\Lq$ then requires careful handling, and for our purposes there is little gain in doing so.}
Results given below are averaged over $\sim 2000$ disorder realisations, and we also average over a substantial fraction of eigenstates in the band (which improves 
statistics and does not materially affect the results shown).

The analytically-based results of Sec.\ \ref{section:analyticstree} are well
corroborated by ED. Fig.\ \ref{fig:L2-dist} shows the resultant ED distribution 
$P_{\ltwo}(\ltwo)$ of $\ltwo$ (here illustrated by disorder strength $W=1$, with 
$K=2$). Since Eq.\ \ref{eq:Calc15}  gives $P_{\Lq}(\Lq) \propto 1/(\Lq L)$,
we plot $L\times P_{\mathcal{L}_{2}}(\ltwo)$ \emph{vs} $\ltwo$, to 
expose cleanly the predicted power-law decay 
$P_{\mathcal{L}_{2}}(\mathcal{L}_{2})\propto 1/\mathcal{L}_{2}$; 
and which behaviour is clearly seen in Fig.\ \ref{fig:L2-dist}.

\begin{figure}
\includegraphics[width=\linewidth]{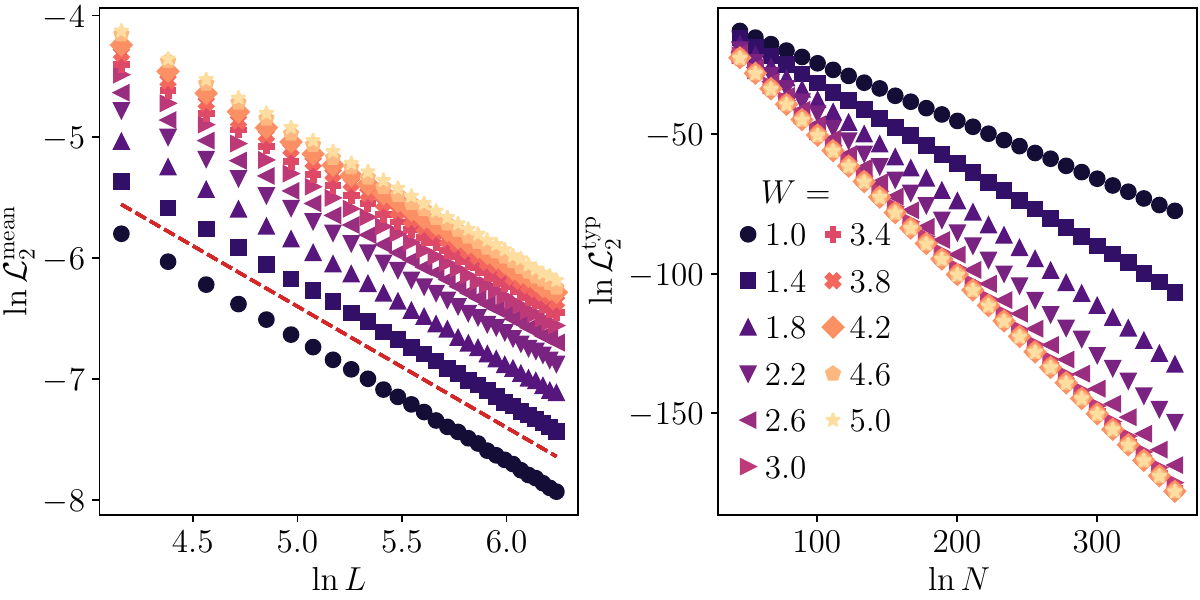}
\caption{Mean $\ltwomean$  (left) and typical $\ltwotyp$ (right) IPR obtained numerically from ED for different disorder strengths $W$ (with $K=2$). 
$\ltwomean$ is plotted against the total number of tree generations, $L$, whereas 
 $\ltwotyp$ is plotted against the total number of sites on the tree, 
$N \propto K^{L}$ (Eq.\ \ref{eq:Calc8}). Red dashed line in left panel for $\ltwomean$ indicates a slope of $-1$. 
}
\label{fig:numerical}
\end{figure}

Fig.\ \ref{fig:numerical} shows ED results for $\ltwomean$ \emph{vs} $L$
and $\ltwotyp$ \emph{vs} $N \propto K^{L}$, for the range of disorder strengths $W$ indicated. These indeed confirm that $\ltwomean \propto 1/L$, with an overall magnitude that increases  with increasing $W$ ($\equiv$ decreasing $\xi$); and that 
$\ltwotyp\propto N^{-\tau_{2}}$ with a fractal exponent $\tau_{2}$ which 
\emph{in}creases with increasing $W$. These results should be compared to their analytical counterparts shown in Fig.\ \ref{fig:analytical}, where
the same quantities were plotted for a range of $\xi$,  remembering that the 
effective $\xi$ will decrease with increasing $W$.
While there is naturally no direct mapping between the disorder strength $W$, 
and $\xi$ appearing in the analytical treatment, the agreement between the two is self-evident. 

\begin{figure}
\includegraphics[width=\linewidth]{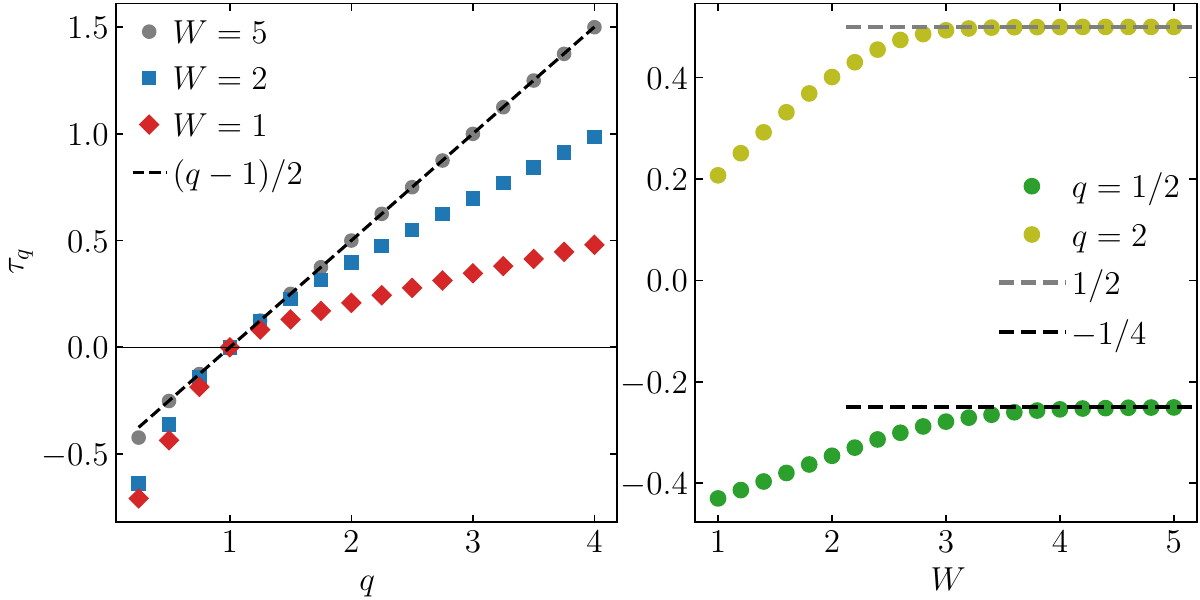}
\caption{
ED results for multifractal exponents $\tau_{q}$.
\emph{Left panel}: $\tau_{q}$ \emph{vs} $q$ for disorder strengths $W=1,2,5$. Black dashed line shows $\tfrac{1}{2}(q-1)$. Corresponding  analytical results for different $\xi$ are shown in Fig.\ \ref{fig:tau2-analytical} (left panel). Discussion in text. \emph{Right panel}: $\tau_{q}$ \emph{vs} $W$, shown for $q=2$ and $1/2$. Dashed lines show the asymptotes $\tfrac{1}{2}(q-1)$. For comparison, analytical results for  $\tau_{q}$ \emph{vs} $\xi$ 
are shown in Fig.\ \ref{fig:tau2-analytical} (right panel). 
}
\label{fig:tauq-anal+ED}
\end{figure}

Fig.\ \ref{fig:tauq-anal+ED} pursues further the behaviour of the multifractal exponents $\tau_{q}$ obtained  from ED.  As in Fig.\ \ref{fig:numerical} for $q=2$,
plots of $\ln\ltyp$ \emph{vs} $\ln N$ are found to show pristine linearity for any
$q$,  enabling $\tau_{q}=-\ln\ltyp/\ln N$ to be read off.
For $q=2$ and $1/2$, the right panel of Fig.\ \ref{fig:tauq-anal+ED} shows the
evolution of the resultant $\tau_{q}$ as a function of $W$; analytical results for the $\xi$-dependence of $\tau_{q}$ are given in Fig.\ \ref{fig:tau2-analytical} 
(right panel). Given that the effective $\xi$ decreases with increasing $W$, the agreement between the two is evident. In particular, the large-$W$ behaviour 
$\tau_{q}=\tfrac{1}{2}(q-1)$  obtained via ED is precisely that arising from the small-$\xi$ analytical results  (Eqs.\ \ref{eq:mf4},\ref{eq:mf5}).

ED results for $\tau_{q}$ \emph{vs} $q$ are shown in the left panel of
Fig.\ \ref{fig:tauq-anal+ED}, for $W=1,2,5$; corresponding  analytical results for different $\xi$'s are shown in Fig.\ \ref{fig:tau2-analytical} (left panel).
For strong disorder, $W=5$, the ED results indeed show the behaviour 
$\tau_{q}=\tfrac{1}{2}(q-1)$ over the wide $q$-range shown, just as they do for 
the $\xi =0.1$ case in the left panel of Fig.\ \ref{fig:tau2-analytical} (as
per Eqs.\ \ref{eq:mf3}). And for weaker disorder, e.g.\ $W=2$, one sees 
three regimes of linear $q$-dependence, as arises for e.g.\ the $\xi=3$ case
shown in the left panel of Fig.\ \ref{fig:tau2-analytical} (and reflecting 
Eq.\ \ref{eq:mf1}).

%%%%%%%%%%%%%%%%%%%%%%%%%%%%%%%%%%%%%%%%%%%%%%%%%%%%%%%%%%%%%%%%%%%%%%%%%%%

\section{Radial Aubry-Andr\'e tree}
\label{section:AAtree}

Quenched disorder is not of course the only mechanism that can
induce localisation: incommensurability 
is another, see e.g.\ ~\cite{aubry1980analyticity,Harper_1955,thouless1983bandwidths,prange1983wave,sarma1988mobility,boers2007mobility,biddle2009localization,ganeshan2015nearest,YaoQuasiPRL2019,wang2020onedimensional,DuthieSelfConPRB2021}.
We consider now what we refer to as a radial Aubry-Andr\'e (AA) model.
Again, the site-energies for sites $r_{n}$ on the Cayley tree satisfy 
$\mathcal{E}_{r_{n}}=\er$ independent of the particular sites $n$ on 
row/generation $r$. In this case they are given by the AA potential~\cite{aubry1980analyticity} $\er= V\mathrm{cos}(2\pi \kappa r +\phi)$, with $\kappa$ an irrational 
-- chosen for ED calculations as $\Phi -1$ with $\Phi$ the golden mean -- and $\phi$ a random phase (averaging over which is the analogue of disorder 
averaging). In the totally symmetric basis $\{|\tilde{r}_{1}\rangle\}$ 
(Eq.\ \ref{eq:mtil1basis}), the associated 1d Krylov Hamiltonian $\hhcal$ again has precisely the form Eq.\ \ref{eq:corr1dtbm}.

In contrast to 1d problems with uncorrelated quenched disorder, the 1d 
AA model possesses a transition between a delocalised and a
localised phase~\cite{aubry1980analyticity}. By a familiar duality argument~\cite{aubry1980analyticity}, this occurs when the potential strength $V$ is precisely twice the magnitude of the hopping matrix element; so for Eq.\ \ref{eq:corr1dtbm} the critical $\vc=2\sqrt{K}\Gamma$. For $V>\vc$, all states are exponentially localised, while for $V<\vc$ all eigenstates are delocalised (there being no mobility edge in the AA model).

%%%%%%%%%%%%%%%%%%%%%%%%%%%%%%%%%%%%%%%%%%%%%%%%%%%%%%%%%%%%%%%%%%%%%%%%%%%
\subsection{Analytical considerations}
\label{section:AACalcs}

Consider first the behaviour of the system in the localised phase of the 1d model,
$V>\vc =2\sqrt{K}\Gamma$. Here, as a function of the localisation length $\xi$, the properties of the system  are again just as given in 
Sec.\ \ref{section:analyticstree}.  The difference in the present case is of course that $\xi$ diverges as $V$ approaches from above the \emph{finite} $\vc$ at which the 
transition occurs. The form of that divergence is known exactly~\cite{thouless1983bandwidths}, $\xi \propto (V-\vc)^{-1}$.

The IPRs $\mathcal{L}_{q}$ in the original tree basis remain given by
Eq.\ \ref{eq:Lqdef2}. As a function of $\xi$, their properties are
just as determined in Secs.\ \ref{section:analyticstree}, \ref{section:tauq};
and the points made about them there hold  equally here.
In this regime multifractality is ubiquitous, and two points about the behaviour of the IPR exponent $\tau_{q}$ should be noted. First,  sufficiently deep in the multifractal phase where $\xi \ll 1$, the behaviour $\tau_{q}=\tfrac{1}{2}(q-1)$ 
(Eqs.\ \ref{eq:Calc12},\ref{eq:mf4},\ref{eq:mf5}) will again arise; 
as too will the behaviour $\tau_{q} \propto 1/\xi$ for $\xi \gg 1$
and any $q>1$ (Eq.\ \ref{eq:mf4}). Second, since 
$1/\xi \propto (V-\vc)$, the latter implies 
\begin{equation}
\label{eq:AA2}
\tau_{q}^{\pd} ~\overset{V \to \vc +}{\sim} ~ (V-V_{\mathrm{c}}^{\pd}) 
~~~~~~:~V_{\mathrm{c}}^{\pd} =2\sqrt{K}\Gamma ~.
\end{equation}
$\tau_{q}$ must thus vanish on approach to the transition from the multifractal/localised side, with a critical 
exponent of $1$ (from Eq.\ \ref{eq:mf5}, the analogue of this for any $q\in (0,1)$ is $\tau_{q}-(q-1) \sim (V-\vc)$ as $V\to \vc+$).

For $V<\vc$,  the 1d model has a delocalised phase
(of which there is  no counterpart in the radially disordered tree 
of the previous sections). Here we can say relatively little \emph{a priori}, save for deep in the small-$V\ll \vc$ regime where essentially uniform delocalisation is expected,  with $|A_{E,r}|^{2} \simeq 1/L$ for all $r$. $\Lq(r_{E})$ then follows from Eq.\ \ref{eq:Lqdef2} (with $N_{r}=K^{r}$), and does not depend on $r_{E}$. So from Eq.\ \ref{eq:Calc7}, $\ltyp \equiv \lmean$ -- here denoted simply by $\Lq$ --
follows for any $q>1$ (on which we focus in the following) 
as
\begin{equation}
\label{eq:AA4}
\Lqp
~\overset{V\to 0}{\sim} ~\frac{1}{[1-K^{-(q-1)}]}\times \frac{1}{L^{q}}~; 
\end{equation}
which we examine numerically below. Let us also remark here that this 
$V\to 0$ $L$-dependence, $\Lq \sim L^{-q}$, 
is unsurprisingly different from that of the IPR 
$L_{q}=\sum_{r=0}^{L}|A_{E,r}|^{2q}$ in the 1d $\tilde{r}_{1}$-basis; which,
with $|A_{E,r}|^{2} \simeq 1/L$, instead gives $L_{q}\sim L^{-(q-1)}$.

%%%%%%%%%%%%%%%%%%%%%%%%%%%%%%%%%%%%%%%%%%%%%%%%%%%%%%%%%%%%%%%%%%%%%%%%%%%
\subsection{Numerical results: radial AA tree}
\label{section:AAnumerics}

\begin{figure} 
\includegraphics[width=\linewidth]{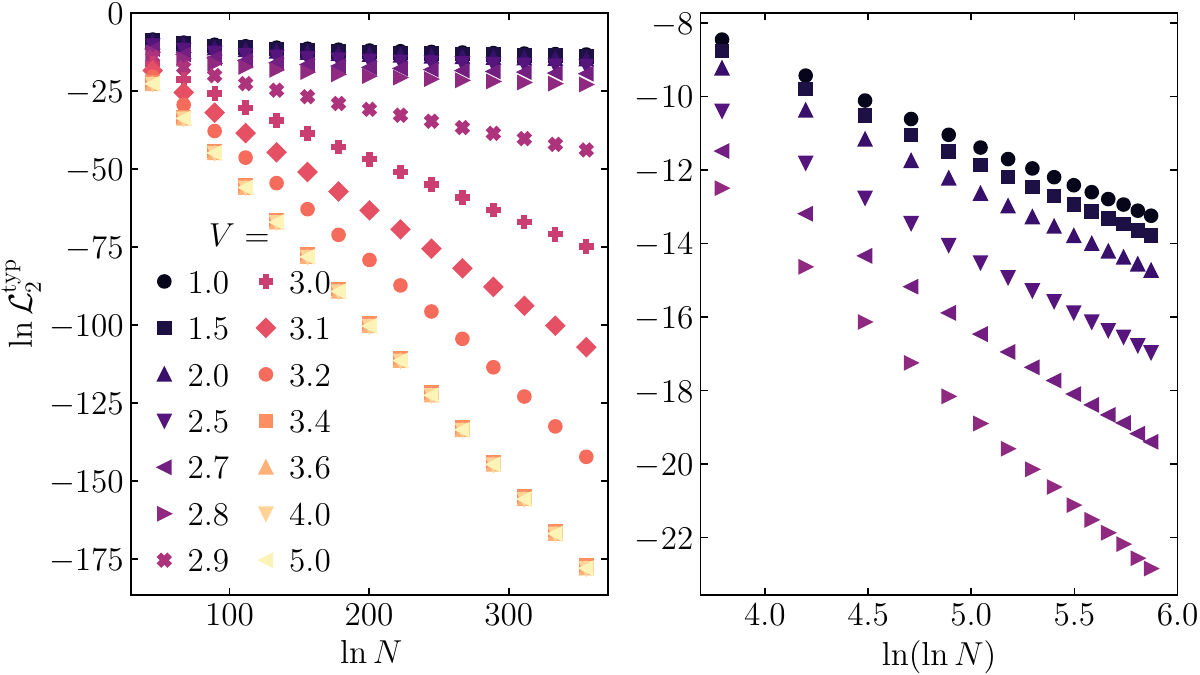}
\caption{Radial AA tree. The $\phi$-averaged typical IPR $\ln \ltwotyp$
is shown for the range of $V$ indicated, with $K=2$.
Left panel shows $\ln \ltwotyp$ \emph{vs} $\ln N \propto L$. 
Right panel shows $\ln \ltwotyp$ \emph{vs} $\ln(\ln N) \propto \ln L$, for
values of $V<\vc$ ($\vc =2\sqrt{K}$  $\simeq 2.8$ for $K=2$).
$\Gamma \equiv 1$ is taken. Discussion in text.
}
\label{fig:NumericsAAtypK23}
\end{figure}

We turn now to numerical results obtained via ED (with data 
averaged over randomly chosen values of $\phi\in [0,\pi]$).

The left panel in Fig.\ \ref{fig:NumericsAAtypK23} shows $\ln \ltwotyp$ \emph{vs} $\ln N$ ($\propto L$), for $K=2$.  The critical $\vc =2\sqrt{K}$ $\simeq 2.8$ (taking $\Gamma \equiv 1$). From this it is seen  that for $V>\vc$, clear multifractal behaviour $\ltwotyp \propto N^{-\tau_{2}}$ indeed arises; and that $\tau_{2}$ in practice saturates by $V\simeq 3.4$,  only modestly in excess of the critical $\vc$.

\begin{figure} 
\includegraphics[width=\linewidth]{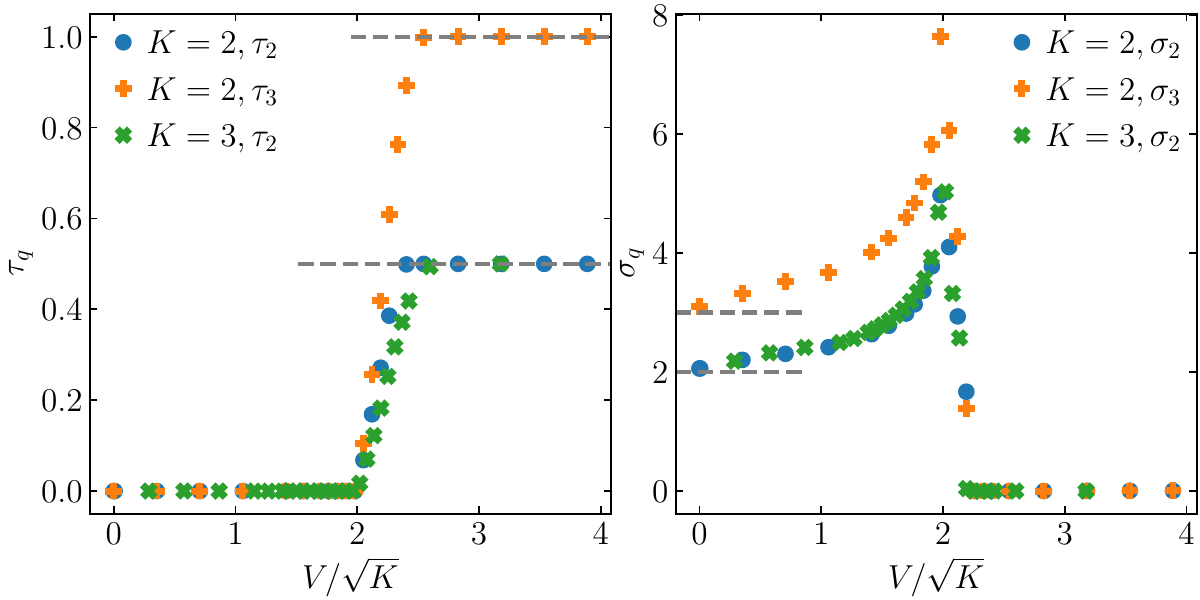}
\caption{Radial AA tree.
\emph{Left panel}: multifractal exponent $\tau_{q}$ \emph{vs} $V/\sqrt{K}$ ($\Gamma \equiv 1$), shown for $q=2,3$ and $K=2, 3$, as indicated. 
In all cases, $\tau_{q}\propto (V-\vc)$ vanishes linearly as $V\to \vc+$,
with the $K$-independent $\vc/\sqrt{K} =2$. Dashed lines denote the large-$V$ asymptotic behaviour $\tau_{q}=\tfrac{1}{2}(q-1)$ (Eq.\ \ref{eq:mf4}).
\emph{Right panel}: $\sigma_{q}$ \emph{vs} $V/\sqrt{K}$ for $q,K$ as indicated; showing $\sigma_{q}\to q$ as $V\to 0$ (dashed lines).
}
\label{fig:NumericsAAtau2sig2} 
\end{figure}

For $V <\vc$ by contrast, $\ln \ltwotyp$ in Fig.\ \ref{fig:NumericsAAtypK23} (left panel) appears scarcely to depend on $\ln N$. What is occurring in this regime is evident in the right panel of Fig.\ \ref{fig:NumericsAAtypK23}, where $\ln \ltwotyp$ is instead plotted \emph{vs} $\ln(\ln N) \propto \ln L$. These plots  are clearly linear, indicating the form $\ltwotyp \propto L^{-\sigma_{2}}$; and which behaviour is consistent with the argument  above, giving 
$\ltwo \propto L^{-2}$ as $V\to 0$ (Eq.\ \ref{eq:AA4}).

To examine the transition between the two distinct behaviours arising for $V\gtrless\vc$, we fit the ED data  to the form
\begin{equation}
\label{eq:AA5}
\ln \mathcal{L}_{q}^{\mathrm{typ}} ~=~-\tau_{q}^{\pd}\ln N -\sigma_{q}^{\pd} \ln\ln N 
+c_{q}^{\pd}.
\end{equation}
The outcome of this is shown in Fig.\ \ref{fig:NumericsAAtau2sig2} where,
for $q=2,3$, the resultant $\tau_{q}$ and $\sigma_{q}$ 
are shown as a function of $V/(\sqrt{K}\Gamma)$ ($\Gamma \equiv 1$). These are given for both $K=2,3$ as indicated, such that the transition should occur for 
$V/\sqrt{K}=2$ in either case; which it evidently does. The following points should be noted:
\begin{enumerate}
\item 
For the multifractal phase, $\tau_{q}$ agrees well with the analytic results of 
Sec.\ \ref{section:AACalcs}. It is indeed seen to saturate for large $V/\sqrt{K}$ 
 to the predicted value $\tau_{q}=\tfrac{1}{2}(q-1)$ (Eq.\ \ref{eq:mf4}).
$\tau_{q}$ decreases with decreasing $V$, vanishes on approach to the transition from the multifractal phase $V\to \vc+$, and does so linearly in $(V-\vc)$, in accordance with  Eq.\ \ref{eq:AA2}. 
\item 
For $V<\vc$ ($V/\sqrt{K} <2$) by contrast, $\tau_{q}=0$.  So in this regime one has the behaviour
\begin{equation}
\label{eq:AA6}
\ltyp ~ \sim ~ \big(\ln N\big)^{-\sigma_{q}^{\pd}}~ \propto ~L^{-\sigma_{q}^{\pd}}
~~~~:~ V <\vc,
\end{equation}
as seen explicitly in Fig.\ \ref{fig:NumericsAAtypK23} (right panel) for $q=2$.
 This is consistent with the deduction in Eq.\ \ref{eq:AA4} that 
$\ltyp \propto L^{-q}$ as $V\to 0$, and from Fig.\ \ref{fig:NumericsAAtau2sig2}
$\sigma_{q}\to q$ as $V\to 0$ is indeed seen to arise.
 On increasing $V$, the $\sigma_{q}$ (Fig.\ \ref{fig:NumericsAAtau2sig2} right) 
are seen to increase  steadily as $V$ approaches $\vc$, then sharply (but we believe 
continuously) decrease in the multifractal phase above $\vc$ (where $\tau_{q}$ itself is non-zero), to an essentially negligible value for $V$ very 
slightly in excess of $\vc$.
\end{enumerate}
We add that the $V<\vc$ behaviour of  $\ltyp$ (Eq.\ \ref{eq:AA6}) is an 
unusually weak form of multifractality, scaling as it does with a power of
$\ln N$ (rather than a power of $N$, as in the multifractal phase $V>\vc$).

\begin{figure} 
\includegraphics[width=\linewidth]{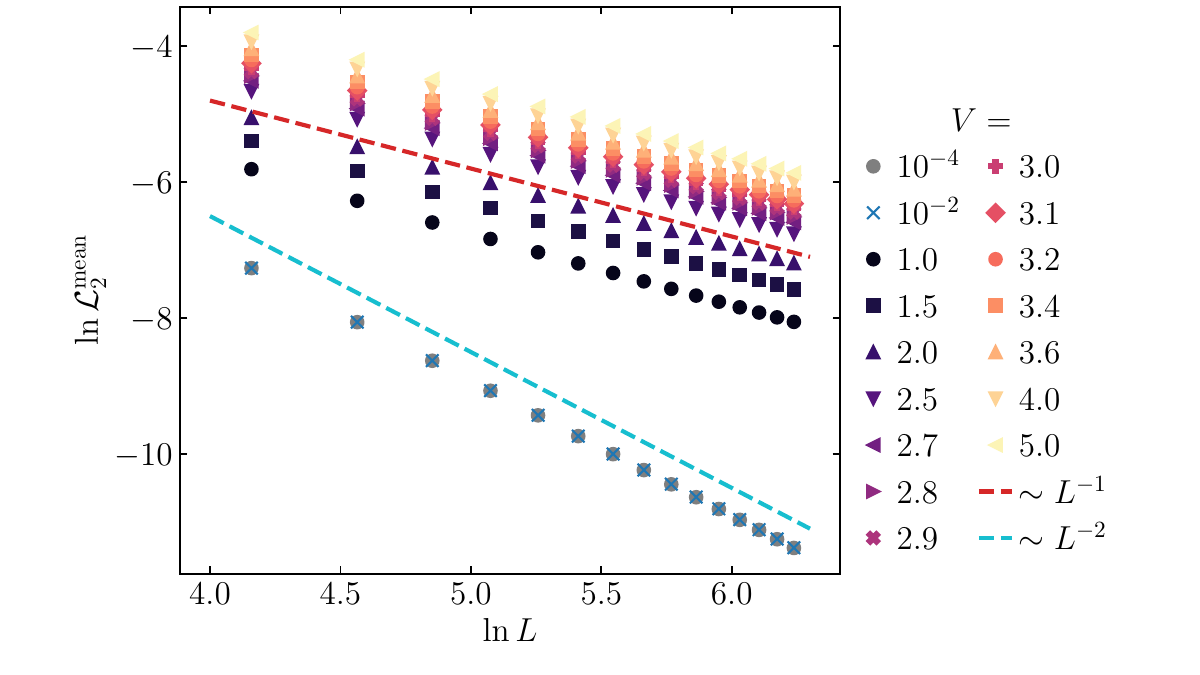}
\caption{Radial AA tree. Mean IPR (for $q=2$ and $K=2$), shown as $\ln\ltwomean$ \emph{vs} $\ln L$, for the wide range of $V$ indicated. For the smallest $V\ll \vc$ shown, $\ltwomean \propto L^{-2}$, in accordance with Eq.\ \ref{eq:AA4} (the data for $V=10^{-4}$ and $10^{-2}$ are indistinguishable); the blue dashed line indicating a slope of $-2$. For larger $V$ by contrast, $\ltwomean \propto L^{-1}$ is found, the slope of $-1$ being indicated by the red dashed line.
}
\label{fig:NumericsAAiprmean} 
\end{figure}

Finally, while our natural focus above has been on the  typical IPR,
 we comment briefly on  the mean IPR; as illustrated in 
Fig.\ \ref{fig:NumericsAAiprmean}, where $\ln \ltwomean$ is shown as a function
of $\ln L$ for a wide range of $V$.
Writing $\ltwomean \propto L^{-\sigma^{\prime}_{2}}$,  Eq.\ \ref{eq:AA4} indeed holds
for the very smallest values of $V\ll \vc$, namely $\sigma_{2}^{\prime}=2 =q$
(with the distribution of $\ltwo$ sufficiently narrow that its mean and typical values are essentially the same). This is evident in 
Fig.\ \ref{fig:NumericsAAiprmean} for the two lowest $V$ shown, $V=10^{-4}$ and 
$10^{-2}$. But on increasing $V$ further -- still deep in the delocalised regime 
$V\ll \vc$ -- the distribution of $\ltwo$ broadens such that $\ltwomean$ and 
$\ltwotyp$ differ appreciably; and $\sigma_{2}^{\prime}$ rapidly crosses over with
increasing $V$ to $\sigma_{2}^{\prime}=1$. It remains at this value across the transition at $V=\vc$, and throughout the  multifractal phase (as known from the considerations of Sec.\ \ref{section:CTs}).
As such, and in marked contrast to the typical $\ltyp$, the behaviour of 
$\lmean$ is in essence `blind' to the transition, reflecting  the  atypicality 
of the mean IPR as a measure of its distribution, 
as discussed extensively in  Sec.\ \ref{section:CTs}.

%%%%%%%%%%%%%%%%%%%%%%%%%%%%%%%%%%%%%%%%%%%%%%%%%%%%%%%%%%%%%%%%%%%%%%%%%%%

\section{Hypercube graph}
\label{section:hcgraph}

So far we have considered Cayley trees, with radial disorder or
analogous incommensurability. Now we turn to the $L$-dimensional hypercube
graph (see Fig.\ \ref{fig:hypercubegraph}), the essential properties 
of which have been summarised in Sec.\ \ref{section:Graphs}.

Before proceeding, let us recall one example of the TBM (Eq.\ \ref{eq:HTBM}) 
on this graph \cite{roy2020fock}; namely the random energy limit, in which all $N=2^{L}$ site energies $\{\mathcal{E}_{i}\}$ are treated as completely uncorrelated 
i.i.d.\  random variables (with  an $N$-dependent variance chosen  to match that of the density of states). This is an example of a quantum random energy model, 
the states of which are known to be invariably delocalised/ergodic~\cite{laumann2014many,baldwin2016manybody,roy2020fock}. Such behaviour is  almost the antithesis of the radial disorder considered here, underscoring the key role correlations in the 
$\{\mathcal{E}_{i}\}$ may play.

As for the Cayley tree of Sec.\ \ref{section:CTs}, we
look first at the case $\hh_{0}\equiv 0$,  so that $\hh \equiv \hhgam$ involves solely the hopping parts of the TBM.
We first show that, as for the tree (Sec.\ \ref{section:radialtree}),
only the totally symmetric combination $\{|\tilde{r}_{1}\rangle \}$ of orbitals 
-- again given by Eq.\ \ref{eq:mtil1basis}, now with $N_{r}=\binom{L}{r}$ -- are connected under $\hhgam$ starting from  the apical `root' site $|0\rangle$ of the graph.

As the topology of the hypercube is quite different from that of a tree,
this is not entirely obvious at first sight. To show it,
it is convenient to split $\hh_{\Gamma}$ (Eq.\ \ref{eq:HTBM}) into 
$\hh_{\Gamma}=\hh_{\Gamma}^{+}+\hh_{\Gamma}^{-}$, where $\hh_{\Gamma}^{\pm}$  connects any basis state $|r_{n}\rangle$ in row $r$ to states in rows $r\pm 1$ 
respectively.  Consider $\hh_{\Gamma}^{+}\sum_{n=1}^{N_{r}} |r_{n}\rangle$.
As noted in Sec.\ \ref{section:Graphs}, each $|r_{n}\rangle$ connects under 
$\hh_{\Gamma}^{+}$ to $(L-r)$ distinct basis states $|(r+1)_{n}\rangle$, and each such $|(r+1)_{n}\rangle$ occurs $(r+1)$ times in 
$\hh_{\Gamma}^{+}\sum_{n=1}^{N_{r}} |r_{n}\rangle$;
so $\hh_{\Gamma}^{+}\sum_{n=1}^{N_{r}} |r_{n}\rangle = \Gamma (r+1)\sum_{n=1}^{N_{r+1}}|(r+1)_{n}\rangle$. 
By the same token, each $|r_{n}\rangle$ connects under $\hh_{\Gamma}^{-}$ to $r$ distinct basis states $|(r-1)_{n}\rangle$, and each such state
occurs $(L-[r-1])$ times in   $\hh_{\Gamma}^{-}\sum_{n=1}^{N_{r}} |r_{n}\rangle$, so
$\hh_{\Gamma}^{-}\sum_{n=1}^{N_{r}} |r_{n}\rangle = \Gamma (L-[r-1])\sum_{n=1}^{N_{r-1}}|(r-1)_{n}\rangle$. 
From these, $\hh_{\Gamma}\sum_{n=1}^{N_{r}} |r_{n}\rangle$ follows.
Using the definition 
$|\tilde{r}_{1}\rangle =N_{r}^{-1/2}\sum_{n=1}^{N_{r}}|r_{n}\rangle$
(Eq.\ \ref{eq:mtil1basis}) of the totally symmetric $|\tilde{r}_{1}\rangle$ 
 then gives
\begin{equation}
\label{eq:rodney1}
\begin{split}
\hh_{\Gamma}^{\pd}|\tilde{r}_{1}^{\pd}\rangle ~=&~
\Gamma \sqrt{r(L-[r-1])}~\big{|}(\widetilde{r-1})_{1}^{\pd}\big\rangle
\\
&~+~
\Gamma \sqrt{(r+1)(L-r)}~\big{|}(\widetilde{r+1})_{1}^{\pd}\big\rangle ;
\end{split}
\end{equation}
holding for any $r=0,1,\cdots L$ (and with $|\tilde{0}_{1}\rangle =|0\rangle$
the apex orbital). Solely the totally symmetric linear combinations of basis states on any row of the hypercube graph are thus indeed coupled under $\hh_{\Gamma}$, starting from $|0\rangle$.

Comparison to Eq.\ \ref{eq:Krylovrecurr} in turn shows Eq.\ \ref{eq:rodney1}
to be precisely the  tridiagonal Krylov Hamiltonian, with Krylov basis 
$|k_{r}\rangle =|\tilde{r}_{1}\rangle$  generated under 
$\hh \equiv \hhgam$ from $|k_{0}\rangle =|0\rangle$; and with 
$b_{r}=\Gamma\sqrt{r(L-[r-1])}$ and $a_{r}=0$.
As for the Cayley tree the hypercube graph thus fragments
and, to be consistent with the notation of Sec.\ \ref{section:CTs},
we denote by $\hhcal_{\Gamma}$ the Krylov Hamiltonian in the basis 
$\mathcal{K}(\hh,|0\rangle)=
\mathrm{span}\{ |\tilde{r}_{1}\rangle: r=0,1,2,\cdots L\}$, viz.\
\begin{equation}
\label{eq:Hgamma0orbTFI}
\hhcal_{\Gamma}^{\pd} =
\Gamma  \sum_{r=1}^{L} \Big(\sqrt{r (L-[r-1])}~
|\widetilde{(r-1)}_{1}^{\pd}\rangle \langle \tilde{r}_{1}^{\pd}| +\mathrm{h.c.} 
\Big).
\end{equation}

For use below, let us comment briefly on the density of states (DoS) of this 
tridiagonal $\hhcal_{\Gamma}$; denoted by $D_{\Gamma}(\omega)$ (and normalised
such that $\int d\omega~ D_{\Gamma}(\omega)=1$).  
The properties of $D_{\Gamma}(\omega)$ we want can be obtained in several ways, but here it suffices to use a result from~\cite{VBalasubEtAlPRD2023} which relates the DoS to the Lanczos coefficients $b_{r}=\Gamma\sqrt{r(L-[r-1])}$.
Defining $x=r/L$ ($\in [0,1]$), and regarding $b(x) \equiv b_{r=xL}$ as a continuous function of $x$ as $L\to \infty$, then~\cite{VBalasubEtAlPRD2023}
\begin{equation}
\label{eq:DosHypKryl1}
D_{\Gamma}(\omega)=\int_{0}^{1}dx~\frac{\Theta\big(4 b^{2}(x) -\omega^{2}\big)}{\pi\sqrt{4 b^{2}(x) -\omega^{2}}}.
\end{equation}
In the present case $b(x) \equiv L\Gamma x(1-x)$, and evaluation of
Eq.\ \ref{eq:DosHypKryl1} gives
\begin{equation}
\label{eq:DosHypKryl2}
D_{\Gamma}(\omega)=\frac{\Theta(\Gamma L- |\omega|)}{2\Gamma L},
\end{equation}
which has a full width $2\Gamma L$, a vanishing mean and a variance
$\int d\omega~\omega^{2} D_{\Gamma}(\omega) = \tfrac{1}{3}\Gamma^{2}L^{2}$
$\propto L^{2}$.\footnote{We add that Eq.\ \ref{eq:DosHypKryl1} also encompasses 
$D_{\Gamma}(\omega)$ for the Krylov $\hhcal_{\Gamma}$ (Eq.\ \ref{eq:Hgamma0orb}) arising for the  tree as $L\to \infty$. In that case, $b_{r}=\sqrt{K}\Gamma$ 
is $r$-independent (Sec.\ \ref{sec:Krylov1}), so $b(x) =\sqrt{K}\Gamma$ is 
$x$-independent. Eq.\ \ref{eq:DosHypKryl1} then (trivially) yields precisely 
the DoS of an infinite 1d chain with a nearest neighbour hopping of 
$\sqrt{K}\Gamma$~\cite{Economoubook};
as it should by the arguments given in  Sec.\ \ref{section:Puretree}.}

Now consider the effect of $\hh_{0}$. As for the Cayley tree, this 
is expressible in the row-resolved form Eq.\ \ref{eq:H0rowresolve}. And by 
the same argument given at the start of Sec.\ \ref{section:radialtree},
the fragmentation arising under $\hhgam$ alone  persists in the presence of radial disorder ($\mathcal{E}_{r_{n}}=\er$).  In particular, the separable part of the Hamiltonian which contains the root site,  again denoted by 
$\hhcal =\hhcal_{0}+\hhcal_{\Gamma}$,  is
\begin{equation}
\label{eq:corr1dtbmTFI}
\hhcal ~=~ \sum_{r=0}^{L}\erpd 
|\tilde{r}_{1}^{\pd}\rangle\langle\tilde{r}_{1}^{\pd}|
 ~+~\hhcal_{\Gamma}^{\pd}.
\end{equation}
This is the full Krylov Hamiltonian, on which we focus
(now with $a_{r}=\er$ and  $b_{r}=\Gamma\sqrt{r(L-[r-1])}$.

As for the tree case, we consider the $L+1$ site-energies $\{\er\}$ to
be independent random variables. Since, as above, the hoppings in $\hhcal_{\Gamma}$ generate a spectral density $D_{\Gamma}(\omega)$ whose variance is $\propto L^{2}$, then to ensure the hopping $\hhcal_{\Gamma}$ does not trivially overwhelm the disorder as $L\to \infty$, the distribution $P_{\er}(\er)$ of $\er$ must have a variance  which essentially matches this behaviour.
While there are many possible choices for this, for our ED calculations 
(Sec.\ \ref{section:Numericshyper}) we consider in practice the distribution 
$P_{\er}(\er)$ to be normal, with zero mean and a  variance of 
$W^{2}r(L-[r-1])\equiv W^{2}L^{2}x(1-x)$
(the $r$-dependence of which matches that of the hoppings $b_{r}^{2}$).
With this, we expect (and find) exponential localisation in the 
$\tilde{r}_{1}$-basis for non-zero disorder strengths $W$;
i.e., just as in Eq.\ \ref{eq:Calc1}, 
$|A_{E,r}|^{2} =\mathcal{N}^{-1}\exp(-|r-r_{E}|/\xi)$,
with localisation centre $r_{E}$ and localisation length $\xi$.
Given this, without regard to the precise form of the disorder distribution,
 we can of course proceed to analytical considerations analogous to those given
in Sec.\ \ref{section:analyticstree} for the tree graph.

%%%%%%%%%%%%%%%%%%%%%%%%%%%%%%%%%%%%%%%%%%%%%%%%%%%%%%%%%%%%%%%%%%%%%%%%%%%
\subsection{Analytical considerations}
\label{section:HypercubeCalcs}

In considering the hypercube, we will focus for brevity on $q\geq 1$.
Note also that Eqs.\ \ref{eq:eigexp1}-\ref{eq:Lqdef2} of Sec.\ \ref{section:radialtree} again hold, now with the number of sites on row $r$ of the graph given by $N_{r}=\binom{L}{r}$. In particular, the IPRs in the original hypercube basis are given by Eq.\ \ref{eq:Lqdef2};
i.e.\ with Eq.\ \ref{eq:Calc1} for $|A_{E,r}|^{2}$,
\begin{equation}
\label{eq:hyp1}
\mathcal{N}^{q}\Lqp(r_{E}^{\pd}) =
\sum_{r=0}^{L}\frac{1}{\binom{L}{r}^{(q-1)}}e^{-q|r-r_{E}^{\pd}|/\xi}.
\end{equation}
Separating this $r$-sum into $r< r_{E}$ and $r>r_{E}$, and using 
$\binom{L}{r}=\binom{L}{L-r}$, shows the physically obvious symmetry
\begin{equation}
\label{eq:hyp4}
\Lqp(r_{E}^{\pd}) =\Lqp(L-r_{E}^{\pd}).
\end{equation}
Only $r_{E}\leq L/2$ need then be considered, with the typical and mean values of
the distribution $P_{\Lq}(\Lq)$ given from Eq.\ \ref{eq:Calc7} as
\begin{equation}
\label{eq:hyp7}
\begin{split}
\ln \ltyp =&\frac{1}{L+1}\Bigg[2\sum_{r_{E}=0}^{\tfrac{L}{2}}
\ln \mathcal{L}_{q}^{\pd} (r_{E}^{\pd})
-\ln\mathcal{L}_{q}^{\pd} \big(\tfrac{L}{2} \big)
\Bigg]
\\
\lmean =& \frac{1}{L+1}\Bigg[2\sum_{r_{E}=0}^{\frac{L}{2}}\mathcal{L}_{q}^{\pd} (r_{E}^{\pd})
-\mathcal{L}_{q}^{\pd} \big(\tfrac{L}{2} \big)
\Bigg].
\end{split}
\end{equation}

In parallel to Sec.\ \ref{section:analyticstree} for the tree,
let us first look at this in the two cases of $\xi \ll 1$ and $\xi \gg 1$, again considering $r_{E}$ to be $\propto L$ with probability unity. For $\xi \ll 1$, 
Eq.\ \ref{eq:hyp1} is asymptotically dominated by the term with $r=r_{E}$, so
\begin{equation}
\label{eq:hyp5}
\Lqp(r_{E}^{\pd})~\overset{\xi \ll 1}{\sim}~
\binom{L}{r_{E}}^{-(q-1)}
\end{equation}
(using also $\mathcal{N} \sim 1$ for $\xi \ll 1$, see Eq.\ \ref{eq:Calc2}).
For $\xi \gg 1$ by contrast (and with $r_{E}\leq L/2$),
the $r=0$ term in Eq.\ \ref{eq:hyp1} dominates, to give
\begin{equation}
\label{eq:hyp6}
\mathcal{N}^{q}\Lqp(r_{E}^{\pd})~\overset{\xi \gg 1}{\sim}~
e^{-q r_{E}^{\pd}/\xi} 
\end{equation}
(with $\mathcal{N} \sim 2\xi$ in this case).

\begin{figure} 
\includegraphics[width=\linewidth]{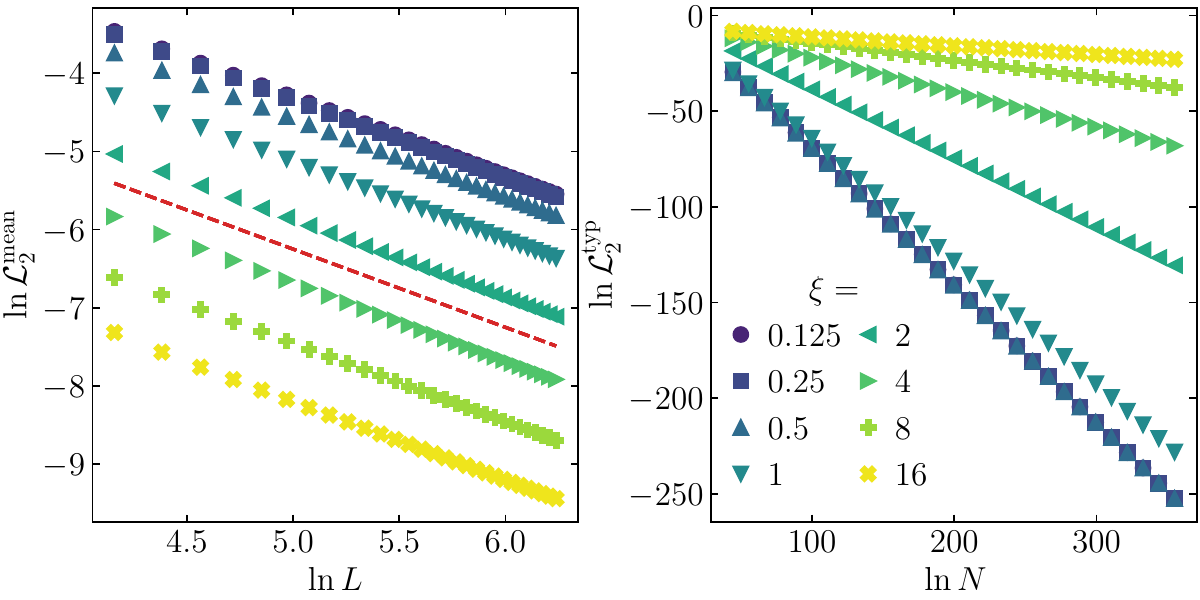}
\caption{Hypercube graph. Mean $\ltwomean$ (left) and typical $\ltwotyp$  
(right) IPR for $q=2$, computed from the full analytical expressions 
Eqs.~\ref{eq:hyp1},\ref{eq:hyp7}, for the range of $\xi$ indicated.
$\ln \ltwomean$ is plotted \emph{vs} $\ln L$, while $\ln \ltwotyp$ is plotted against $\ln N  =L \ln 2$. Red dashed line in the $\ln \ltwomean$ plot indicates a slope of $-1$.  
}
\label{fig:analythyper}
\end{figure}

Using the \emph{full} Eq.\ \ref{eq:hyp1} for $\Lq(r_{E})$, together with 
Eq.\ \ref{eq:hyp7}, Fig.\ \ref{fig:analythyper}
shows the mean and typical IPRs for $q = 2$ over a wide range of $\xi$; 
with $\ltwomean$ shown as a function of $L$ and $\ltwotyp$
as a function of the total number of sites on the graph, $N=2^{L}$.
Precisely the same three qualitative features stand out as were apparent 
in Fig.\ \ref{fig:analytical} for the Cayley tree:
that $\ltwomean \propto 1/L$; that $\ltwotyp$ is many orders of magnitude smaller than $\ltwomean$; 
and that $\ltwotyp \propto N^{-\tau_{2}}$ where the multifractal exponent 
$\tau_{2}$ decreases with increasing $\xi$ (i.e.\ decreasing disorder).

To understand these features, we again analyse the behaviour in the two 
limits of large and small $\xi$.  First, consider $\lmean$.
For $\xi \ll 1$, Eq.\ \ref{eq:hyp5} used in Eq.\ \ref{eq:hyp7} shows
that the latter sum is dominated by $r_{E}=0$ for $L\gg 1$, giving
\begin{equation}
\label{eq:hyp8}
\lmean ~\overset{L \gg 1}{\sim}~ \frac{2}{L} ~~~~~~:~ \xi \ll 1  .
\end{equation}
For $\xi \gg 1$, Eqs. \ref{eq:hyp6},\ref{eq:hyp7} correspondingly give
\begin{equation}
\label{eq:hyp9}
\lmean ~\overset{L \gg 1}{\sim}~\frac{1}{q(2\xi)^{q-1}}
\frac{1}{L }
~~~~~~:~\xi \gg 1 .
\end{equation}
So just as for the Cayley tree, whether $\xi \gg 1$ or $\xi \ll 1$,
one indeed finds $\lmean \propto 1/L$.

Now turn to $\ltyp$, considering first $\xi \gg 1$ for which 
Eqs.\ \ref{eq:hyp6},\ref{eq:hyp7} give
\begin{equation}
\label{eq:hyp10}
\ltyp\overset{L\gg 1}{\sim}\frac{1}{(2\xi)^{q}}\exp\Big[ -\frac{q L}{4\xi}\Big]
\sim~ N^{-\frac{q}{4\xi \ln 2}}
~~~~:~\xi \gg 1 
\end{equation}
with the power-law decay in $N$ reflecting multifractality.
From the definition  of the multifractal exponent (Eq.\ \ref{eq:Calc9}),
$\tau_{q}=-\ln\ltyp/\ln N$, so
\begin{equation}
\label{eq:hyp11}
\tau_{q}^{\pd} ~=~\frac{q}{4 \xi \ln 2} ~~~~:~ \xi \gg 1.
\end{equation}
For $\xi \ll 1$, Eq.\ \ref{eq:hyp5} used in Eq.\ \ref{eq:hyp7} gives
\begin{equation}
\label{eq:hyp12}
\begin{split}
\ln \ltyp \overset{\xi \ll 1}{\sim}
&-\frac{(q-1)}{L+1}\sum_{r_{E}=0}^{L} \ln\binom{L}{r_{E}}
\\
&=-\frac{(q-1)}{L+1}\ln\Bigg[\frac{\big(\Gamma(1+L)\big)^{L}}{G(1+L)G(2+L)}
\Bigg]
\end{split}
\end{equation}
with $G(z)$ the Barnes G-function; from the large-$L$ asymptotics of which
$\tau_{q}=-\ln \ltyp/L\ln 2$ follows as
\begin{equation}
\label{eq:hyp15}
\tau_{q}^{\pd} =\frac{(q-1)}{2\ln 2}\left[ 1 -\frac{\ln L}{L} +\mathcal{O}(L^{-1})\right]
~~~~:~ \xi \ll 1 .
\end{equation}
Eqs.\ \ref{eq:hyp11},\ref{eq:hyp15} for $\tau_{q}(\xi)$
have of course been obtained asymptotically, for large- and small-$\xi$.
As with their counterparts for the Cayley tree (Sec.\ \ref{section:tauq}), however, and for the same basic reasons, the $\tau_{q}$ for $L\to \infty$ can 
be determined exactly within the framework employed.  We first simply state the outcome of the latter, before returning to Eqs.\ \ref{eq:hyp11},\ref{eq:hyp15}.

%%%%%%%%%%%%%%%%%%%%%%%%%%%%%%%%%%%%%%%%%%%%%%%%%%%%%%%%%%%%%%%%%%%%%%%%%%%
\subsubsection{Multifractal exponents $\tau_{q}$}
\label{section:tauqhyper}

\begin{figure}
\includegraphics[width=\linewidth]{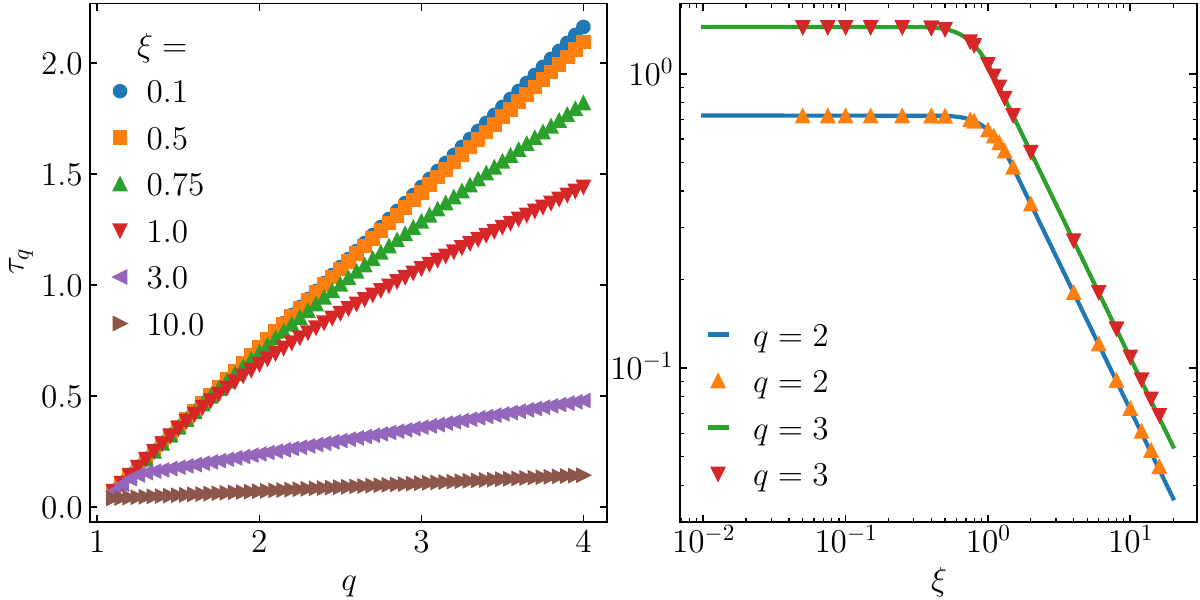}
\caption{Hypercube graph. Multifractal exponents $\tau_{q}$, given by  
Eqs.\ \ref{eq:hyp16},\ref{eq:hyp17}.
\emph{Left panel}: $\tau_{q}$ \emph{vs} $q$, shown for the range of $\xi$
indicated. 
\emph{Right panel}: $\tau_{q}$ \emph{vs} $\xi$, shown for $q=2$ and $3$. 
The small-$\xi$ asymptote $\tau_{q}=(q-1)/(2\ln2)$ (Eq.\ \ref{eq:hyp15}) is evident, as is the behaviour $\tau_{q}=q/(4\xi\ln2)$ occurring for all $\xi\geq \xi_{c}$ 
(Eq.\ \ref{eq:hyp16}). Solid lines denote results from Eqs.\ \ref{eq:hyp16},\ref{eq:hyp17} (which are exact as $L\to \infty$); points  show $\tau_{q}=-\ln\ltyp/L\ln2$
calculated numerically using  Eqs.~\ref{eq:hyp1},\ref{eq:hyp7} for $L=512$.
}
\label{fig:tau2-analythyper}
\end{figure}

For $\xi \geq \xi_{c}$, $\tau_{q}$ as a function of $\xi$ is given by
\begin{equation}
\label{eq:hyp16}
\tau_{q}^{\pd} ~=~ \frac{q}{4\xi \ln 2}
~~~~:~\xi \geq \xi_{c}^{\pd}:=\frac{q}{(q-1)2\ln 2}
\end{equation}
(with $\xi_{c}$ thus defined). This is the large-$\xi$ form Eq.\ \ref{eq:hyp11} again, but in fact holds for all $\xi\geq \xi_{c}$.  For $\xi\leq \xi_{c}$, the 
$\xi$-dependence of $\tau_{q}$ is given for any chosen $q>1$ by solution of the parametric equations 
\begin{subequations}
\label{eq:hyp17}
\begin{align}
\tau_{q}^{\pd}=&\frac{(q-1)}{\ln 2}\Big[
\frac{1}{2}- y -(1-y)\ln(1-y)\Big]
\label{eq:hyp17a}
\\
\xi =&-\frac{q}{(q-1)}
\Big[\ln y +\big(\tfrac{1}{y}-1\big)\ln(1-y)\Big]^{-1}
\label{eq:hyp17b}
\end{align}
\end{subequations}
with $y$ in the interval $[0,\tfrac{1}{2}]$.
As $y\to \tfrac{1}{2}-$, one sees that $\xi\to\xi_{c}-$ and 
$\tau_{q}(\xi) \to q/(4\xi_{c}\ln 2)$ ($\equiv (q-1)/2$) thus connects continuously to
Eq.\ \ref{eq:hyp16}.  $\tau_{q}(\xi)$ increases continuously with decreasing 
$\xi<\xi_{c}$, which corresponds to $y<1/2$ in Eqs.\ \ref{eq:hyp17}.
In particular, $y\ll 1/2$ corresponds to $\xi \ll \xi_{c}$, for which
Eq.\ \ref{eq:hyp17a} gives $\tau_{q}\to (q-1)/(2\ln 2)$ -- indeed recovering
Eq.\ \ref{eq:hyp15}.

These same equations naturally determine the $q$-dependence of $\tau_{q}$ for
some given $\xi$.
Eq.\ \ref{eq:hyp16} holds for all $q\geq q_{*}:=[1-\tfrac{1}{2\xi\ln 2}]^{-1}$,
which necessitates $2\xi\ln 2>1$. For $q\leq q_{*}$ (and any chosen $\xi$), 
$\tau_{q}$ is again determined parametrically by Eqs.\ \ref{eq:hyp17}. Note also that as $q \to 1$, Eqs.\ \ref{eq:hyp17} are governed by $y\to 0$, 
Eq.\ \ref{eq:hyp17a} yielding $\tau_{q}\to (q-1)/(2\ln 2)$ \emph{independently} of 
$\xi$ (and recovering correctly $\tau_{1}=0$, as mandated trivially by wavefunction normalisation).

The results above are  exemplified in Fig.\ \ref{fig:tau2-analythyper},
the right panel of which shows the $\xi$-dependence of $\tau_{q}$ for
$q=2$ and $q=3$. Solid lines show the requisite $L\to \infty$  results from 
Eqs.\ \ref{eq:hyp16},\ref{eq:hyp17}, while points  show $\tau_{q}=-\ln\ltyp/L\ln2$
calculated numerically using  Eqs.~\ref{eq:hyp1},\ref{eq:hyp7} for $L=512$.
As they should, the two evidently agree very well (though the  $\mathcal{O}(\ln L/L)$ finite-size corrections of Eq.\ \ref{eq:hyp15} for $\xi \ll 1$ are just about discernible). The figure also shows clearly both the large-$\xi$ behaviour
$\tau_{q}\propto 1/\xi$ (Eqs.\ \ref{eq:hyp11},\ref{eq:hyp16}), 
and the small-$\xi$ limiting behaviour $\tau_{q}=(q-1)/(2\ln 2)$ 
(Eq.\ \ref{eq:hyp15}). The left panel of Fig.\ \ref{fig:tau2-analythyper} 
shows the corresponding $q$-dependence of $\tau_{q}$ obtained from 
Eqs.\ \ref{eq:hyp16},\ref{eq:hyp17} for a wide range of $\xi$, from which the non-monotonicity in $q$ that is symptomatic of multifractality is evident.
Only for the smallest $\xi=0.1$ is the behaviour $\tau_{q}=(q-1)/(2\ln 2)$
seen over the entire $q$-range shown in practice.

%%%%%%%%%%%%%%%%%%%%%%%%%%%%%%%%%%%%%%%%%%%%%%%%%%%%%%%%%%%%%%%%%%%%%%%%
\subsection{Numerical results: hypercube graph}
\label{section:Numericshyper}

\begin{figure}
\includegraphics[width=\linewidth]{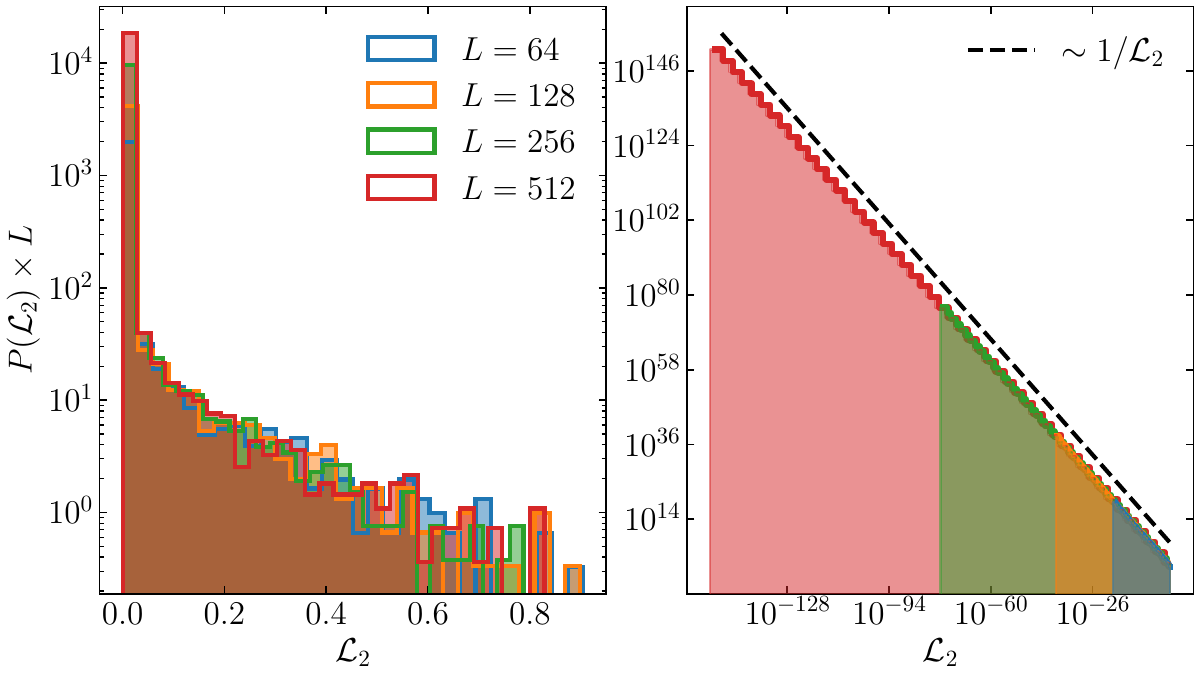}
\caption{
Distribution $P_{\mathcal{L}_{2}}(\mathcal{L}_{2})$ of ${\cal L}_2$ for the hypercube graph, obtained from ED (with $W=1$), for the linear system sizes indicated
(with $N = 2^{L}$). Just as in Fig.\ \ref{fig:L2-dist} for the tree, 
$L\times P_{\ltwo}(\ltwo)$ \emph{vs} $\ltwo$ is plotted. Left panel shows data with linear bins,  right panel with logarithmic bins. In right panel, the black dashed line
shows the $P_{\mathcal{L}_{2}}(\mathcal{L}_{2})\propto 1/{\cal L}_2$ behaviour.
}
\label{fig:L2distribution-ED-hypercube}
\end{figure}

The results above are well supported  by ED calculations, obtained in the
same way as for the tree graph (Sec.\ \ref{section:TreeNumerics}), and
with the normal distribution for $P_{\er}(\er)$ as specified after 
Eq.\ \ref{eq:corr1dtbmTFI} above.

We have not yet commented on the distribution $P_{\Lq}(\Lq)$ for the
hypercube graph. For $q=2$, Fig.\ \ref{fig:L2distribution-ED-hypercube} shows 
$L\times P_{\mathcal{L}_{2}}(\mathcal{L}_{2})$ \emph{vs} $\ltwo$ 
obtained from ED (as in Fig.\ \ref{fig:L2-dist} for the tree). Just as for
the tree case the distribution is heavy tailed, and the same power-law decay 
$P_{\mathcal{L}_{2}}(\mathcal{L}_{2})\propto 1/\mathcal{L}_{2}$ is clearly seen 
(with the same implications discussed at the end of 
Sec.\ \ref{section:analyticstree}). One can also deduce this form analytically, in direct parallel to the calculation for the tree in 
Eq.\ \ref{eq:Calc13} ff for the case $\xi \gg 1$, where (Eq.\ \ref{eq:hyp6}) 
$\mathcal{L}_{q}(r_{E}) \equiv c_{1}e^{-c_{2}r_{E}}$ has the same form as in 
Eq.\ \ref{eq:Calc13}. For $\xi \ll 1$, $\mathcal{L}_{q}(r_{E})$ has a different form (Eq.\ \ref{eq:hyp5}); but here too it can be shown that $P_{\mathcal{L}_{2}}(\mathcal{L}_{2})\propto 1/\mathcal{L}_{2}$.

\begin{figure}
\includegraphics[width=\linewidth]{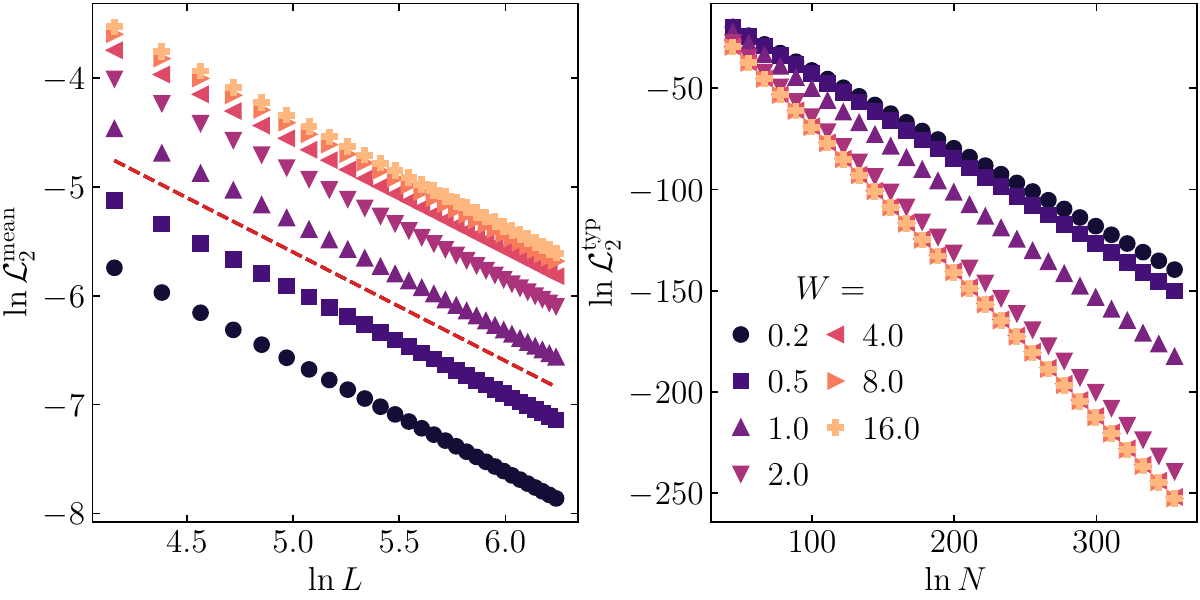}
\caption{For the hypercube graph, $\ln\ltwomean$  (left) and $\ln\ltwotyp$ (right), obtained from ED for different disorder strengths $W$ as indicated. 
$\ln\ltwomean$ is plotted against  $\ln L$, whereas $\ln\ltwotyp$ is plotted against $\ln N = L \ln 2$.  Red dashed line in the left panel for $\ln \ltwomean$ indicates a slope of $-1$. 
}
\label{fig:numericalhyper}
\end{figure}

ED results for $\ltwomean$ \emph{vs} $L$ and $\ltwotyp$ \emph{vs} $N$
are shown in Fig.\ \ref{fig:numericalhyper}, for a range of disorder strengths 
$W$ as indicated. These  confirm that $\ltwomean \propto 1/L$, and that 
$\ltwotyp\propto N^{-\tau_{2}}$ with a fractal exponent $\tau_{2}$ which increases with increasing $W$; and can be compared to their analytical counterparts 
shown in Fig.\ \ref{fig:analythyper}.

\begin{figure}
\includegraphics[width=\linewidth]{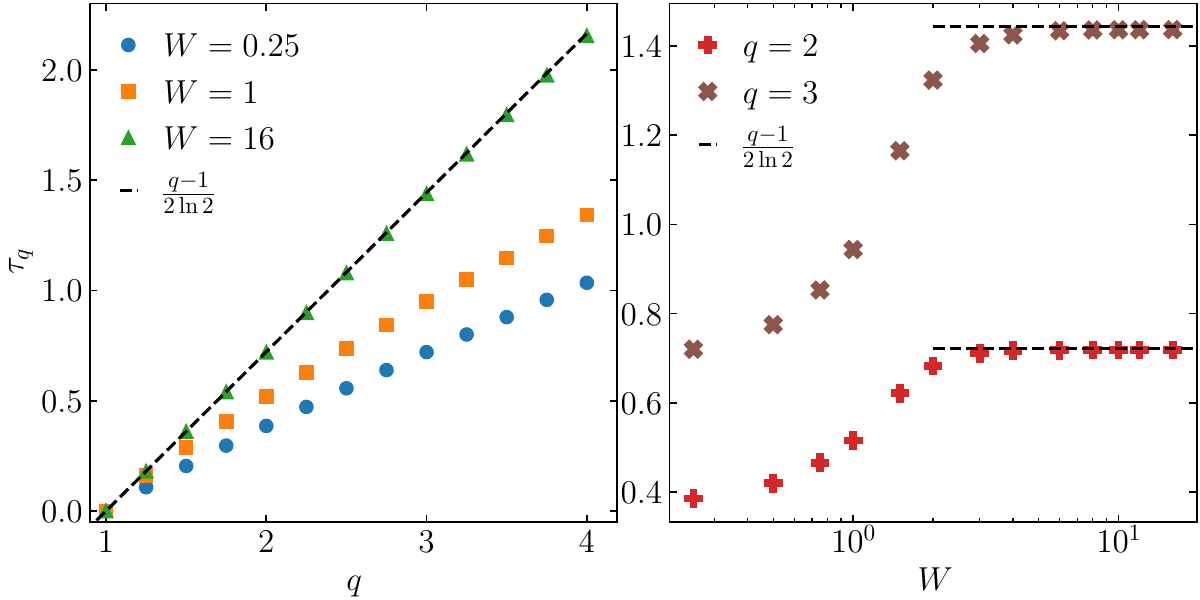}
\caption{
ED results for multifractal exponents $\tau_{q}$ of the hypercube graph. 
\emph{Left panel}: $\tau_{q}$ \emph{vs} $q$ for $W=0.25,1,16$. 
For strong disorder, $W=16$, the dashed line shows the asymptotic behaviour
$(q-1)/(2\ln 2)$. Corresponding analytic results for different $\xi$ are given
in Fig.\ \ref{fig:tau2-analythyper} (left panel).
\emph{Right panel}: $\tau_{q}$ \emph{vs} $W$, shown for $q=2$ and $q=3$.  Dashed lines show the asymptote $(q-1)/(2\ln 2)$ (Eq.\ \ref{eq:hyp15}). Analytic results 
for $\tau_{q}$ \emph{vs} $\xi$  with $q=2,4$ are given in 
Fig.\ \ref{fig:tau2-analythyper} right panel. 
}
\label{fig:tauq-anal+ED-hypercube}
\end{figure}

Multifractal exponents $\tau_{q}$ obtained  from ED are shown in
Fig.\ \ref{fig:tauq-anal+ED-hypercube}. As in Fig.\ \ref{fig:numericalhyper} 
for $q=2$, plots of $\ln\ltyp$ \emph{vs} $\ln N$ show clear linearity for any
$q$,  enabling $\tau_{q}=-\ln\ltyp/\ln N$ to be read off.
For $q=2$ and $3$, the right panel of Fig.\ \ref{fig:tauq-anal+ED-hypercube} shows the
evolution of $\tau_{q}$ as a function of $W$; analytical results for the 
$\xi$-dependence of $\tau_{q}$ are shown in Fig.\ \ref{fig:tau2-analythyper} 
(right panel). For small values of $\xi$, Eq.\ \ref{eq:hyp15} predicts the 
asymptotic behaviour $\tau_{q}=(q-1)/[2\ln 2]$. That behaviour is indeed seen in the full ED results for strong disorder (reflecting the fact that 
$\tau_{q}=(q-1)/[2\ln 2]$  does not \emph{per se}  depend on $\xi$, so should hold for sufficiently large disorder strengths $W$ for which $\xi \ll 1$).
ED results for $\tau_{q}$ \emph{vs} $q$ are shown in the left panel of
Fig.\ \ref{fig:tauq-anal+ED-hypercube}, for $W=0.25,1,16$; corresponding  analytical results for different $\xi$'s are shown in Fig.\ \ref{fig:tau2-analythyper} (left panel). For strong disorder, $W=16$, ED results indeed show the asymptotic behaviour 
$\tau_{q}=(q-1)/(2\ln 2)$ over the wide $q$-range shown, just as they do for 
the $\xi =0.1$ case in the left panel of Fig.\ \ref{fig:tau2-analythyper}.

%%%%%%%%%%%%%%%%%%%%%%%%%%%%%%%%%%%%%%%%%%%%%%%%%%%%%%%%%%%%%%%%%%%%%%%%
\section{Concluding remarks \label{sec:conclusion}}
\label{section:summary}

We close with a brief summary, and a few remarks by way of outlook. The central 
outcome of this work is a concrete demonstration -- through a combination of analytical and numerical results -- of the emergence of robust multifractality in eigenstates of high-dimensional graphs with correlated disorder. Focusing on rooted trees and hypercubes, we introduced a class of disorder correlations, 
termed radial disorder, in which the site energies are identical for all sites at a given distance from the root, but vary independently across different distances. This correlation pattern was shown to lead to an emergent fragmentation of the graph into one-dimensional chains with uncorrelated disorder. The chain emanating from the root can be naturally identified as the Krylov chain generated from  the root-localised initial state.

This structure makes the mechanism underlying the multifractality analytically transparent. The interplay between conventional Anderson localisation along the one-dimensional chain, and the exponential growth in the number of sites with distance from the root (whose linear combinations form the effective chain sites), leads to the multifractality. Since both ingredients are inherently robust, so too is the multifractality. A particularly notable feature is the exceptionally broad distribution of IPRs, which leads to a marked difference between the scalings of the mean and typical IPRs. We find that the mean and typical IPRs scale as power laws with the number of generations and the total number of sites, respectively; implying that it is the latter which defines a multifractal exponent. In short, radial disorder transforms high-dimensional graphs into a natural stage for multifractality to emerge, persist, and be understood.

That being said, it is important to mention that the current work itself introduces the idea and a model of radial disorder in high-dimensional graphs. The results suggest it is likely to provide fertile ground for investigating 
%various
further aspects of localisation physics thereon, such as the effects on dynamics and transport due to radial disorder. On a more fundamental level, however, it would be of interest to identify and classify the broader class of high-dimensional 
%expander 
graphs that can host this phenomenon, and to uncover the deeper mathematical reasons behind it. In the shorter term, it will be instructive to study the impact of perturbatively small additional disorder along the rows of the graph, on top of the radial disorder, by mapping the problem to weakly coupled one-dimensional chains.

% \comment{
% Titles of figs in text:
% Fig 1     tree-schematic.pdf
% Fig 2     hypercube.pdf
% Fig 3     tree-mean-typ-irp-analytical.pdf
% Fig 4     tree-tauq-analytical.pdf
% Fig 5     tree-IPR-dist-K2-W1.pdf
% Fig 6     tree-mean-typ-ipr-ED.pdf
% Fig 7     tree-tauq-EED.pdf
% Fig 8     AA-iprs-typ.pdf
% Fig 9     AA-tauq-sigq.pdf
% Fig10     AA-iprs-mean.pdf
% Fig11     hypercube-mean-typ-ipr-analytical.pdf
% Fig12     hypercube-tauq-analytical.pdf
% Fig13     hypercube-ipr-dist-W1.pdf
% Fig14     hypercube-mean-typ-ipr-ED.pdf
% Fig15     hypercube-tauq-ED.pdf
% }

\begin{acknowledgements}
DEL expresses his thanks for the hospitality of both the International Centre for Theoretical Sciences (ICTS-TIFR), and the Physics Department, Indian
Institute of Science (IISc), Bengaluru. SR acknowledges support from SERB-DST, Government of India, under Grant No. SRG/2023/000858; from the Department of Atomic Energy under Project No. RTI4001; and by a Max Planck Partner Group grant between ICTS-TIFR, Bengaluru and MPIPKS, Dresden.
\end{acknowledgements}

%%%%%%%%%%%%%%%%%%%%%%%%%%%%%%%%%%%%%%%%%%%%%%%%%%%%%%%%%%%%%%%%%%%%%%%%%%%%%%%%%

%%%%%%%%%%%%%%%%%%%%%%%%%%%%%%%%%%%%%%%%%%%%%%%%%%%%%%%%%%%%%%%%%%%%%%%%%%%%%%%%%%%%%%%%%

\bibliography{refs}

\begin{thebibliography}{87}%
\makeatletter
\providecommand \@ifxundefined [1]{%
 \@ifx{#1\undefined}
}%
\providecommand \@ifnum [1]{%
 \ifnum #1\expandafter \@firstoftwo
 \else \expandafter \@secondoftwo
 \fi
}%
\providecommand \@ifx [1]{%
 \ifx #1\expandafter \@firstoftwo
 \else \expandafter \@secondoftwo
 \fi
}%
\providecommand \natexlab [1]{#1}%
\providecommand \enquote  [1]{``#1''}%
\providecommand \bibnamefont  [1]{#1}%
\providecommand \bibfnamefont [1]{#1}%
\providecommand \citenamefont [1]{#1}%
\providecommand \href@noop [0]{\@secondoftwo}%
\providecommand \href [0]{\begingroup \@sanitize@url \@href}%
\providecommand \@href[1]{\@@startlink{#1}\@@href}%
\providecommand \@@href[1]{\endgroup#1\@@endlink}%
\providecommand \@sanitize@url [0]{\catcode `\\12\catcode `\$12\catcode
  `\&12\catcode `\#12\catcode `\^12\catcode `\_12\catcode `\%12\relax}%
\providecommand \@@startlink[1]{}%
\providecommand \@@endlink[0]{}%
\providecommand \url  [0]{\begingroup\@sanitize@url \@url }%
\providecommand \@url [1]{\endgroup\@href {#1}{\urlprefix }}%
\providecommand \urlprefix  [0]{URL }%
\providecommand \Eprint [0]{\href }%
\providecommand \doibase [0]{https://doi.org/}%
\providecommand \selectlanguage [0]{\@gobble}%
\providecommand \bibinfo  [0]{\@secondoftwo}%
\providecommand \bibfield  [0]{\@secondoftwo}%
\providecommand \translation [1]{[#1]}%
\providecommand \BibitemOpen [0]{}%
\providecommand \bibitemStop [0]{}%
\providecommand \bibitemNoStop [0]{.\EOS\space}%
\providecommand \EOS [0]{\spacefactor3000\relax}%
\providecommand \BibitemShut  [1]{\csname bibitem#1\endcsname}%
\let\auto@bib@innerbib\@empty
%</preamble>
\bibitem [{\citenamefont {Anderson}(1958)}]{anderson1958absence}%
  \BibitemOpen
  \bibfield  {author} {\bibinfo {author} {\bibfnamefont {P.~W.}\ \bibnamefont
  {Anderson}},\ }\bibfield  {title} {\bibinfo {title} {Absence of diffusion in
  certain random lattices},\ }\href {https://doi.org/10.1103/PhysRev.109.1492}
  {\bibfield  {journal} {\bibinfo  {journal} {Phys. Rev.}\ }\textbf {\bibinfo
  {volume} {109}},\ \bibinfo {pages} {1492} (\bibinfo {year}
  {1958})}\BibitemShut {NoStop}%
\bibitem [{\citenamefont {Abou-Chacra}\ \emph {et~al.}(1973)\citenamefont
  {Abou-Chacra}, \citenamefont {Anderson},\ and\ \citenamefont
  {Thouless}}]{abou-chacra1973self}%
  \BibitemOpen
  \bibfield  {author} {\bibinfo {author} {\bibfnamefont {R.}~\bibnamefont
  {Abou-Chacra}}, \bibinfo {author} {\bibfnamefont {P.~W.}\ \bibnamefont
  {Anderson}},\ and\ \bibinfo {author} {\bibfnamefont {D.~J.}\ \bibnamefont
  {Thouless}},\ }\bibfield  {title} {\bibinfo {title} {A self-consistent theory
  of localization},\ }\href {https://doi.org/10.1088/0022-3719/6/10/009}
  {\bibfield  {journal} {\bibinfo  {journal} {J. Phys. C}\ }\textbf {\bibinfo
  {volume} {6}},\ \bibinfo {pages} {1734} (\bibinfo {year} {1973})}\BibitemShut
  {NoStop}%
\bibitem [{\citenamefont {Abrahams}\ \emph {et~al.}(1979)\citenamefont
  {Abrahams}, \citenamefont {Anderson}, \citenamefont {Licciardello},\ and\
  \citenamefont {Ramakrishnan}}]{abrahams1979scaling}%
  \BibitemOpen
  \bibfield  {author} {\bibinfo {author} {\bibfnamefont {E.}~\bibnamefont
  {Abrahams}}, \bibinfo {author} {\bibfnamefont {P.~W.}\ \bibnamefont
  {Anderson}}, \bibinfo {author} {\bibfnamefont {D.~C.}\ \bibnamefont
  {Licciardello}},\ and\ \bibinfo {author} {\bibfnamefont {T.~V.}\ \bibnamefont
  {Ramakrishnan}},\ }\bibfield  {title} {\bibinfo {title} {Scaling theory of
  localization: Absence of quantum diffusion in two dimensions},\ }\href
  {https://doi.org/10.1103/PhysRevLett.42.673} {\bibfield  {journal} {\bibinfo
  {journal} {Phys. Rev. Lett.}\ }\textbf {\bibinfo {volume} {42}},\ \bibinfo
  {pages} {673} (\bibinfo {year} {1979})}\BibitemShut {NoStop}%
\bibitem [{\citenamefont {Lee}\ and\ \citenamefont
  {Ramakrishnan}(1985)}]{lee1985disordered}%
  \BibitemOpen
  \bibfield  {author} {\bibinfo {author} {\bibfnamefont {P.~A.}\ \bibnamefont
  {Lee}}\ and\ \bibinfo {author} {\bibfnamefont {T.~V.}\ \bibnamefont
  {Ramakrishnan}},\ }\bibfield  {title} {\bibinfo {title} {Disordered
  electronic systems},\ }\href {https://doi.org/10.1103/RevModPhys.57.287}
  {\bibfield  {journal} {\bibinfo  {journal} {Rev. Mod. Phys.}\ }\textbf
  {\bibinfo {volume} {57}},\ \bibinfo {pages} {287} (\bibinfo {year}
  {1985})}\BibitemShut {NoStop}%
\bibitem [{\citenamefont {Evers}\ and\ \citenamefont
  {Mirlin}(2008)}]{evers2008anderson}%
  \BibitemOpen
  \bibfield  {author} {\bibinfo {author} {\bibfnamefont {F.}~\bibnamefont
  {Evers}}\ and\ \bibinfo {author} {\bibfnamefont {A.~D.}\ \bibnamefont
  {Mirlin}},\ }\bibfield  {title} {\bibinfo {title} {Anderson transitions},\
  }\href {https://doi.org/10.1103/RevModPhys.80.1355} {\bibfield  {journal}
  {\bibinfo  {journal} {Rev. Mod. Phys.}\ }\textbf {\bibinfo {volume} {80}},\
  \bibinfo {pages} {1355} (\bibinfo {year} {2008})}\BibitemShut {NoStop}%
\bibitem [{\citenamefont {Garc\'{\i}a-Mata}\ \emph {et~al.}(2017)\citenamefont
  {Garc\'{\i}a-Mata}, \citenamefont {Giraud}, \citenamefont {Georgeot},
  \citenamefont {Martin}, \citenamefont {Dubertrand},\ and\ \citenamefont
  {Lemari\'e}}]{garciamata2017scaling}%
  \BibitemOpen
  \bibfield  {author} {\bibinfo {author} {\bibfnamefont {I.}~\bibnamefont
  {Garc\'{\i}a-Mata}}, \bibinfo {author} {\bibfnamefont {O.}~\bibnamefont
  {Giraud}}, \bibinfo {author} {\bibfnamefont {B.}~\bibnamefont {Georgeot}},
  \bibinfo {author} {\bibfnamefont {J.}~\bibnamefont {Martin}}, \bibinfo
  {author} {\bibfnamefont {R.}~\bibnamefont {Dubertrand}},\ and\ \bibinfo
  {author} {\bibfnamefont {G.}~\bibnamefont {Lemari\'e}},\ }\bibfield  {title}
  {\bibinfo {title} {Scaling theory of the {A}nderson transition in random
  graphs: Ergodicity and universality},\ }\href
  {https://doi.org/10.1103/PhysRevLett.118.166801} {\bibfield  {journal}
  {\bibinfo  {journal} {Phys. Rev. Lett.}\ }\textbf {\bibinfo {volume} {118}},\
  \bibinfo {pages} {166801} (\bibinfo {year} {2017})}\BibitemShut {NoStop}%
\bibitem [{\citenamefont {Garc\'{\i}a-Mata}\ \emph {et~al.}(2020)\citenamefont
  {Garc\'{\i}a-Mata}, \citenamefont {Martin}, \citenamefont {Dubertrand},
  \citenamefont {Giraud}, \citenamefont {Georgeot},\ and\ \citenamefont
  {Lemari\'e}}]{garciamata2020two}%
  \BibitemOpen
  \bibfield  {author} {\bibinfo {author} {\bibfnamefont {I.}~\bibnamefont
  {Garc\'{\i}a-Mata}}, \bibinfo {author} {\bibfnamefont {J.}~\bibnamefont
  {Martin}}, \bibinfo {author} {\bibfnamefont {R.}~\bibnamefont {Dubertrand}},
  \bibinfo {author} {\bibfnamefont {O.}~\bibnamefont {Giraud}}, \bibinfo
  {author} {\bibfnamefont {B.}~\bibnamefont {Georgeot}},\ and\ \bibinfo
  {author} {\bibfnamefont {G.}~\bibnamefont {Lemari\'e}},\ }\bibfield  {title}
  {\bibinfo {title} {Two critical localization lengths in the {A}nderson
  transition on random graphs},\ }\href
  {https://doi.org/10.1103/PhysRevResearch.2.012020} {\bibfield  {journal}
  {\bibinfo  {journal} {Phys. Rev. Research}\ }\textbf {\bibinfo {volume}
  {2}},\ \bibinfo {pages} {012020} (\bibinfo {year} {2020})}\BibitemShut
  {NoStop}%
\bibitem [{\citenamefont {Garc\'{\i}a-Mata}\ \emph {et~al.}(2022)\citenamefont
  {Garc\'{\i}a-Mata}, \citenamefont {Martin}, \citenamefont {Giraud},
  \citenamefont {Georgeot}, \citenamefont {Dubertrand},\ and\ \citenamefont
  {Lemari\'e}}]{garciamata2022critical}%
  \BibitemOpen
  \bibfield  {author} {\bibinfo {author} {\bibfnamefont {I.}~\bibnamefont
  {Garc\'{\i}a-Mata}}, \bibinfo {author} {\bibfnamefont {J.}~\bibnamefont
  {Martin}}, \bibinfo {author} {\bibfnamefont {O.}~\bibnamefont {Giraud}},
  \bibinfo {author} {\bibfnamefont {B.}~\bibnamefont {Georgeot}}, \bibinfo
  {author} {\bibfnamefont {R.}~\bibnamefont {Dubertrand}},\ and\ \bibinfo
  {author} {\bibfnamefont {G.}~\bibnamefont {Lemari\'e}},\ }\bibfield  {title}
  {\bibinfo {title} {Critical properties of the {A}nderson transition on random
  graphs: Two-parameter scaling theory, {K}osterlitz-{T}houless type flow, and
  many-body localization},\ }\href
  {https://doi.org/10.1103/PhysRevB.106.214202} {\bibfield  {journal} {\bibinfo
   {journal} {Phys. Rev. B}\ }\textbf {\bibinfo {volume} {106}},\ \bibinfo
  {pages} {214202} (\bibinfo {year} {2022})}\BibitemShut {NoStop}%
\bibitem [{\citenamefont {Vanoni}\ \emph {et~al.}(2024)\citenamefont {Vanoni},
  \citenamefont {Altshuler}, \citenamefont {Kravtsov},\ and\ \citenamefont
  {Scardicchio}}]{vanoni2024renormalisation}%
  \BibitemOpen
  \bibfield  {author} {\bibinfo {author} {\bibfnamefont {C.}~\bibnamefont
  {Vanoni}}, \bibinfo {author} {\bibfnamefont {B.~L.}\ \bibnamefont
  {Altshuler}}, \bibinfo {author} {\bibfnamefont {V.~E.}\ \bibnamefont
  {Kravtsov}},\ and\ \bibinfo {author} {\bibfnamefont {A.}~\bibnamefont
  {Scardicchio}},\ }\bibfield  {title} {\bibinfo {title} {Renormalization group
  analysis of the {A}nderson model on random regular graphs},\ }\href
  {https://doi.org/10.1073/pnas.2401955121} {\bibfield  {journal} {\bibinfo
  {journal} {Proceedings of the National Academy of Sciences}\ }\textbf
  {\bibinfo {volume} {121}},\ \bibinfo {pages} {e2401955121} (\bibinfo {year}
  {2024})}\BibitemShut {NoStop}%
\bibitem [{\citenamefont {Altshuler}\ \emph {et~al.}(2024)\citenamefont
  {Altshuler}, \citenamefont {Kravtsov}, \citenamefont {Scardicchio},
  \citenamefont {Sierant},\ and\ \citenamefont
  {Vanoni}}]{altshuler2024renormalization}%
  \BibitemOpen
  \bibfield  {author} {\bibinfo {author} {\bibfnamefont {B.~L.}\ \bibnamefont
  {Altshuler}}, \bibinfo {author} {\bibfnamefont {V.~E.}\ \bibnamefont
  {Kravtsov}}, \bibinfo {author} {\bibfnamefont {A.}~\bibnamefont
  {Scardicchio}}, \bibinfo {author} {\bibfnamefont {P.}~\bibnamefont
  {Sierant}},\ and\ \bibinfo {author} {\bibfnamefont {C.}~\bibnamefont
  {Vanoni}},\ }\href {https://arxiv.org/abs/2403.01974} {\bibinfo {title}
  {Renormalization group for {A}nderson localization on high-dimensional
  lattices}} (\bibinfo {year} {2024}),\ \Eprint
  {https://arxiv.org/abs/2403.01974} {arXiv:2403.01974 [cond-mat.dis-nn]}
  \BibitemShut {NoStop}%
\bibitem [{\citenamefont {Basko}\ \emph {et~al.}(2006)\citenamefont {Basko},
  \citenamefont {Aleiner},\ and\ \citenamefont {Altshuler}}]{basko2006metal}%
  \BibitemOpen
  \bibfield  {author} {\bibinfo {author} {\bibfnamefont {D.~M.}\ \bibnamefont
  {Basko}}, \bibinfo {author} {\bibfnamefont {I.~L.}\ \bibnamefont {Aleiner}},\
  and\ \bibinfo {author} {\bibfnamefont {B.~L.}\ \bibnamefont {Altshuler}},\
  }\bibfield  {title} {\bibinfo {title} {Metal--insulator transition in a
  weakly interacting many-electron system with localized single-particle
  states},\ }\href
  {http://www.sciencedirect.com/science/article/pii/S0003491605002630}
  {\bibfield  {journal} {\bibinfo  {journal} {Annals of {P}hysics}\ }\textbf
  {\bibinfo {volume} {321}},\ \bibinfo {pages} {1126} (\bibinfo {year}
  {2006})}\BibitemShut {NoStop}%
\bibitem [{\citenamefont {Gornyi}\ \emph {et~al.}(2005)\citenamefont {Gornyi},
  \citenamefont {Mirlin},\ and\ \citenamefont
  {Polyakov}}]{gornyi2005interacting}%
  \BibitemOpen
  \bibfield  {author} {\bibinfo {author} {\bibfnamefont {I.~V.}\ \bibnamefont
  {Gornyi}}, \bibinfo {author} {\bibfnamefont {A.~D.}\ \bibnamefont {Mirlin}},\
  and\ \bibinfo {author} {\bibfnamefont {D.~G.}\ \bibnamefont {Polyakov}},\
  }\bibfield  {title} {\bibinfo {title} {Interacting electrons in disordered
  wires: Anderson localization and low-${T}$ transport},\ }\href
  {https://doi.org/10.1103/PhysRevLett.95.206603} {\bibfield  {journal}
  {\bibinfo  {journal} {Phys. Rev. Lett.}\ }\textbf {\bibinfo {volume} {95}},\
  \bibinfo {pages} {206603} (\bibinfo {year} {2005})}\BibitemShut {NoStop}%
\bibitem [{\citenamefont {Oganesyan}\ and\ \citenamefont
  {Huse}(2007)}]{oganesyan2007localisation}%
  \BibitemOpen
  \bibfield  {author} {\bibinfo {author} {\bibfnamefont {V.}~\bibnamefont
  {Oganesyan}}\ and\ \bibinfo {author} {\bibfnamefont {D.~A.}\ \bibnamefont
  {Huse}},\ }\bibfield  {title} {\bibinfo {title} {Localization of interacting
  fermions at high temperature},\ }\href
  {https://doi.org/10.1103/PhysRevB.75.155111} {\bibfield  {journal} {\bibinfo
  {journal} {Phys. Rev. B}\ }\textbf {\bibinfo {volume} {75}},\ \bibinfo
  {pages} {155111} (\bibinfo {year} {2007})}\BibitemShut {NoStop}%
\bibitem [{\citenamefont {Nandkishore}\ and\ \citenamefont
  {Huse}(2015)}]{nandkishore2015many}%
  \BibitemOpen
  \bibfield  {author} {\bibinfo {author} {\bibfnamefont {R.}~\bibnamefont
  {Nandkishore}}\ and\ \bibinfo {author} {\bibfnamefont {D.~A.}\ \bibnamefont
  {Huse}},\ }\bibfield  {title} {\bibinfo {title} {Many-body localization and
  thermalization in quantum statistical mechanics},\ }\href
  {https://doi.org/10.1146/annurev-conmatphys-031214-014726} {\bibfield
  {journal} {\bibinfo  {journal} {Annu. Rev. Condens. Matter Phys.}\ }\textbf
  {\bibinfo {volume} {6}},\ \bibinfo {pages} {15} (\bibinfo {year}
  {2015})}\BibitemShut {NoStop}%
\bibitem [{\citenamefont {Abanin}\ and\ \citenamefont
  {Papi{\'c}}(2017)}]{abanin2017recent}%
  \BibitemOpen
  \bibfield  {author} {\bibinfo {author} {\bibfnamefont {D.~A.}\ \bibnamefont
  {Abanin}}\ and\ \bibinfo {author} {\bibfnamefont {Z.}~\bibnamefont
  {Papi{\'c}}},\ }\bibfield  {title} {\bibinfo {title} {Recent progress in
  many-body localization},\ }\href {http://dx.doi.org/10.1002/andp.201700169}
  {\bibfield  {journal} {\bibinfo  {journal} {Annalen der Physik}\ }\textbf
  {\bibinfo {volume} {529}},\ \bibinfo {pages} {1700169} (\bibinfo {year}
  {2017})}\BibitemShut {NoStop}%
\bibitem [{\citenamefont {Abanin}\ \emph {et~al.}(2019)\citenamefont {Abanin},
  \citenamefont {Altman}, \citenamefont {Bloch},\ and\ \citenamefont
  {Serbyn}}]{abanin2019colloquium}%
  \BibitemOpen
  \bibfield  {author} {\bibinfo {author} {\bibfnamefont {D.~A.}\ \bibnamefont
  {Abanin}}, \bibinfo {author} {\bibfnamefont {E.}~\bibnamefont {Altman}},
  \bibinfo {author} {\bibfnamefont {I.}~\bibnamefont {Bloch}},\ and\ \bibinfo
  {author} {\bibfnamefont {M.}~\bibnamefont {Serbyn}},\ }\bibfield  {title}
  {\bibinfo {title} {Colloquium: Many-body localization, thermalization, and
  entanglement},\ }\href {https://doi.org/10.1103/RevModPhys.91.021001}
  {\bibfield  {journal} {\bibinfo  {journal} {Rev. Mod. Phys.}\ }\textbf
  {\bibinfo {volume} {91}},\ \bibinfo {pages} {021001} (\bibinfo {year}
  {2019})}\BibitemShut {NoStop}%
\bibitem [{\citenamefont {Sierant}\ \emph {et~al.}(2025)\citenamefont
  {Sierant}, \citenamefont {Lewenstein}, \citenamefont {Scardicchio},
  \citenamefont {Vidmar},\ and\ \citenamefont
  {Zakrzewski}}]{SierantReview2025}%
  \BibitemOpen
  \bibfield  {author} {\bibinfo {author} {\bibfnamefont {P.}~\bibnamefont
  {Sierant}}, \bibinfo {author} {\bibfnamefont {M.}~\bibnamefont {Lewenstein}},
  \bibinfo {author} {\bibfnamefont {A.}~\bibnamefont {Scardicchio}}, \bibinfo
  {author} {\bibfnamefont {L.}~\bibnamefont {Vidmar}},\ and\ \bibinfo {author}
  {\bibfnamefont {J.}~\bibnamefont {Zakrzewski}},\ }\bibfield  {title}
  {\bibinfo {title} {Many-body localization in the age of classical
  computing},\ }\href {https://doi.org/10.1088/1361-6633/ad9756} {\bibfield
  {journal} {\bibinfo  {journal} {Reports on Progress in Physics}\ }\textbf
  {\bibinfo {volume} {88}},\ \bibinfo {pages} {026502} (\bibinfo {year}
  {2025})}\BibitemShut {NoStop}%
\bibitem [{\citenamefont {Logan}\ and\ \citenamefont
  {Wolynes}(1990)}]{logan1990quantum}%
  \BibitemOpen
  \bibfield  {author} {\bibinfo {author} {\bibfnamefont {D.~E.}\ \bibnamefont
  {Logan}}\ and\ \bibinfo {author} {\bibfnamefont {P.~G.}\ \bibnamefont
  {Wolynes}},\ }\bibfield  {title} {\bibinfo {title} {Quantum localization and
  energy flow in many-dimensional {F}ermi resonant systems},\ }\href
  {https://aip.scitation.org/doi/10.1063/1.458637} {\bibfield  {journal}
  {\bibinfo  {journal} {J. Chem. Phys.}\ }\textbf {\bibinfo {volume} {93}},\
  \bibinfo {pages} {4994} (\bibinfo {year} {1990})}\BibitemShut {NoStop}%
\bibitem [{\citenamefont {Altshuler}\ \emph {et~al.}(1997)\citenamefont
  {Altshuler}, \citenamefont {Gefen}, \citenamefont {Kamenev},\ and\
  \citenamefont {Levitov}}]{altshuler1997quasiparticle}%
  \BibitemOpen
  \bibfield  {author} {\bibinfo {author} {\bibfnamefont {B.~L.}\ \bibnamefont
  {Altshuler}}, \bibinfo {author} {\bibfnamefont {Y.}~\bibnamefont {Gefen}},
  \bibinfo {author} {\bibfnamefont {A.}~\bibnamefont {Kamenev}},\ and\ \bibinfo
  {author} {\bibfnamefont {L.~S.}\ \bibnamefont {Levitov}},\ }\bibfield
  {title} {\bibinfo {title} {Quasiparticle lifetime in a finite system: A
  nonperturbative approach},\ }\href
  {https://doi.org/10.1103/PhysRevLett.78.2803} {\bibfield  {journal} {\bibinfo
   {journal} {Phys. Rev. Lett.}\ }\textbf {\bibinfo {volume} {78}},\ \bibinfo
  {pages} {2803} (\bibinfo {year} {1997})}\BibitemShut {NoStop}%
\bibitem [{\citenamefont {Monthus}\ and\ \citenamefont
  {Garel}(2010)}]{MonthusGarel2010PRB}%
  \BibitemOpen
  \bibfield  {author} {\bibinfo {author} {\bibfnamefont {C.}~\bibnamefont
  {Monthus}}\ and\ \bibinfo {author} {\bibfnamefont {T.}~\bibnamefont
  {Garel}},\ }\bibfield  {title} {\bibinfo {title} {Many-body localization
  transition in a lattice model of interacting fermions: {S}tatistics of
  renormalized hoppings in configuration space},\ }\href
  {https://doi.org/10.1103/PhysRevB.81.134202} {\bibfield  {journal} {\bibinfo
  {journal} {Phys. Rev. B}\ }\textbf {\bibinfo {volume} {81}},\ \bibinfo
  {pages} {134202} (\bibinfo {year} {2010})}\BibitemShut {NoStop}%
\bibitem [{\citenamefont {Pietracaprina}\ \emph {et~al.}(2016)\citenamefont
  {Pietracaprina}, \citenamefont {Ros},\ and\ \citenamefont
  {Scardicchio}}]{pietracaprina2016forward}%
  \BibitemOpen
  \bibfield  {author} {\bibinfo {author} {\bibfnamefont {F.}~\bibnamefont
  {Pietracaprina}}, \bibinfo {author} {\bibfnamefont {V.}~\bibnamefont {Ros}},\
  and\ \bibinfo {author} {\bibfnamefont {A.}~\bibnamefont {Scardicchio}},\
  }\bibfield  {title} {\bibinfo {title} {Forward approximation as a mean-field
  approximation for the {A}nderson and many-body localization transitions},\
  }\href {https://doi.org/10.1103/PhysRevB.93.054201} {\bibfield  {journal}
  {\bibinfo  {journal} {Phys. Rev. B}\ }\textbf {\bibinfo {volume} {93}},\
  \bibinfo {pages} {054201} (\bibinfo {year} {2016})}\BibitemShut {NoStop}%
\bibitem [{\citenamefont {Biroli}\ and\ \citenamefont
  {Tarzia}(2017{\natexlab{a}})}]{biroli2017delocalized}%
  \BibitemOpen
  \bibfield  {author} {\bibinfo {author} {\bibfnamefont {G.}~\bibnamefont
  {Biroli}}\ and\ \bibinfo {author} {\bibfnamefont {M.}~\bibnamefont
  {Tarzia}},\ }\bibfield  {title} {\bibinfo {title} {Delocalized glassy
  dynamics and many-body localization},\ }\href
  {https://doi.org/10.1103/PhysRevB.96.201114} {\bibfield  {journal} {\bibinfo
  {journal} {Phys. Rev. B}\ }\textbf {\bibinfo {volume} {96}},\ \bibinfo
  {pages} {201114} (\bibinfo {year} {2017}{\natexlab{a}})}\BibitemShut
  {NoStop}%
\bibitem [{\citenamefont {Altland}\ and\ \citenamefont
  {Micklitz}(2017)}]{altland2017field}%
  \BibitemOpen
  \bibfield  {author} {\bibinfo {author} {\bibfnamefont {A.}~\bibnamefont
  {Altland}}\ and\ \bibinfo {author} {\bibfnamefont {T.}~\bibnamefont
  {Micklitz}},\ }\bibfield  {title} {\bibinfo {title} {{Field Theory Approach
  to Many-Body Localization}},\ }\href
  {https://doi.org/10.1103/PhysRevLett.118.127202} {\bibfield  {journal}
  {\bibinfo  {journal} {Phys. Rev. Lett.}\ }\textbf {\bibinfo {volume} {118}},\
  \bibinfo {pages} {127202} (\bibinfo {year} {2017})}\BibitemShut {NoStop}%
\bibitem [{\citenamefont {Logan}\ and\ \citenamefont
  {Welsh}(2019)}]{logan2019many}%
  \BibitemOpen
  \bibfield  {author} {\bibinfo {author} {\bibfnamefont {D.~E.}\ \bibnamefont
  {Logan}}\ and\ \bibinfo {author} {\bibfnamefont {S.}~\bibnamefont {Welsh}},\
  }\bibfield  {title} {\bibinfo {title} {Many-body localization in {F}ock
  space: {A} local perspective},\ }\href
  {https://doi.org/10.1103/PhysRevB.99.045131} {\bibfield  {journal} {\bibinfo
  {journal} {Phys. Rev. B}\ }\textbf {\bibinfo {volume} {99}},\ \bibinfo
  {pages} {045131} (\bibinfo {year} {2019})}\BibitemShut {NoStop}%
\bibitem [{\citenamefont {Roy}\ and\ \citenamefont
  {Logan}(2020{\natexlab{a}})}]{roy2020fock}%
  \BibitemOpen
  \bibfield  {author} {\bibinfo {author} {\bibfnamefont {S.}~\bibnamefont
  {Roy}}\ and\ \bibinfo {author} {\bibfnamefont {D.~E.}\ \bibnamefont
  {Logan}},\ }\bibfield  {title} {\bibinfo {title} {Fock-space correlations and
  the origins of many-body localization},\ }\href
  {https://doi.org/10.1103/PhysRevB.101.134202} {\bibfield  {journal} {\bibinfo
   {journal} {Phys. Rev. B}\ }\textbf {\bibinfo {volume} {101}},\ \bibinfo
  {pages} {134202} (\bibinfo {year} {2020}{\natexlab{a}})}\BibitemShut
  {NoStop}%
\bibitem [{\citenamefont {Tikhonov}\ and\ \citenamefont
  {Mirlin}(2021{\natexlab{a}})}]{tikhonov2021eigenstate}%
  \BibitemOpen
  \bibfield  {author} {\bibinfo {author} {\bibfnamefont {K.~S.}\ \bibnamefont
  {Tikhonov}}\ and\ \bibinfo {author} {\bibfnamefont {A.~D.}\ \bibnamefont
  {Mirlin}},\ }\bibfield  {title} {\bibinfo {title} {Eigenstate correlations
  around the many-body localization transition},\ }\href
  {https://doi.org/10.1103/PhysRevB.103.064204} {\bibfield  {journal} {\bibinfo
   {journal} {Phys. Rev. B}\ }\textbf {\bibinfo {volume} {103}},\ \bibinfo
  {pages} {064204} (\bibinfo {year} {2021}{\natexlab{a}})}\BibitemShut
  {NoStop}%
\bibitem [{\citenamefont {De~Tomasi}\ \emph {et~al.}(2021)\citenamefont
  {De~Tomasi}, \citenamefont {Khaymovich}, \citenamefont {Pollmann},\ and\
  \citenamefont {Warzel}}]{detomasi2020rare}%
  \BibitemOpen
  \bibfield  {author} {\bibinfo {author} {\bibfnamefont {G.}~\bibnamefont
  {De~Tomasi}}, \bibinfo {author} {\bibfnamefont {I.~M.}\ \bibnamefont
  {Khaymovich}}, \bibinfo {author} {\bibfnamefont {F.}~\bibnamefont
  {Pollmann}},\ and\ \bibinfo {author} {\bibfnamefont {S.}~\bibnamefont
  {Warzel}},\ }\bibfield  {title} {\bibinfo {title} {Rare thermal bubbles at
  the many-body localization transition from the {F}ock space point of view},\
  }\href {https://doi.org/10.1103/PhysRevB.104.024202} {\bibfield  {journal}
  {\bibinfo  {journal} {Phys. Rev. B}\ }\textbf {\bibinfo {volume} {104}},\
  \bibinfo {pages} {024202} (\bibinfo {year} {2021})}\BibitemShut {NoStop}%
\bibitem [{\citenamefont {Roy}\ and\ \citenamefont
  {Logan}(2021)}]{roy2021fockspace}%
  \BibitemOpen
  \bibfield  {author} {\bibinfo {author} {\bibfnamefont {S.}~\bibnamefont
  {Roy}}\ and\ \bibinfo {author} {\bibfnamefont {D.~E.}\ \bibnamefont
  {Logan}},\ }\bibfield  {title} {\bibinfo {title} {Fock-space anatomy of
  eigenstates across the many-body localization transition},\ }\href
  {https://doi.org/10.1103/PhysRevB.104.174201} {\bibfield  {journal} {\bibinfo
   {journal} {Phys. Rev. B}\ }\textbf {\bibinfo {volume} {104}},\ \bibinfo
  {pages} {174201} (\bibinfo {year} {2021})}\BibitemShut {NoStop}%
\bibitem [{\citenamefont {Tarzia}(2020)}]{tarzia2020manybody}%
  \BibitemOpen
  \bibfield  {author} {\bibinfo {author} {\bibfnamefont {M.}~\bibnamefont
  {Tarzia}},\ }\bibfield  {title} {\bibinfo {title} {Many-body localization
  transition in {H}ilbert space},\ }\href
  {https://doi.org/10.1103/PhysRevB.102.014208} {\bibfield  {journal} {\bibinfo
   {journal} {Phys. Rev. B}\ }\textbf {\bibinfo {volume} {102}},\ \bibinfo
  {pages} {014208} (\bibinfo {year} {2020})}\BibitemShut {NoStop}%
\bibitem [{\citenamefont {Sutradhar}\ \emph {et~al.}(2022)\citenamefont
  {Sutradhar}, \citenamefont {Ghosh}, \citenamefont {Roy}, \citenamefont
  {Logan}, \citenamefont {Mukerjee},\ and\ \citenamefont
  {Banerjee}}]{sutradhar2022scaling}%
  \BibitemOpen
  \bibfield  {author} {\bibinfo {author} {\bibfnamefont {J.}~\bibnamefont
  {Sutradhar}}, \bibinfo {author} {\bibfnamefont {S.}~\bibnamefont {Ghosh}},
  \bibinfo {author} {\bibfnamefont {S.}~\bibnamefont {Roy}}, \bibinfo {author}
  {\bibfnamefont {D.~E.}\ \bibnamefont {Logan}}, \bibinfo {author}
  {\bibfnamefont {S.}~\bibnamefont {Mukerjee}},\ and\ \bibinfo {author}
  {\bibfnamefont {S.}~\bibnamefont {Banerjee}},\ }\bibfield  {title} {\bibinfo
  {title} {Scaling of the {F}ock-space propagator and multifractality across
  the many-body localization transition},\ }\href
  {https://doi.org/10.1103/PhysRevB.106.054203} {\bibfield  {journal} {\bibinfo
   {journal} {Phys. Rev. B}\ }\textbf {\bibinfo {volume} {106}},\ \bibinfo
  {pages} {054203} (\bibinfo {year} {2022})}\BibitemShut {NoStop}%
\bibitem [{\citenamefont {Roy}(2023)}]{roy2023anatomy}%
  \BibitemOpen
  \bibfield  {author} {\bibinfo {author} {\bibfnamefont {S.}~\bibnamefont
  {Roy}},\ }\bibfield  {title} {\bibinfo {title} {{Anatomy of localisation
  protected quantum order on Hilbert space}},\ }\href
  {https://doi.org/10.1088/1361-648x/ace413} {\bibfield  {journal} {\bibinfo
  {journal} {Journal of Physics: Condensed Matter}\ }\textbf {\bibinfo {volume}
  {35}},\ \bibinfo {pages} {415601} (\bibinfo {year} {2023})}\BibitemShut
  {NoStop}%
\bibitem [{\citenamefont {Roy}\ \emph {et~al.}(2023)\citenamefont {Roy},
  \citenamefont {Sutradhar},\ and\ \citenamefont
  {Banerjee}}]{roy2023diagnostics}%
  \BibitemOpen
  \bibfield  {author} {\bibinfo {author} {\bibfnamefont {N.}~\bibnamefont
  {Roy}}, \bibinfo {author} {\bibfnamefont {J.}~\bibnamefont {Sutradhar}},\
  and\ \bibinfo {author} {\bibfnamefont {S.}~\bibnamefont {Banerjee}},\
  }\bibfield  {title} {\bibinfo {title} {Diagnostics of nonergodic extended
  states and many body localization proximity effect through real-space and
  {F}ock-space excitations},\ }\href
  {https://doi.org/10.1103/PhysRevB.107.115155} {\bibfield  {journal} {\bibinfo
   {journal} {Phys. Rev. B}\ }\textbf {\bibinfo {volume} {107}},\ \bibinfo
  {pages} {115155} (\bibinfo {year} {2023})}\BibitemShut {NoStop}%
\bibitem [{\citenamefont {Herre}\ \emph {et~al.}(2023)\citenamefont {Herre},
  \citenamefont {Karcher}, \citenamefont {Tikhonov},\ and\ \citenamefont
  {Mirlin}}]{herre2023ergodicity}%
  \BibitemOpen
  \bibfield  {author} {\bibinfo {author} {\bibfnamefont {J.-N.}\ \bibnamefont
  {Herre}}, \bibinfo {author} {\bibfnamefont {J.~F.}\ \bibnamefont {Karcher}},
  \bibinfo {author} {\bibfnamefont {K.~S.}\ \bibnamefont {Tikhonov}},\ and\
  \bibinfo {author} {\bibfnamefont {A.~D.}\ \bibnamefont {Mirlin}},\ }\bibfield
   {title} {\bibinfo {title} {Ergodicity-to-localization transition on random
  regular graphs with large connectivity and in many-body quantum dots},\
  }\href {https://doi.org/10.1103/PhysRevB.108.014203} {\bibfield  {journal}
  {\bibinfo  {journal} {Phys. Rev. B}\ }\textbf {\bibinfo {volume} {108}},\
  \bibinfo {pages} {014203} (\bibinfo {year} {2023})}\BibitemShut {NoStop}%
\bibitem [{\citenamefont {Schir\'o}\ and\ \citenamefont
  {Tarzia}(2020)}]{schiro2020toy}%
  \BibitemOpen
  \bibfield  {author} {\bibinfo {author} {\bibfnamefont {M.}~\bibnamefont
  {Schir\'o}}\ and\ \bibinfo {author} {\bibfnamefont {M.}~\bibnamefont
  {Tarzia}},\ }\bibfield  {title} {\bibinfo {title} {Toy model for anomalous
  transport and {G}riffiths effects near the many-body localization
  transition},\ }\href {https://doi.org/10.1103/PhysRevB.101.014203} {\bibfield
   {journal} {\bibinfo  {journal} {Phys. Rev. B}\ }\textbf {\bibinfo {volume}
  {101}},\ \bibinfo {pages} {014203} (\bibinfo {year} {2020})}\BibitemShut
  {NoStop}%
\bibitem [{\citenamefont {Scoquart}\ \emph {et~al.}(2024)\citenamefont
  {Scoquart}, \citenamefont {Gornyi},\ and\ \citenamefont
  {Mirlin}}]{scoquart2024role}%
  \BibitemOpen
  \bibfield  {author} {\bibinfo {author} {\bibfnamefont {T.}~\bibnamefont
  {Scoquart}}, \bibinfo {author} {\bibfnamefont {I.~V.}\ \bibnamefont
  {Gornyi}},\ and\ \bibinfo {author} {\bibfnamefont {A.~D.}\ \bibnamefont
  {Mirlin}},\ }\bibfield  {title} {\bibinfo {title} {{Role of Fock-space
  correlations in many-body localization}},\ }\href
  {https://doi.org/10.1103/PhysRevB.109.214203} {\bibfield  {journal} {\bibinfo
   {journal} {Phys. Rev. B}\ }\textbf {\bibinfo {volume} {109}},\ \bibinfo
  {pages} {214203} (\bibinfo {year} {2024})}\BibitemShut {NoStop}%
\bibitem [{\citenamefont {Ghosh}\ \emph {et~al.}(2024)\citenamefont {Ghosh},
  \citenamefont {Sutradhar}, \citenamefont {Mukerjee},\ and\ \citenamefont
  {Banerjee}}]{ghosh2024scaling}%
  \BibitemOpen
  \bibfield  {author} {\bibinfo {author} {\bibfnamefont {S.}~\bibnamefont
  {Ghosh}}, \bibinfo {author} {\bibfnamefont {J.}~\bibnamefont {Sutradhar}},
  \bibinfo {author} {\bibfnamefont {S.}~\bibnamefont {Mukerjee}},\ and\
  \bibinfo {author} {\bibfnamefont {S.}~\bibnamefont {Banerjee}},\ }\href@noop
  {} {\bibinfo {title} {Scaling of {F}ock space propagator in quasiperiodic
  many-body localizing systems}} (\bibinfo {year} {2024}),\ \Eprint
  {https://arxiv.org/abs/2401.03027} {arXiv:2401.03027 [cond-mat.dis-nn]}
  \BibitemShut {NoStop}%
\bibitem [{\citenamefont {Tikhonov}\ and\ \citenamefont
  {Mirlin}(2021{\natexlab{b}})}]{tikhonov2021anderson}%
  \BibitemOpen
  \bibfield  {author} {\bibinfo {author} {\bibfnamefont {K.}~\bibnamefont
  {Tikhonov}}\ and\ \bibinfo {author} {\bibfnamefont {A.}~\bibnamefont
  {Mirlin}},\ }\bibfield  {title} {\bibinfo {title} {From {A}nderson
  localization on random regular graphs to many-body localization},\ }\href
  {https://doi.org/https://doi.org/10.1016/j.aop.2021.168525} {\bibfield
  {journal} {\bibinfo  {journal} {Annals of Physics}\ }\textbf {\bibinfo
  {volume} {435}},\ \bibinfo {pages} {168525} (\bibinfo {year}
  {2021}{\natexlab{b}})}\BibitemShut {NoStop}%
\bibitem [{\citenamefont {Roy}\ and\ \citenamefont
  {Logan}(2024)}]{roy2024fock}%
  \BibitemOpen
  \bibfield  {author} {\bibinfo {author} {\bibfnamefont {S.}~\bibnamefont
  {Roy}}\ and\ \bibinfo {author} {\bibfnamefont {D.~E.}\ \bibnamefont
  {Logan}},\ }\bibfield  {title} {\bibinfo {title} {{The Fock-space landscape
  of many-body localisation}},\ }\href
  {https://doi.org/10.1088/1361-648x/ad94c3} {\bibfield  {journal} {\bibinfo
  {journal} {Journal of Physics: Condensed Matter}\ }\textbf {\bibinfo {volume}
  {37}},\ \bibinfo {pages} {073003} (\bibinfo {year} {2024})}\BibitemShut
  {NoStop}%
\bibitem [{\citenamefont {De~Luca}\ and\ \citenamefont
  {Scardicchio}(2013)}]{deluca2013ergodicity}%
  \BibitemOpen
  \bibfield  {author} {\bibinfo {author} {\bibfnamefont {A.}~\bibnamefont
  {De~Luca}}\ and\ \bibinfo {author} {\bibfnamefont {A.}~\bibnamefont
  {Scardicchio}},\ }\bibfield  {title} {\bibinfo {title} {Ergodicity breaking
  in a model showing many-body localization},\ }\href
  {https://doi.org/10.1209/0295-5075/101/37003} {\bibfield  {journal} {\bibinfo
   {journal} {Europhys. Lett.}\ }\textbf {\bibinfo {volume} {101}},\ \bibinfo
  {pages} {37003} (\bibinfo {year} {2013})}\BibitemShut {NoStop}%
\bibitem [{\citenamefont {Biroli}\ and\ \citenamefont
  {Tarzia}(2017{\natexlab{b}})}]{biroli2017delocalised}%
  \BibitemOpen
  \bibfield  {author} {\bibinfo {author} {\bibfnamefont {G.}~\bibnamefont
  {Biroli}}\ and\ \bibinfo {author} {\bibfnamefont {M.}~\bibnamefont
  {Tarzia}},\ }\bibfield  {title} {\bibinfo {title} {Delocalized glassy
  dynamics and many-body localization},\ }\href
  {https://doi.org/10.1103/PhysRevB.96.201114} {\bibfield  {journal} {\bibinfo
  {journal} {Phys. Rev. B}\ }\textbf {\bibinfo {volume} {96}},\ \bibinfo
  {pages} {201114} (\bibinfo {year} {2017}{\natexlab{b}})}\BibitemShut
  {NoStop}%
\bibitem [{\citenamefont {Biroli}\ and\ \citenamefont
  {Tarzia}(2020)}]{biroli2020anomalous}%
  \BibitemOpen
  \bibfield  {author} {\bibinfo {author} {\bibfnamefont {G.}~\bibnamefont
  {Biroli}}\ and\ \bibinfo {author} {\bibfnamefont {M.}~\bibnamefont
  {Tarzia}},\ }\bibfield  {title} {\bibinfo {title} {Anomalous dynamics on the
  ergodic side of the many-body localization transition and the glassy phase of
  directed polymers in random media},\ }\href
  {https://doi.org/10.1103/PhysRevB.102.064211} {\bibfield  {journal} {\bibinfo
   {journal} {Phys. Rev. B}\ }\textbf {\bibinfo {volume} {102}},\ \bibinfo
  {pages} {064211} (\bibinfo {year} {2020})}\BibitemShut {NoStop}%
\bibitem [{\citenamefont {Roy}\ and\ \citenamefont
  {Logan}(2020{\natexlab{b}})}]{roy2020localisation}%
  \BibitemOpen
  \bibfield  {author} {\bibinfo {author} {\bibfnamefont {S.}~\bibnamefont
  {Roy}}\ and\ \bibinfo {author} {\bibfnamefont {D.~E.}\ \bibnamefont
  {Logan}},\ }\bibfield  {title} {\bibinfo {title} {Localization on {C}ertain
  {G}raphs with {S}trongly {C}orrelated {D}isorder},\ }\href
  {https://doi.org/10.1103/PhysRevLett.125.250402} {\bibfield  {journal}
  {\bibinfo  {journal} {Phys. Rev. Lett.}\ }\textbf {\bibinfo {volume} {125}},\
  \bibinfo {pages} {250402} (\bibinfo {year} {2020}{\natexlab{b}})}\BibitemShut
  {NoStop}%
\bibitem [{\citenamefont {Biroli}\ \emph {et~al.}(2024)\citenamefont {Biroli},
  \citenamefont {Hartmann},\ and\ \citenamefont
  {Tarzia}}]{biroli2024largedeviation}%
  \BibitemOpen
  \bibfield  {author} {\bibinfo {author} {\bibfnamefont {G.}~\bibnamefont
  {Biroli}}, \bibinfo {author} {\bibfnamefont {A.~K.}\ \bibnamefont
  {Hartmann}},\ and\ \bibinfo {author} {\bibfnamefont {M.}~\bibnamefont
  {Tarzia}},\ }\bibfield  {title} {\bibinfo {title} {Large-deviation analysis
  of rare resonances for the many-body localization transition},\ }\href
  {https://doi.org/10.1103/PhysRevB.110.014205} {\bibfield  {journal} {\bibinfo
   {journal} {Phys. Rev. B}\ }\textbf {\bibinfo {volume} {110}},\ \bibinfo
  {pages} {014205} (\bibinfo {year} {2024})}\BibitemShut {NoStop}%
\bibitem [{\citenamefont {Luitz}\ \emph {et~al.}(2015)\citenamefont {Luitz},
  \citenamefont {Laflorencie},\ and\ \citenamefont {Alet}}]{luitz2015many}%
  \BibitemOpen
  \bibfield  {author} {\bibinfo {author} {\bibfnamefont {D.~J.}\ \bibnamefont
  {Luitz}}, \bibinfo {author} {\bibfnamefont {N.}~\bibnamefont {Laflorencie}},\
  and\ \bibinfo {author} {\bibfnamefont {F.}~\bibnamefont {Alet}},\ }\bibfield
  {title} {\bibinfo {title} {Many-body localization edge in the random-field
  {H}eisenberg chain},\ }\href {https://doi.org/10.1103/PhysRevB.91.081103}
  {\bibfield  {journal} {\bibinfo  {journal} {Phys. Rev. B}\ }\textbf {\bibinfo
  {volume} {91}},\ \bibinfo {pages} {081103} (\bibinfo {year}
  {2015})}\BibitemShut {NoStop}%
\bibitem [{\citenamefont {Mac\'e}\ \emph {et~al.}(2019)\citenamefont {Mac\'e},
  \citenamefont {Alet},\ and\ \citenamefont
  {Laflorencie}}]{mace2019multifractal}%
  \BibitemOpen
  \bibfield  {author} {\bibinfo {author} {\bibfnamefont {N.}~\bibnamefont
  {Mac\'e}}, \bibinfo {author} {\bibfnamefont {F.}~\bibnamefont {Alet}},\ and\
  \bibinfo {author} {\bibfnamefont {N.}~\bibnamefont {Laflorencie}},\
  }\bibfield  {title} {\bibinfo {title} {Multifractal scalings across the
  many-body localization transition},\ }\href
  {https://doi.org/10.1103/PhysRevLett.123.180601} {\bibfield  {journal}
  {\bibinfo  {journal} {Phys. Rev. Lett.}\ }\textbf {\bibinfo {volume} {123}},\
  \bibinfo {pages} {180601} (\bibinfo {year} {2019})}\BibitemShut {NoStop}%
\bibitem [{\citenamefont {Rosenzweig}\ and\ \citenamefont
  {Porter}(1960)}]{rosezweig1960repulsion}%
  \BibitemOpen
  \bibfield  {author} {\bibinfo {author} {\bibfnamefont {N.}~\bibnamefont
  {Rosenzweig}}\ and\ \bibinfo {author} {\bibfnamefont {C.~E.}\ \bibnamefont
  {Porter}},\ }\bibfield  {title} {\bibinfo {title} {Repulsion of energy levels
  in complex atomic spectra},\ }\href
  {https://doi.org/10.1103/PhysRev.120.1698} {\bibfield  {journal} {\bibinfo
  {journal} {Phys. Rev.}\ }\textbf {\bibinfo {volume} {120}},\ \bibinfo {pages}
  {1698} (\bibinfo {year} {1960})}\BibitemShut {NoStop}%
\bibitem [{\citenamefont {Mirlin}\ \emph {et~al.}(1996)\citenamefont {Mirlin},
  \citenamefont {Fyodorov}, \citenamefont {Dittes}, \citenamefont {Quezada},\
  and\ \citenamefont {Seligman}}]{mirlin1996transition}%
  \BibitemOpen
  \bibfield  {author} {\bibinfo {author} {\bibfnamefont {A.~D.}\ \bibnamefont
  {Mirlin}}, \bibinfo {author} {\bibfnamefont {Y.~V.}\ \bibnamefont
  {Fyodorov}}, \bibinfo {author} {\bibfnamefont {F.-M.}\ \bibnamefont
  {Dittes}}, \bibinfo {author} {\bibfnamefont {J.}~\bibnamefont {Quezada}},\
  and\ \bibinfo {author} {\bibfnamefont {T.~H.}\ \bibnamefont {Seligman}},\
  }\bibfield  {title} {\bibinfo {title} {Transition from localized to extended
  eigenstates in the ensemble of power-law random banded matrices},\ }\href
  {https://doi.org/10.1103/PhysRevE.54.3221} {\bibfield  {journal} {\bibinfo
  {journal} {Phys. Rev. E}\ }\textbf {\bibinfo {volume} {54}},\ \bibinfo
  {pages} {3221} (\bibinfo {year} {1996})}\BibitemShut {NoStop}%
\bibitem [{\citenamefont {Kravtsov}\ \emph {et~al.}(2015)\citenamefont
  {Kravtsov}, \citenamefont {Khaymovich}, \citenamefont {Cuevas},\ and\
  \citenamefont {Amini}}]{kravtsov2015random}%
  \BibitemOpen
  \bibfield  {author} {\bibinfo {author} {\bibfnamefont {V.}~\bibnamefont
  {Kravtsov}}, \bibinfo {author} {\bibfnamefont {I.}~\bibnamefont
  {Khaymovich}}, \bibinfo {author} {\bibfnamefont {E.}~\bibnamefont {Cuevas}},\
  and\ \bibinfo {author} {\bibfnamefont {M.}~\bibnamefont {Amini}},\ }\bibfield
   {title} {\bibinfo {title} {A random matrix model with localization and
  ergodic transitions},\ }\href@noop {} {\bibfield  {journal} {\bibinfo
  {journal} {New J. Phys.}\ }\textbf {\bibinfo {volume} {17}},\ \bibinfo
  {pages} {122002} (\bibinfo {year} {2015})}\BibitemShut {NoStop}%
\bibitem [{\citenamefont {Cugliandolo}\ \emph {et~al.}(2024)\citenamefont
  {Cugliandolo}, \citenamefont {Schehr}, \citenamefont {Tarzia},\ and\
  \citenamefont {Venturelli}}]{CugliandoloPRB2024}%
  \BibitemOpen
  \bibfield  {author} {\bibinfo {author} {\bibfnamefont {L.~F.}\ \bibnamefont
  {Cugliandolo}}, \bibinfo {author} {\bibfnamefont {G.}~\bibnamefont {Schehr}},
  \bibinfo {author} {\bibfnamefont {M.}~\bibnamefont {Tarzia}},\ and\ \bibinfo
  {author} {\bibfnamefont {D.}~\bibnamefont {Venturelli}},\ }\bibfield  {title}
  {\bibinfo {title} {Multifractal phase in the weighted adjacency matrices of
  random {E}rd\"os-{R}\'enyi graphs},\ }\href
  {https://doi.org/10.1103/PhysRevB.110.174202} {\bibfield  {journal} {\bibinfo
   {journal} {Phys. Rev. B}\ }\textbf {\bibinfo {volume} {110}},\ \bibinfo
  {pages} {174202} (\bibinfo {year} {2024})}\BibitemShut {NoStop}%
\bibitem [{\citenamefont {Chalker}(1988)}]{chalker1988scaling}%
  \BibitemOpen
  \bibfield  {author} {\bibinfo {author} {\bibfnamefont {J.~T.}\ \bibnamefont
  {Chalker}},\ }\bibfield  {title} {\bibinfo {title} {Scaling and correlations
  at a mobility edge in two dimensions},\ }\href@noop {} {\bibfield  {journal}
  {\bibinfo  {journal} {Journal of Physics C: Solid State Physics}\ }\textbf
  {\bibinfo {volume} {21}},\ \bibinfo {pages} {L119} (\bibinfo {year}
  {1988})}\BibitemShut {NoStop}%
\bibitem [{\citenamefont {Chalker}(1990)}]{chalker1990scaling}%
  \BibitemOpen
  \bibfield  {author} {\bibinfo {author} {\bibfnamefont {J.~T.}\ \bibnamefont
  {Chalker}},\ }\bibfield  {title} {\bibinfo {title} {Scaling and eigenfunction
  correlations near a mobility edge},\ }\href@noop {} {\bibfield  {journal}
  {\bibinfo  {journal} {Physica A: Statistical Mechanics and its Applications}\
  }\textbf {\bibinfo {volume} {167}},\ \bibinfo {pages} {253} (\bibinfo {year}
  {1990})}\BibitemShut {NoStop}%
\bibitem [{\citenamefont {Ludwig}\ \emph {et~al.}(1994)\citenamefont {Ludwig},
  \citenamefont {Fisher}, \citenamefont {Shankar},\ and\ \citenamefont
  {Grinstein}}]{ludwig1994integer}%
  \BibitemOpen
  \bibfield  {author} {\bibinfo {author} {\bibfnamefont {A.~W.~W.}\
  \bibnamefont {Ludwig}}, \bibinfo {author} {\bibfnamefont {M.~P.~A.}\
  \bibnamefont {Fisher}}, \bibinfo {author} {\bibfnamefont {R.}~\bibnamefont
  {Shankar}},\ and\ \bibinfo {author} {\bibfnamefont {G.}~\bibnamefont
  {Grinstein}},\ }\bibfield  {title} {\bibinfo {title} {{Integer quantum Hall
  transition: An alternative approach and exact results}},\ }\href
  {https://doi.org/10.1103/PhysRevB.50.7526} {\bibfield  {journal} {\bibinfo
  {journal} {Phys. Rev. B}\ }\textbf {\bibinfo {volume} {50}},\ \bibinfo
  {pages} {7526} (\bibinfo {year} {1994})}\BibitemShut {NoStop}%
\bibitem [{\citenamefont {Huckestein}(1995)}]{huckenstein1995scaling}%
  \BibitemOpen
  \bibfield  {author} {\bibinfo {author} {\bibfnamefont {B.}~\bibnamefont
  {Huckestein}},\ }\bibfield  {title} {\bibinfo {title} {{Scaling theory of the
  integer quantum Hall effect}},\ }\href
  {https://doi.org/10.1103/RevModPhys.67.357} {\bibfield  {journal} {\bibinfo
  {journal} {Rev. Mod. Phys.}\ }\textbf {\bibinfo {volume} {67}},\ \bibinfo
  {pages} {357} (\bibinfo {year} {1995})}\BibitemShut {NoStop}%
\bibitem [{\citenamefont {Evers}\ \emph {et~al.}(2001)\citenamefont {Evers},
  \citenamefont {Mildenberger},\ and\ \citenamefont
  {Mirlin}}]{evers2001multifractality}%
  \BibitemOpen
  \bibfield  {author} {\bibinfo {author} {\bibfnamefont {F.}~\bibnamefont
  {Evers}}, \bibinfo {author} {\bibfnamefont {A.}~\bibnamefont
  {Mildenberger}},\ and\ \bibinfo {author} {\bibfnamefont {A.~D.}\ \bibnamefont
  {Mirlin}},\ }\bibfield  {title} {\bibinfo {title} {Multifractality of wave
  functions at the quantum {H}all transition revisited},\ }\href
  {https://doi.org/10.1103/PhysRevB.64.241303} {\bibfield  {journal} {\bibinfo
  {journal} {Phys. Rev. B}\ }\textbf {\bibinfo {volume} {64}},\ \bibinfo
  {pages} {241303} (\bibinfo {year} {2001})}\BibitemShut {NoStop}%
\bibitem [{\citenamefont {Mirlin}\ and\ \citenamefont
  {Evers}(2000)}]{mirlin2000multifractality}%
  \BibitemOpen
  \bibfield  {author} {\bibinfo {author} {\bibfnamefont {A.~D.}\ \bibnamefont
  {Mirlin}}\ and\ \bibinfo {author} {\bibfnamefont {F.}~\bibnamefont {Evers}},\
  }\bibfield  {title} {\bibinfo {title} {Multifractality and critical
  fluctuations at the {A}nderson transition},\ }\href
  {https://doi.org/10.1103/PhysRevB.62.7920} {\bibfield  {journal} {\bibinfo
  {journal} {Phys. Rev. B}\ }\textbf {\bibinfo {volume} {62}},\ \bibinfo
  {pages} {7920} (\bibinfo {year} {2000})}\BibitemShut {NoStop}%
\bibitem [{\citenamefont {Tikhonov}\ and\ \citenamefont
  {Mirlin}(2019)}]{tikhonov2019critical}%
  \BibitemOpen
  \bibfield  {author} {\bibinfo {author} {\bibfnamefont {K.~S.}\ \bibnamefont
  {Tikhonov}}\ and\ \bibinfo {author} {\bibfnamefont {A.~D.}\ \bibnamefont
  {Mirlin}},\ }\bibfield  {title} {\bibinfo {title} {Critical behavior at the
  localization transition on random regular graphs},\ }\href
  {https://doi.org/10.1103/PhysRevB.99.214202} {\bibfield  {journal} {\bibinfo
  {journal} {Phys. Rev. B}\ }\textbf {\bibinfo {volume} {99}},\ \bibinfo
  {pages} {214202} (\bibinfo {year} {2019})}\BibitemShut {NoStop}%
\bibitem [{\citenamefont {Nosov}\ \emph {et~al.}(2019)\citenamefont {Nosov},
  \citenamefont {Khaymovich},\ and\ \citenamefont
  {Kravtsov}}]{nosov2018correlation}%
  \BibitemOpen
  \bibfield  {author} {\bibinfo {author} {\bibfnamefont {P.~A.}\ \bibnamefont
  {Nosov}}, \bibinfo {author} {\bibfnamefont {I.~M.}\ \bibnamefont
  {Khaymovich}},\ and\ \bibinfo {author} {\bibfnamefont {V.~E.}\ \bibnamefont
  {Kravtsov}},\ }\bibfield  {title} {\bibinfo {title} {Correlation-induced
  localization},\ }\href {https://doi.org/10.1103/PhysRevB.99.104203}
  {\bibfield  {journal} {\bibinfo  {journal} {Phys. Rev. B}\ }\textbf {\bibinfo
  {volume} {99}},\ \bibinfo {pages} {104203} (\bibinfo {year}
  {2019})}\BibitemShut {NoStop}%
\bibitem [{\citenamefont {Duthie}\ \emph {et~al.}(2022)\citenamefont {Duthie},
  \citenamefont {Roy},\ and\ \citenamefont {Logan}}]{duthie2022anomalous}%
  \BibitemOpen
  \bibfield  {author} {\bibinfo {author} {\bibfnamefont {A.}~\bibnamefont
  {Duthie}}, \bibinfo {author} {\bibfnamefont {S.}~\bibnamefont {Roy}},\ and\
  \bibinfo {author} {\bibfnamefont {D.~E.}\ \bibnamefont {Logan}},\ }\bibfield
  {title} {\bibinfo {title} {Anomalous multifractality in quantum chains with
  strongly correlated disorder},\ }\href
  {https://doi.org/10.1103/PhysRevB.106.L020201} {\bibfield  {journal}
  {\bibinfo  {journal} {Phys. Rev. B}\ }\textbf {\bibinfo {volume} {106}},\
  \bibinfo {pages} {L020201} (\bibinfo {year} {2022})}\BibitemShut {NoStop}%
\bibitem [{\citenamefont {Zirnbauer}(1986)}]{zirnbauer1986localisation}%
  \BibitemOpen
  \bibfield  {author} {\bibinfo {author} {\bibfnamefont {M.~R.}\ \bibnamefont
  {Zirnbauer}},\ }\bibfield  {title} {\bibinfo {title} {{Localization
  transition on the Bethe lattice}},\ }\href
  {https://doi.org/10.1103/PhysRevB.34.6394} {\bibfield  {journal} {\bibinfo
  {journal} {Phys. Rev. B}\ }\textbf {\bibinfo {volume} {34}},\ \bibinfo
  {pages} {6394} (\bibinfo {year} {1986})}\BibitemShut {NoStop}%
\bibitem [{\citenamefont {Chalker}\ and\ \citenamefont
  {Siak}(1990)}]{chalker1990anderson}%
  \BibitemOpen
  \bibfield  {author} {\bibinfo {author} {\bibfnamefont {J.~T.}\ \bibnamefont
  {Chalker}}\ and\ \bibinfo {author} {\bibfnamefont {S.}~\bibnamefont {Siak}},\
  }\bibfield  {title} {\bibinfo {title} {Anderson localisation on a {C}ayley
  tree: a new model with a simple solution},\ }\href
  {https://doi.org/10.1088/0953-8984/2/11/011} {\bibfield  {journal} {\bibinfo
  {journal} {J. Phys.: Cond. Matt.}\ }\textbf {\bibinfo {volume} {2}},\
  \bibinfo {pages} {2671} (\bibinfo {year} {1990})}\BibitemShut {NoStop}%
\bibitem [{\citenamefont {Mirlin}\ and\ \citenamefont
  {Fyodorov}(1991)}]{mirlin1991localisation}%
  \BibitemOpen
  \bibfield  {author} {\bibinfo {author} {\bibfnamefont {A.~D.}\ \bibnamefont
  {Mirlin}}\ and\ \bibinfo {author} {\bibfnamefont {Y.~V.}\ \bibnamefont
  {Fyodorov}},\ }\bibfield  {title} {\bibinfo {title} {Localization transition
  in the {A}nderson model on the {B}ethe lattice: Spontaneous symmetry breaking
  and correlation functions},\ }\href
  {https://doi.org/https://doi.org/10.1016/0550-3213(91)90028-V} {\bibfield
  {journal} {\bibinfo  {journal} {Nuclear Physics B}\ }\textbf {\bibinfo
  {volume} {366}},\ \bibinfo {pages} {507} (\bibinfo {year}
  {1991})}\BibitemShut {NoStop}%
\bibitem [{\citenamefont {Derrida}\ and\ \citenamefont
  {Rodgers}(1993)}]{derrida1993anderson}%
  \BibitemOpen
  \bibfield  {author} {\bibinfo {author} {\bibfnamefont {B.}~\bibnamefont
  {Derrida}}\ and\ \bibinfo {author} {\bibfnamefont {G.~J.}\ \bibnamefont
  {Rodgers}},\ }\bibfield  {title} {\bibinfo {title} {{Anderson model on a
  Cayley tree: the density of states}},\ }\href
  {https://doi.org/10.1088/0305-4470/26/9/004} {\bibfield  {journal} {\bibinfo
  {journal} {Journal of Physics A: Mathematical and General}\ }\textbf
  {\bibinfo {volume} {26}},\ \bibinfo {pages} {L457} (\bibinfo {year}
  {1993})}\BibitemShut {NoStop}%
\bibitem [{\citenamefont {Monthus}\ and\ \citenamefont
  {Garel}(2008)}]{monthus2008anderson}%
  \BibitemOpen
  \bibfield  {author} {\bibinfo {author} {\bibfnamefont {C.}~\bibnamefont
  {Monthus}}\ and\ \bibinfo {author} {\bibfnamefont {T.}~\bibnamefont
  {Garel}},\ }\bibfield  {title} {\bibinfo {title} {Anderson transition on the
  {C}ayley tree as a traveling wave critical point for various probability
  distributions},\ }\href {https://doi.org/10.1088/1751-8113/42/7/075002}
  {\bibfield  {journal} {\bibinfo  {journal} {Journal of Physics A:
  Mathematical and Theoretical}\ }\textbf {\bibinfo {volume} {42}},\ \bibinfo
  {pages} {075002} (\bibinfo {year} {2008})}\BibitemShut {NoStop}%
\bibitem [{\citenamefont {Monthus}\ and\ \citenamefont
  {Garel}(2011)}]{monthus2011anderson}%
  \BibitemOpen
  \bibfield  {author} {\bibinfo {author} {\bibfnamefont {C.}~\bibnamefont
  {Monthus}}\ and\ \bibinfo {author} {\bibfnamefont {T.}~\bibnamefont
  {Garel}},\ }\bibfield  {title} {\bibinfo {title} {{Anderson localization on
  the Cayley tree: multifractal statistics of the transmission at criticality
  and off criticality}},\ }\href@noop {} {\bibfield  {journal} {\bibinfo
  {journal} {J. Phys. A}\ }\textbf {\bibinfo {volume} {44}},\ \bibinfo {pages}
  {145001} (\bibinfo {year} {2011})}\BibitemShut {NoStop}%
\bibitem [{\citenamefont {Tikhonov}\ and\ \citenamefont
  {Mirlin}(2016)}]{Tikhonov2016CayleyTree}%
  \BibitemOpen
  \bibfield  {author} {\bibinfo {author} {\bibfnamefont {K.~S.}\ \bibnamefont
  {Tikhonov}}\ and\ \bibinfo {author} {\bibfnamefont {A.~D.}\ \bibnamefont
  {Mirlin}},\ }\bibfield  {title} {\bibinfo {title} {Fractality of wave
  functions on a {C}ayley tree: Difference between tree and locally treelike
  graph without boundary},\ }\href {https://doi.org/10.1103/PhysRevB.94.184203}
  {\bibfield  {journal} {\bibinfo  {journal} {Phys. Rev. B}\ }\textbf {\bibinfo
  {volume} {94}},\ \bibinfo {pages} {184203} (\bibinfo {year}
  {2016})}\BibitemShut {NoStop}%
\bibitem [{\citenamefont {Kravtsov}\ \emph {et~al.}(2018)\citenamefont
  {Kravtsov}, \citenamefont {Altshuler},\ and\ \citenamefont
  {Ioffe}}]{kravtsov2018nonergodic}%
  \BibitemOpen
  \bibfield  {author} {\bibinfo {author} {\bibfnamefont {V.}~\bibnamefont
  {Kravtsov}}, \bibinfo {author} {\bibfnamefont {B.}~\bibnamefont
  {Altshuler}},\ and\ \bibinfo {author} {\bibfnamefont {L.}~\bibnamefont
  {Ioffe}},\ }\bibfield  {title} {\bibinfo {title} {{Non-ergodic delocalized
  phase in Anderson model on Bethe lattice and regular graph}},\ }\href
  {https://doi.org/https://doi.org/10.1016/j.aop.2017.12.009} {\bibfield
  {journal} {\bibinfo  {journal} {Annals of Physics}\ }\textbf {\bibinfo
  {volume} {389}},\ \bibinfo {pages} {148} (\bibinfo {year}
  {2018})}\BibitemShut {NoStop}%
\bibitem [{\citenamefont {Savitz}\ \emph {et~al.}(2019)\citenamefont {Savitz},
  \citenamefont {Peng},\ and\ \citenamefont {Refael}}]{savitz2019anderson}%
  \BibitemOpen
  \bibfield  {author} {\bibinfo {author} {\bibfnamefont {S.}~\bibnamefont
  {Savitz}}, \bibinfo {author} {\bibfnamefont {C.}~\bibnamefont {Peng}},\ and\
  \bibinfo {author} {\bibfnamefont {G.}~\bibnamefont {Refael}},\ }\bibfield
  {title} {\bibinfo {title} {Anderson localization on the {B}ethe lattice using
  cages and the {W}egner flow},\ }\href
  {https://doi.org/10.1103/PhysRevB.100.094201} {\bibfield  {journal} {\bibinfo
   {journal} {Phys. Rev. B}\ }\textbf {\bibinfo {volume} {100}},\ \bibinfo
  {pages} {094201} (\bibinfo {year} {2019})}\BibitemShut {NoStop}%
\bibitem [{\citenamefont {Parisi}\ \emph {et~al.}(2019)\citenamefont {Parisi},
  \citenamefont {Pascazio}, \citenamefont {Pietracaprina}, \citenamefont
  {Ros},\ and\ \citenamefont {Scardicchio}}]{parisi2020anderson}%
  \BibitemOpen
  \bibfield  {author} {\bibinfo {author} {\bibfnamefont {G.}~\bibnamefont
  {Parisi}}, \bibinfo {author} {\bibfnamefont {S.}~\bibnamefont {Pascazio}},
  \bibinfo {author} {\bibfnamefont {F.}~\bibnamefont {Pietracaprina}}, \bibinfo
  {author} {\bibfnamefont {V.}~\bibnamefont {Ros}},\ and\ \bibinfo {author}
  {\bibfnamefont {A.}~\bibnamefont {Scardicchio}},\ }\bibfield  {title}
  {\bibinfo {title} {{Anderson transition on the Bethe lattice: an approach
  with real energies}},\ }\href {https://doi.org/10.1088/1751-8121/ab56e8}
  {\bibfield  {journal} {\bibinfo  {journal} {Journal of Physics A:
  Mathematical and Theoretical}\ }\textbf {\bibinfo {volume} {53}},\ \bibinfo
  {pages} {014003} (\bibinfo {year} {2019})}\BibitemShut {NoStop}%
\bibitem [{\citenamefont {Derrida}(1980)}]{derrida1980random}%
  \BibitemOpen
  \bibfield  {author} {\bibinfo {author} {\bibfnamefont {B.}~\bibnamefont
  {Derrida}},\ }\bibfield  {title} {\bibinfo {title} {Random-energy model:
  Limit of a family of disordered models},\ }\href
  {https://doi.org/10.1103/PhysRevLett.45.79} {\bibfield  {journal} {\bibinfo
  {journal} {Phys. Rev. Lett.}\ }\textbf {\bibinfo {volume} {45}},\ \bibinfo
  {pages} {79} (\bibinfo {year} {1980})}\BibitemShut {NoStop}%
\bibitem [{\citenamefont {Laumann}\ \emph {et~al.}(2014)\citenamefont
  {Laumann}, \citenamefont {Pal},\ and\ \citenamefont
  {Scardicchio}}]{laumann2014many}%
  \BibitemOpen
  \bibfield  {author} {\bibinfo {author} {\bibfnamefont {C.~R.}\ \bibnamefont
  {Laumann}}, \bibinfo {author} {\bibfnamefont {A.}~\bibnamefont {Pal}},\ and\
  \bibinfo {author} {\bibfnamefont {A.}~\bibnamefont {Scardicchio}},\
  }\bibfield  {title} {\bibinfo {title} {Many-body mobility edge in a
  mean-field quantum spin glass},\ }\href
  {https://doi.org/10.1103/PhysRevLett.113.200405} {\bibfield  {journal}
  {\bibinfo  {journal} {Phys. Rev. Lett.}\ }\textbf {\bibinfo {volume} {113}},\
  \bibinfo {pages} {200405} (\bibinfo {year} {2014})}\BibitemShut {NoStop}%
\bibitem [{\citenamefont {Baldwin}\ \emph {et~al.}(2016)\citenamefont
  {Baldwin}, \citenamefont {Laumann}, \citenamefont {Pal},\ and\ \citenamefont
  {Scardicchio}}]{baldwin2016manybody}%
  \BibitemOpen
  \bibfield  {author} {\bibinfo {author} {\bibfnamefont {C.~L.}\ \bibnamefont
  {Baldwin}}, \bibinfo {author} {\bibfnamefont {C.~R.}\ \bibnamefont
  {Laumann}}, \bibinfo {author} {\bibfnamefont {A.}~\bibnamefont {Pal}},\ and\
  \bibinfo {author} {\bibfnamefont {A.}~\bibnamefont {Scardicchio}},\
  }\bibfield  {title} {\bibinfo {title} {The many-body localized phase of the
  quantum random energy model},\ }\href
  {https://doi.org/10.1103/PhysRevB.93.024202} {\bibfield  {journal} {\bibinfo
  {journal} {Phys. Rev. B}\ }\textbf {\bibinfo {volume} {93}},\ \bibinfo
  {pages} {024202} (\bibinfo {year} {2016})}\BibitemShut {NoStop}%
\bibitem [{\citenamefont {Aubry}\ and\ \citenamefont
  {Andr{\'e}}(1980)}]{aubry1980analyticity}%
  \BibitemOpen
  \bibfield  {author} {\bibinfo {author} {\bibfnamefont {S.}~\bibnamefont
  {Aubry}}\ and\ \bibinfo {author} {\bibfnamefont {G.}~\bibnamefont
  {Andr{\'e}}},\ }\bibfield  {title} {\bibinfo {title} {Analyticity breaking
  and {A}nderson localization in incommensurate lattices},\ }\href@noop {}
  {\bibfield  {journal} {\bibinfo  {journal} {Ann. Israel Phys. Soc}\ }\textbf
  {\bibinfo {volume} {3}},\ \bibinfo {pages} {18} (\bibinfo {year}
  {1980})}\BibitemShut {NoStop}%
\bibitem [{\citenamefont {Balasubramanian}\ \emph {et~al.}(2022)\citenamefont
  {Balasubramanian}, \citenamefont {Caputa}, \citenamefont {Magan},\ and\
  \citenamefont {Wu}}]{bala2022quantum}%
  \BibitemOpen
  \bibfield  {author} {\bibinfo {author} {\bibfnamefont {V.}~\bibnamefont
  {Balasubramanian}}, \bibinfo {author} {\bibfnamefont {P.}~\bibnamefont
  {Caputa}}, \bibinfo {author} {\bibfnamefont {J.~M.}\ \bibnamefont {Magan}},\
  and\ \bibinfo {author} {\bibfnamefont {Q.}~\bibnamefont {Wu}},\ }\bibfield
  {title} {\bibinfo {title} {Quantum chaos and the complexity of spread of
  states},\ }\href {https://doi.org/10.1103/PhysRevD.106.046007} {\bibfield
  {journal} {\bibinfo  {journal} {Phys. Rev. D}\ }\textbf {\bibinfo {volume}
  {106}},\ \bibinfo {pages} {046007} (\bibinfo {year} {2022})}\BibitemShut
  {NoStop}%
\bibitem [{\citenamefont {Balasubramanian}\ \emph {et~al.}(2023)\citenamefont
  {Balasubramanian}, \citenamefont {Magan},\ and\ \citenamefont
  {Wu}}]{VBalasubEtAlPRD2023}%
  \BibitemOpen
  \bibfield  {author} {\bibinfo {author} {\bibfnamefont {V.}~\bibnamefont
  {Balasubramanian}}, \bibinfo {author} {\bibfnamefont {J.~M.}\ \bibnamefont
  {Magan}},\ and\ \bibinfo {author} {\bibfnamefont {Q.}~\bibnamefont {Wu}},\
  }\bibfield  {title} {\bibinfo {title} {Tridiagonalizing random matrices},\
  }\href {https://doi.org/10.1103/PhysRevD.107.126001} {\bibfield  {journal}
  {\bibinfo  {journal} {Phys. Rev. D}\ }\textbf {\bibinfo {volume} {107}},\
  \bibinfo {pages} {126001} (\bibinfo {year} {2023})}\BibitemShut {NoStop}%
\bibitem [{\citenamefont {Gautam}\ \emph {et~al.}(2024)\citenamefont {Gautam},
  \citenamefont {Pal}, \citenamefont {Pal}, \citenamefont {Gill}, \citenamefont
  {Jaiswal},\ and\ \citenamefont {Sarkar}}]{gautam2024spread}%
  \BibitemOpen
  \bibfield  {author} {\bibinfo {author} {\bibfnamefont {M.}~\bibnamefont
  {Gautam}}, \bibinfo {author} {\bibfnamefont {K.}~\bibnamefont {Pal}},
  \bibinfo {author} {\bibfnamefont {K.}~\bibnamefont {Pal}}, \bibinfo {author}
  {\bibfnamefont {A.}~\bibnamefont {Gill}}, \bibinfo {author} {\bibfnamefont
  {N.}~\bibnamefont {Jaiswal}},\ and\ \bibinfo {author} {\bibfnamefont
  {T.}~\bibnamefont {Sarkar}},\ }\bibfield  {title} {\bibinfo {title} {Spread
  complexity evolution in quenched interacting quantum systems},\ }\href
  {https://doi.org/10.1103/PhysRevB.109.014312} {\bibfield  {journal} {\bibinfo
   {journal} {Phys. Rev. B}\ }\textbf {\bibinfo {volume} {109}},\ \bibinfo
  {pages} {014312} (\bibinfo {year} {2024})}\BibitemShut {NoStop}%
\bibitem [{\citenamefont {Nandy}\ \emph {et~al.}(2025)\citenamefont {Nandy},
  \citenamefont {Matsoukas-Roubeas}, \citenamefont {Martínez-Azcona},
  \citenamefont {Dymarsky},\ and\ \citenamefont {{del Campo}}}]{NANDY2025}%
  \BibitemOpen
  \bibfield  {author} {\bibinfo {author} {\bibfnamefont {P.}~\bibnamefont
  {Nandy}}, \bibinfo {author} {\bibfnamefont {A.~S.}\ \bibnamefont
  {Matsoukas-Roubeas}}, \bibinfo {author} {\bibfnamefont {P.}~\bibnamefont
  {Martínez-Azcona}}, \bibinfo {author} {\bibfnamefont {A.}~\bibnamefont
  {Dymarsky}},\ and\ \bibinfo {author} {\bibfnamefont {A.}~\bibnamefont {{del
  Campo}}},\ }\bibfield  {title} {\bibinfo {title} {Quantum dynamics in
  {K}rylov space: Methods and applications},\ }\href
  {https://doi.org/https://doi.org/10.1016/j.physrep.2025.05.001} {\bibfield
  {journal} {\bibinfo  {journal} {Physics Reports}\ }\textbf {\bibinfo {volume}
  {1125-1128}},\ \bibinfo {pages} {1} (\bibinfo {year} {2025})}\BibitemShut
  {NoStop}%
\bibitem [{\citenamefont {Harper}(1955)}]{Harper_1955}%
  \BibitemOpen
  \bibfield  {author} {\bibinfo {author} {\bibfnamefont {P.~G.}\ \bibnamefont
  {Harper}},\ }\bibfield  {title} {\bibinfo {title} {Single band motion of
  conduction electrons in a uniform magnetic field},\ }\href
  {https://doi.org/10.1088/0370-1298/68/10/304} {\bibfield  {journal} {\bibinfo
   {journal} {Proceedings of the Physical Society. Section A}\ }\textbf
  {\bibinfo {volume} {68}},\ \bibinfo {pages} {874} (\bibinfo {year}
  {1955})}\BibitemShut {NoStop}%
\bibitem [{\citenamefont {Thouless}(1983)}]{thouless1983bandwidths}%
  \BibitemOpen
  \bibfield  {author} {\bibinfo {author} {\bibfnamefont {D.~J.}\ \bibnamefont
  {Thouless}},\ }\bibfield  {title} {\bibinfo {title} {Bandwidths for a
  quasiperiodic tight-binding model},\ }\href
  {https://doi.org/10.1103/PhysRevB.28.4272} {\bibfield  {journal} {\bibinfo
  {journal} {Phys. Rev. B}\ }\textbf {\bibinfo {volume} {28}},\ \bibinfo
  {pages} {4272} (\bibinfo {year} {1983})}\BibitemShut {NoStop}%
\bibitem [{\citenamefont {Prange}\ \emph {et~al.}(1983)\citenamefont {Prange},
  \citenamefont {Grempel},\ and\ \citenamefont {Fishman}}]{prange1983wave}%
  \BibitemOpen
  \bibfield  {author} {\bibinfo {author} {\bibfnamefont {R.~E.}\ \bibnamefont
  {Prange}}, \bibinfo {author} {\bibfnamefont {D.~R.}\ \bibnamefont
  {Grempel}},\ and\ \bibinfo {author} {\bibfnamefont {S.}~\bibnamefont
  {Fishman}},\ }\bibfield  {title} {\bibinfo {title} {Wave functions at a
  mobility edge: An example of a singular continuous spectrum},\ }\href
  {https://doi.org/10.1103/PhysRevB.28.7370} {\bibfield  {journal} {\bibinfo
  {journal} {Phys. Rev. B}\ }\textbf {\bibinfo {volume} {28}},\ \bibinfo
  {pages} {7370} (\bibinfo {year} {1983})}\BibitemShut {NoStop}%
\bibitem [{\citenamefont {Das~Sarma}\ \emph {et~al.}(1988)\citenamefont
  {Das~Sarma}, \citenamefont {He},\ and\ \citenamefont
  {Xie}}]{sarma1988mobility}%
  \BibitemOpen
  \bibfield  {author} {\bibinfo {author} {\bibfnamefont {S.}~\bibnamefont
  {Das~Sarma}}, \bibinfo {author} {\bibfnamefont {S.}~\bibnamefont {He}},\ and\
  \bibinfo {author} {\bibfnamefont {X.~C.}\ \bibnamefont {Xie}},\ }\bibfield
  {title} {\bibinfo {title} {Mobility edge in a model one-dimensional
  potential},\ }\href {https://doi.org/10.1103/PhysRevLett.61.2144} {\bibfield
  {journal} {\bibinfo  {journal} {Phys. Rev. Lett.}\ }\textbf {\bibinfo
  {volume} {61}},\ \bibinfo {pages} {2144} (\bibinfo {year}
  {1988})}\BibitemShut {NoStop}%
\bibitem [{\citenamefont {Boers}\ \emph {et~al.}(2007)\citenamefont {Boers},
  \citenamefont {Goedeke}, \citenamefont {Hinrichs},\ and\ \citenamefont
  {Holthaus}}]{boers2007mobility}%
  \BibitemOpen
  \bibfield  {author} {\bibinfo {author} {\bibfnamefont {D.~J.}\ \bibnamefont
  {Boers}}, \bibinfo {author} {\bibfnamefont {B.}~\bibnamefont {Goedeke}},
  \bibinfo {author} {\bibfnamefont {D.}~\bibnamefont {Hinrichs}},\ and\
  \bibinfo {author} {\bibfnamefont {M.}~\bibnamefont {Holthaus}},\ }\bibfield
  {title} {\bibinfo {title} {Mobility edges in bichromatic optical lattices},\
  }\href {https://doi.org/10.1103/PhysRevA.75.063404} {\bibfield  {journal}
  {\bibinfo  {journal} {Phys. Rev. A}\ }\textbf {\bibinfo {volume} {75}},\
  \bibinfo {pages} {063404} (\bibinfo {year} {2007})}\BibitemShut {NoStop}%
\bibitem [{\citenamefont {Biddle}\ \emph {et~al.}(2009)\citenamefont {Biddle},
  \citenamefont {Wang}, \citenamefont {Priour},\ and\ \citenamefont
  {Das~Sarma}}]{biddle2009localization}%
  \BibitemOpen
  \bibfield  {author} {\bibinfo {author} {\bibfnamefont {J.}~\bibnamefont
  {Biddle}}, \bibinfo {author} {\bibfnamefont {B.}~\bibnamefont {Wang}},
  \bibinfo {author} {\bibfnamefont {D.~J.}\ \bibnamefont {Priour}},\ and\
  \bibinfo {author} {\bibfnamefont {S.}~\bibnamefont {Das~Sarma}},\ }\bibfield
  {title} {\bibinfo {title} {Localization in one-dimensional incommensurate
  lattices beyond the {A}ubry-{A}ndr\'e model},\ }\href
  {https://doi.org/10.1103/PhysRevA.80.021603} {\bibfield  {journal} {\bibinfo
  {journal} {Phys. Rev. A}\ }\textbf {\bibinfo {volume} {80}},\ \bibinfo
  {pages} {021603} (\bibinfo {year} {2009})}\BibitemShut {NoStop}%
\bibitem [{\citenamefont {Ganeshan}\ \emph {et~al.}(2015)\citenamefont
  {Ganeshan}, \citenamefont {Pixley},\ and\ \citenamefont
  {Das~Sarma}}]{ganeshan2015nearest}%
  \BibitemOpen
  \bibfield  {author} {\bibinfo {author} {\bibfnamefont {S.}~\bibnamefont
  {Ganeshan}}, \bibinfo {author} {\bibfnamefont {J.~H.}\ \bibnamefont
  {Pixley}},\ and\ \bibinfo {author} {\bibfnamefont {S.}~\bibnamefont
  {Das~Sarma}},\ }\bibfield  {title} {\bibinfo {title} {Nearest neighbor tight
  binding models with an exact mobility edge in one dimension},\ }\href
  {https://doi.org/10.1103/PhysRevLett.114.146601} {\bibfield  {journal}
  {\bibinfo  {journal} {Phys. Rev. Lett.}\ }\textbf {\bibinfo {volume} {114}},\
  \bibinfo {pages} {146601} (\bibinfo {year} {2015})}\BibitemShut {NoStop}%
\bibitem [{\citenamefont {Yao}\ \emph {et~al.}(2019)\citenamefont {Yao},
  \citenamefont {Khoudli}, \citenamefont {Bresque},\ and\ \citenamefont
  {Sanchez-Palencia}}]{YaoQuasiPRL2019}%
  \BibitemOpen
  \bibfield  {author} {\bibinfo {author} {\bibfnamefont {H.}~\bibnamefont
  {Yao}}, \bibinfo {author} {\bibfnamefont {A.}~\bibnamefont {Khoudli}},
  \bibinfo {author} {\bibfnamefont {L.}~\bibnamefont {Bresque}},\ and\ \bibinfo
  {author} {\bibfnamefont {L.}~\bibnamefont {Sanchez-Palencia}},\ }\bibfield
  {title} {\bibinfo {title} {Critical behavior and fractality in shallow
  one-dimensional quasiperiodic potentials},\ }\href
  {https://doi.org/10.1103/PhysRevLett.123.070405} {\bibfield  {journal}
  {\bibinfo  {journal} {Phys. Rev. Lett.}\ }\textbf {\bibinfo {volume} {123}},\
  \bibinfo {pages} {070405} (\bibinfo {year} {2019})}\BibitemShut {NoStop}%
\bibitem [{\citenamefont {Wang}\ \emph {et~al.}(2020)\citenamefont {Wang},
  \citenamefont {Xia}, \citenamefont {Zhang}, \citenamefont {Yao},
  \citenamefont {Chen}, \citenamefont {You}, \citenamefont {Zhou},\ and\
  \citenamefont {Liu}}]{wang2020onedimensional}%
  \BibitemOpen
  \bibfield  {author} {\bibinfo {author} {\bibfnamefont {Y.}~\bibnamefont
  {Wang}}, \bibinfo {author} {\bibfnamefont {X.}~\bibnamefont {Xia}}, \bibinfo
  {author} {\bibfnamefont {L.}~\bibnamefont {Zhang}}, \bibinfo {author}
  {\bibfnamefont {H.}~\bibnamefont {Yao}}, \bibinfo {author} {\bibfnamefont
  {S.}~\bibnamefont {Chen}}, \bibinfo {author} {\bibfnamefont {J.}~\bibnamefont
  {You}}, \bibinfo {author} {\bibfnamefont {Q.}~\bibnamefont {Zhou}},\ and\
  \bibinfo {author} {\bibfnamefont {X.-J.}\ \bibnamefont {Liu}},\ }\bibfield
  {title} {\bibinfo {title} {One-dimensional quasiperiodic mosaic lattice with
  exact mobility edges},\ }\href
  {https://doi.org/10.1103/PhysRevLett.125.196604} {\bibfield  {journal}
  {\bibinfo  {journal} {Phys. Rev. Lett.}\ }\textbf {\bibinfo {volume} {125}},\
  \bibinfo {pages} {196604} (\bibinfo {year} {2020})}\BibitemShut {NoStop}%
\bibitem [{\citenamefont {Duthie}\ \emph {et~al.}(2021)\citenamefont {Duthie},
  \citenamefont {Roy},\ and\ \citenamefont {Logan}}]{DuthieSelfConPRB2021}%
  \BibitemOpen
  \bibfield  {author} {\bibinfo {author} {\bibfnamefont {A.}~\bibnamefont
  {Duthie}}, \bibinfo {author} {\bibfnamefont {S.}~\bibnamefont {Roy}},\ and\
  \bibinfo {author} {\bibfnamefont {D.~E.}\ \bibnamefont {Logan}},\ }\bibfield
  {title} {\bibinfo {title} {Self-consistent theory of mobility edges in
  quasiperiodic chains},\ }\href {https://doi.org/10.1103/PhysRevB.103.L060201}
  {\bibfield  {journal} {\bibinfo  {journal} {Phys. Rev. B}\ }\textbf {\bibinfo
  {volume} {103}},\ \bibinfo {pages} {L060201} (\bibinfo {year}
  {2021})}\BibitemShut {NoStop}%
\bibitem [{\citenamefont {Economou}(2006)}]{Economoubook}%
  \BibitemOpen
  \bibfield  {author} {\bibinfo {author} {\bibfnamefont {E.~N.}\ \bibnamefont
  {Economou}},\ }\href@noop {} {\emph {\bibinfo {title} {Green's Functions in
  Quantum Physics}}}\ (\bibinfo  {publisher} {Springer},\ \bibinfo {address}
  {Berlin},\ \bibinfo {year} {2006})\BibitemShut {NoStop}%
\end{thebibliography}%
\end{document}